%% file: manuscript_v2a.tex
\documentclass[twocolumn,superscriptaddress,floatfix,preprintnumbers,aps ,nofootinbib,hyperref]{revtex4-2}
\usepackage[utf8]{inputenc}
\usepackage{epsf}
\usepackage{latexsym,amssymb,euscript}
\usepackage[dvips]{graphicx}
\usepackage{amsmath}
\usepackage{nicefrac}
\usepackage{slashed}
\usepackage{booktabs}
\usepackage[linktocpage]{hyperref}
\usepackage{braket}
\usepackage{chngcntr}
\usepackage{bbold}
\usepackage{graphics}
\usepackage{graphicx}
\usepackage{mciteplus}
\usepackage[titletoc]{appendix}
\usepackage[normalem]{ulem}
\graphicspath{{./figures/}}
\hypersetup{
 linktocpage = false,
 urlcolor = urlblue,
 colorlinks = true,
 linkcolor = urlblue,
 anchorcolor = citegreen,
 citecolor = citegreen,
 pdfstartview = {XYZ null null 1.25} 
           }
\usepackage[left=2cm, right=2cm]{geometry}
\usepackage{pstricks}
\usepackage{color}
\usepackage{xcolor}
\definecolor{urlblue}{rgb}{0.2,0.4,0.7}
\definecolor{citegreen}{rgb}{0,0.4,0.2}
\definecolor{linkred}{rgb}{0.9,0.2,0.1}
\usepackage{float}
\usepackage{academicons}
\definecolor{orcidlogocol}{HTML}{A6CE39}
\usepackage{changepage}
\usepackage{fancyhdr}
\usepackage{epsfig}
\usepackage{letltxmacro}
\LetLtxMacro{\oldcite}{\cite}
\renewcommand{\cite}[1]{\mbox{\oldcite{#1}}}

\usepackage{orcidlink}

\newcommand{\drv}{{\rm d}}

\newcommand{\as}{\alpha_s}

\newcommand{\MSb}{\overline{\rm MS}}

\newcommand{\LL}{{\rm LL/LO}}

\newcommand{\NLLp}{{\rm NLL/NLO^+}}

\newcommand{\HENLOp}{{\rm HE}\mbox{-}{\rm NLO^+}}

\newcommand{\DY}{\Delta Y}

\newcommand{\Jpsi}{J/\psi}

\newcommand{\TQQ}{T_{4Q}}
\newcommand{\TQc}{T_{4c}}
\newcommand{\TQcZpp}{T_{4c}(0^{++})}
\newcommand{\TQcOpm}{T_{4c}(1^{+-})}
\newcommand{\TQcTpp}{T_{4c}(2^{++})}
\newcommand{\TQb}{T_{4b}}

\newcommand{\TQbOpm}{T_{4b}(1^{+-})}

\newcommand{{\HFNRevo}}{\tt HF-NRevo}

\newcommand{{\Jethad}}{\tt JETHAD}
\newcommand{{\symJethad}}{\tt symJETHAD}
\newcommand{{\psymJethad}}{\tt (sym)JETHAD}
\newcommand{{\Hell}}{\tt HELL}
\newcommand{{\RadISH}}{\tt RadISH}
\newcommand{{\Pegasus}}{\tt QCD-PEGASUS}
\newcommand{{\HOPPET}}{\tt HOPPET}
\newcommand{{\QCDNUM}}{\tt QCDNUM}
\newcommand{{\APFEL}}{\tt APFEL}
\newcommand{{\APFELpp}}{\tt APFEL++}
\newcommand{{\APFELppp}}{\tt APFEL(++)}
\newcommand{{\EKO}}{\tt EKO}
\newcommand{{\FeynCalc}}{\tt FeynCalc}

\allowdisplaybreaks

\bibpunct{[}{]}{,}{n}{}{,}

\begin{document}

\title{Fragmentation functions for axial-vector heavy tetraquarks: A TQ4Q1.1 update}

\author{Francesco~Giovanni~Celiberto\,\orcidlink{0000-0003-3299-2203}} 
\email{francesco.celiberto@uah.es}
\affiliation{Universidad de Alcal\'a (UAH), Departamento de F\'isica y Matem\'aticas, E-28805 Alcal\'a de Henares, Madrid, Spain}

%==========================

\begin{abstract}
We present and discuss the release of novel sets of collinear, variable-flavor-number-scheme fragmentation functions for axial-vector, fully heavy $T_{4c}(1^{+-})$ and $T_{4b}(1^{+-})$ tetraquarks. 
Working within the single-parton approximation at leading power and employing nonrelativistic QCD factorization adapted for tetraquark Fock states, we model the initial-scale fragmentation input using a recent calculation for the constituent heavy-quark channel.
Standard DGLAP evolution is then applied, ensuring consistent implementation of evolution thresholds.
To support phenomenology, we investigate NLL/NLO$^+$ rates for tetraquark-jet systems at the HL-LHC and FCC from {\psymJethad}. 
This study further integrates exotic matter explorations with precision QCD calculations.
\end{abstract}

\maketitle

%==========================
\textbf{\textit{Introduction.}}
%==========================
Can high-energy precision Quantum Chromodynamics (QCD) provide a coherent framework for describing the production of exotic matter at colliders?
Despite significant progress, the fundamental mechanisms driving exotic matter formation remain an open question. 
However, recent advancements in (all-order) perturbative techniques and QCD factorization could offer new and unexpected perspectives.
Mesons and baryons, made of minimal valence-quark configurations, are \emph{ordinary} hadrons. QCD color neutrality, however, allows bound states with different valence content. When their quantum numbers defy conventional models, they are called \emph{exotics}~\cite{Esposito:2016noz,Olsen:2017bmm,Brambilla:2019esw,Karliner:2017qhf,Lebed:2016hpi}. Their structure is still debated. Two main scenarios exist: lowest Fock states with active gluons (\emph{hybrids}~\cite{Kou:2005gt,Braaten:2013boa,Berwein:2015vca} and \emph{glueballs}~\cite{Minkowski:1998mf,Mathieu:2008me,Chen:2021cjr,D0:2020tig}) and compact \emph{multiquark} states~\cite{Gell-Mann:1964ewy,Jaffe:1976ig,Ader:1981db,Maiani:2015vwa}.
%%%%%%%
The first exotic hadron, $X(3872)$, was discovered by Belle in 2003~\cite{Belle:2003nnu} and later confirmed by multiple experiments. 
Classified as a hidden-flavor state made of heavy-quark pairs~\cite{Chen:2016qju,Liu:2019zoy,Esposito:2020ywk}, its discovery marked the beginning of the Second Quarkonium Revolution, following the first triggered by the $\Jpsi$ in 1974. Although $X(3872)$ has nonexotic quantum numbers, its isospin-violating decays suggest a structure beyond pure quarkonium~\cite{Esposito:2025hlp}. 
Competing interpretations include a meson molecule~\cite{Tornqvist:1993ng,Braaten:2003he,Guo:2013sya,Wang:2014gwa,Mutuk:2022ckn,Esposito:2023mxw}, a double-diquark configuration~\cite{Maiani:2004vq,tHooft:2008rus,Maiani:2017kyi,Wang:2013exa,Grinstein:2024rcu}, or a hadroquarkonium composed of a quarkonium core and an orbiting light meson~\cite{Dubynskiy:2008mq,Voloshin:2013dpa,Guo:2017jvc,Ferretti:2018ojb,Ferretti:2020ewe}.
%%%%%%%
For years, $X(3872)$ was the only exotic hadron observed in prompt proton collisions, until the discovery of the doubly charmed $T_{cc}^+$~\cite{LHCb:2021vvq,LHCb:2021auc}. 
The $X(6900)$ resonance~\cite{LHCb:2020bwg} is considered a candidate for the $[J^{PC}=0^{++}]$ ground or the $[2^{++}]$ radial state of the fully charmed tetraquark $\TQc$~\cite{Chen:2022asf,Belov:2024qyi}. 
Since the heavy-quark mass $m_Q$ exceeds the perturbative threshold, a fully heavy tetraquark, $\TQQ$, can be described as a nonrelativistic $|Q\bar{Q}Q\bar{Q}\rangle$ state, similar to $|Q\bar{Q}\rangle$ quarkonia. 
Thus, methods developed for quarkonia extend naturally to heavy tetraquarks. As charmonia are QCD ``hydrogen atoms''~\cite{Pineda:2011dg}, $\TQc$ states may resemble QCD ``heavier nuclei'' or ``molecules''~\cite{Wu:2016vtq}.
%%%%%%%
Exotic hadron production remains poorly understood, despite progress on their spectra and decays since $X(3872)$. 
Only model-dependent methods, like color evaporation~\cite{Maciula:2020wri}, multiparticle interactions~\cite{Carvalho:2015nqf,Abreu:2023wwg}, and high-energy dynamics~\cite{Cisek:2022uqx}, have been explored.
Large-$p_T$ $X(3872)$ yields at the LHC~\cite{CMS:2013fpt,ATLAS:2016kwu,LHCb:2021ten} suggest that structural scenarios may act as initial-scale proxies for QCD-based mechanisms like single-parton \emph{fragmentation}.
%%%%%%%
A new set of collinear fragmentation functions (FFs), {\tt TQ4Q1.1}, was recently introduced~\cite{Celiberto:2024beg} to describe $S$-wave color-singlet $\TQQ(0^{++})$ and $\TQQ(2^{++})$ production at moderate-to-large $p_T$ within the variable-flavor number scheme (VFNS)~\cite{Mele:1990cw,Cacciari:1993mq}. 
These FFs are based on initial-scale nonrelativistic QCD (NRQCD) inputs for the gluon~\cite{Feng:2020riv} and charm~\cite{Bai:2024ezn} channels~\cite{Caswell:1985ui,Bodwin:1994jh}.
Leveraging key aspects of the newly developed heavy-flavor nonrelativistic evolution ({\HFNRevo}) scheme~\cite{Celiberto:2024mex,Celiberto:2024bxu}, a proper DGLAP evolution of these inputs is performed, ensuring a consistent treatment of all parton thresholds.

In this \textit{Letter}, we present new FFs for axial-vector heavy tetraquarks, $\TQQ(1^{+-})$, extending them to both charmed and bottomed states. 
Tetraquarks with [$J=1^{+-}$] are of special interest due to their structural richness: they may appear as axial-vector states or as mixed configurations involving quarkonia and tetraquarks. 
In the light sector, the active role of chiral symmetry breaking enhances $[1^{++} \leftrightarrow 1^{+-}]$ mixing~\cite{Giacosa:2006tf,Kim:2017yvd,Kim:2018zob,Zhang:2025xee,Swanson:2023zlm}, making pure tetraquark states hard to isolate~\cite{Wang:2008mw,Kim:2022qfj}.
In contrast, $1^{+-}$ fully heavy tetraquarks experience strongly suppressed mixing.  
Heavy-quark spin symmetry (HQSS)~\cite{Isgur:1991wq,Neubert:1993mb} implies weak spin-dependent interactions between charm and bottom quarks, reducing $[1^{++} \leftrightarrow 1^{+-}]$ overlap~\cite{Weng:2020jao}.  
Lattice QCD and potential models confirm a rigid spin structure, with $1^{+-}$ nearly a pure eigenstate~\cite{An:2022qpt}.  
Unlike heavy-light tetraquarks, which mix moderately~\cite{liu:2020eha}, they offer a clean probe of HQSS and exotic production mechanisms.
EIC~\cite{AbdulKhalek:2021gbh} predictions for $\TQcOpm$ photoproduction, based on single-parton~\cite{Bai:2024ezn} or two-parton~\cite{Feng:2023ghc} nonrelativistic QCD (NRQCD) fragmentation, indicate promising yields.  
In hadroproduction, $\TQcOpm$ rates at the LHC are lower than for $\TQcZpp$ and $\TQcTpp$, but still significant~\cite{Feng:2023agq}.  
A preliminary {\tt TQ4Q1.1} update was used to describe indirect production via Higgs and electroweak decays~\cite{Ma:2025ryo}.
A full VFNS set of DGLAP-evolving FFs for $0^{++}$, $1^{+-}$, and $2^{++}$ heavy tetraquarks provides a robust tool to investigate their high-energy production mechanisms.
In this light, the timely release of the {\tt TQ4Q1.1} update~\cite{Celiberto:2025_TQ4Q11_AVT} provides a valuable resource to the Community, driving progress in this rapidly evolving field.
%%%%%%%
Axial-vector fully heavy tetraquarks exhibit unique dynamical and phenomenological features.  
Unlike scalar and tensor states, often affected by mixing with quarkonia or molecular configurations~\cite{Chen:2020xwe,Esposito:2016noz,Kim:2022qfj}, they offer a cleaner probe of multiquark dynamics~\cite{Zhu:2020xni,Weng:2020jao,An:2022qpt}.
Their leading-order production is suppressed by the LandauYang theorem and $C$-parity selection rules~\cite{Karliner:2020dta}, making them ideal to isolate strong-interaction effects.
Theoretical studies~\cite{Becchi:2020mjz,Feng:2023agq,Karliner:2017qjm} further highlight their distinctive production and decay patterns.  
Mixing is reduced in the fully heavy sector due to HQSS~\cite{Kim:2017yvd,Zhu:2020xni}.
This motivates a dedicated exploration of the $1^{+-}$ channel, especially at the High-Luminosity LHC (HL-LHC) and Future Circular Collider (FCC), where rare exotic states may become accessible.  
Accurate theoretical predictions are essential to support these searches.

\vspace{0.10cm}
%==========================
\textbf{\textit{{\tt TQ4Q1.1} functions for $\boldsymbol{T_{4Q}(1^{+-})}$ tetraquarks.}}
%==========================

Heavy-flavored hadron fragmentation differs substantially from the light sector due to the large heavy-quark masses, which lie above the perturbative-QCD threshold.  
While light-hadron FFs are purely nonperturbative, heavy-hadron FFs include perturbative parts~\cite{Cacciari:1993mq,Jaffe:1993ie,Helenius:2018uul,Bonino:2023icn,Cacciari:2024kaa,Czakon:2024tjr}.  
For $D$, $B$, and $\Lambda_{c,b}$ hadrons, fragmentation proceeds in two steps: perturbative parton-to-$Q$ splitting~\cite{Mele:1990cw}, followed by nonperturbative hadronization.  
To build VFNS FFs, DGLAP evolution must be applied to the nonperturbative inputs, yielding $\mu_F$-dependent distributions at fixed accuracy.
%%%%%%%
The fragmentation \emph{Ansatz} for heavy-light hadrons becomes a formal property in quarkonium studies via NRQCD, which factorizes short- and long-distance effects into short-distance coefficients (SDCs) and long-distance matrix elements (LDMEs).  
At high $p_T$, single-parton fragmentation dominates, despite starting at higher order, due to a $[p_T/m_Q]^2$ enhancement~\cite{Braaten:1993rw,Cacciari:1994dr,Zheng:2019dfk,Zhang:2018mlo,Artoisenet:2014lpa}, surpassing two-parton mechanisms~\cite{Fleming:2012wy,Kang:2014tta,Boer:2023zit}.  
NLO NRQCD input enabled the first VFNS FFs for vector quarkonia~\cite{Celiberto:2022dyf} and $B_c(^1S_0,^3S_1)$ mesons~\cite{Celiberto:2022keu,Celiberto:2024omj}.
%%%
\begin{figure}[!t]
\centering
\includegraphics[width=0.330\textwidth]{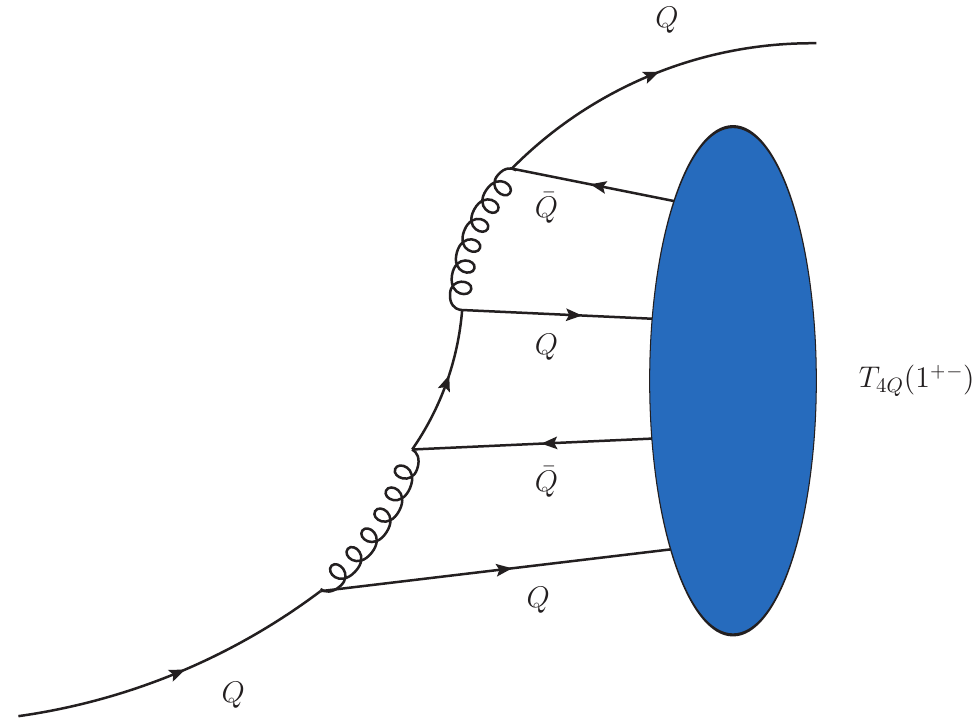}
%\hspace{0.50cm}

\caption{LO diagram for $S$-wave color-singlet $[Q \to \TQQ(1^{+-})]$ fragmentation at $\mu_{F,0}$.
The SDC is shown on the left. The blue oval denotes the LDME.
The $[\bar{Q} \to \TQQ(1^{+-})]$ channel is omitted.}
\label{fig:FF_diagram}
\end{figure}
%%%
NRQCD factorization provides a robust framework to model $\TQQ$ fragmentation~\cite{Zhang:2020hoh,Zhu:2020xni}.  
Initial-scale leading-order inputs for parton-to-$\TQQ$ fragmentation were computed in~\cite{Feng:2020riv,Bai:2024ezn,Bai:2024flh}.  
Using these as model proxies, the first VFNS FFs for $\TQQ(0^{++})$ and $\TQQ(2^{++})$ were released in {\tt TQ4Q1.1}, extending {\tt TQ4Q1.0}~\cite{Celiberto:2024mab}, which was based on the Suzuki model~\cite{Nejad:2021mmp,Celiberto:2023rzw}.
For a fully heavy tetraquark, $\TQQ(J^{PC})$, with total angular momentum, parity, and charge $J^{PC}$,
the LO NRQCD initial-scale input for the $[i \to \TQQ]$ collinear FF reads
\begin{equation}
\begin{split}
 \label{eq:TQQ_FF_initial-scale}
 D^{\TQQ(J^{PC})}_i(z,\mu_{F,0}) \, &= \,
 \frac{1}{m_Q^9}
 \sum_{[\tau]} 
 \mathcal{S}^{(J^{PC})}_i(z,[\tau]) \\
 &\times \, \langle {\cal O}^{\TQQ(J^{PC})}([\tau]) \rangle
 \;,
\end{split}
\end{equation}
with $i$ a generic parton.
Here, $\mathcal{S}^{(J^{PC})}_i(z,[\tau])$ represents the SDC describing the perturbative component of the $[i \to (Q\bar{Q}Q\bar{Q})]$ fragmentation, while $\langle {\cal O}^{\TQQ(J^{PC})}([\tau]) \rangle$ denotes the color-composite LDMEs~\cite{Feng:2020riv} characterizing the purely nonperturbative hadronization of $\TQQ(J^{PC})$.
The composite quantum number $[\tau]$ spans the configurations: $[3,3]$, $[6,6]$, $[3,6]$, and $[6,3]$.
In the diquark-antidiquark basis, the color-singlet tetraquark appears as $[\bar{3} \otimes 3]$ or $[6 \otimes \bar{6}]$.  
Fermi-Dirac statistics allow spin $0$, $1$, or $2$ for $[\bar{3} \otimes 3]$, but only spin $0$ for $[6 \otimes \bar{6}]$.  
Thus, only the $[3,3]$ channel contributes to the fragmentation of the axial-vector $\TQQ(1^{+-})$.
%%%%%%%
The gluon FF is suppressed at leading order (LO).  
Because of the Landau-Yang theorem, an on-shell, color-singlet gluon cannot fragment into a $1^{+-}$ tetraquark, as it cannot decay into two identical massless vector bosons.  
In contrast, transitions into scalar $[0^{++}]$ or tensor $[2^{++}]$ states are allowed.  
The nonconstituent (light or heavy) quark FF is also suppressed at LO, not by Landau-Yang, but by $C$-parity conservation~\cite{Bai:2024flh}.  
Hence, the only allowed LO channel is $[Q \to \TQQ(1^{+-})]$ (Fig.~\ref{fig:FF_diagram}).  
Full symmetry between $Q$ and $\bar{Q}$ channels is assumed.
%%%
\begin{figure*}[!t]
\centering

   \hspace{-0.00cm}
   \includegraphics[scale=0.360,clip]{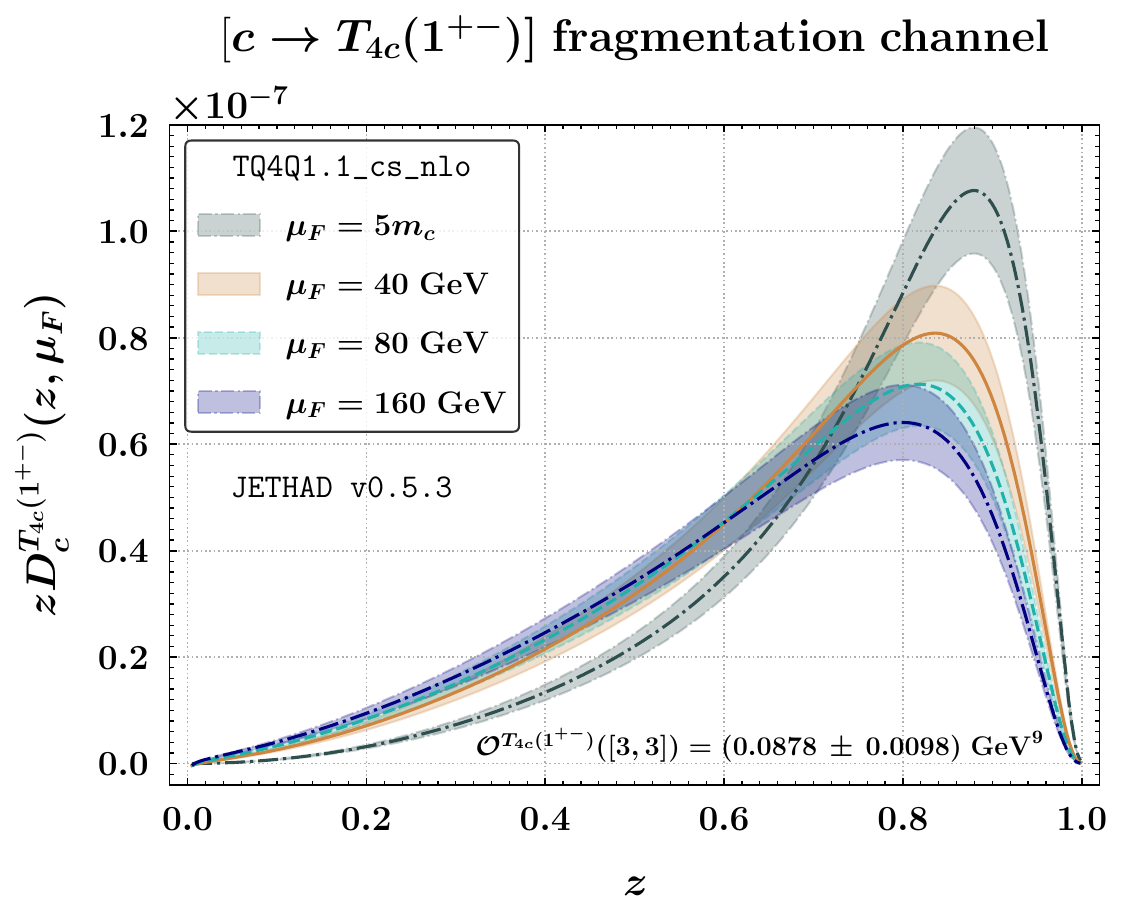}
   \hspace{0.90cm}
   \includegraphics[scale=0.360,clip]{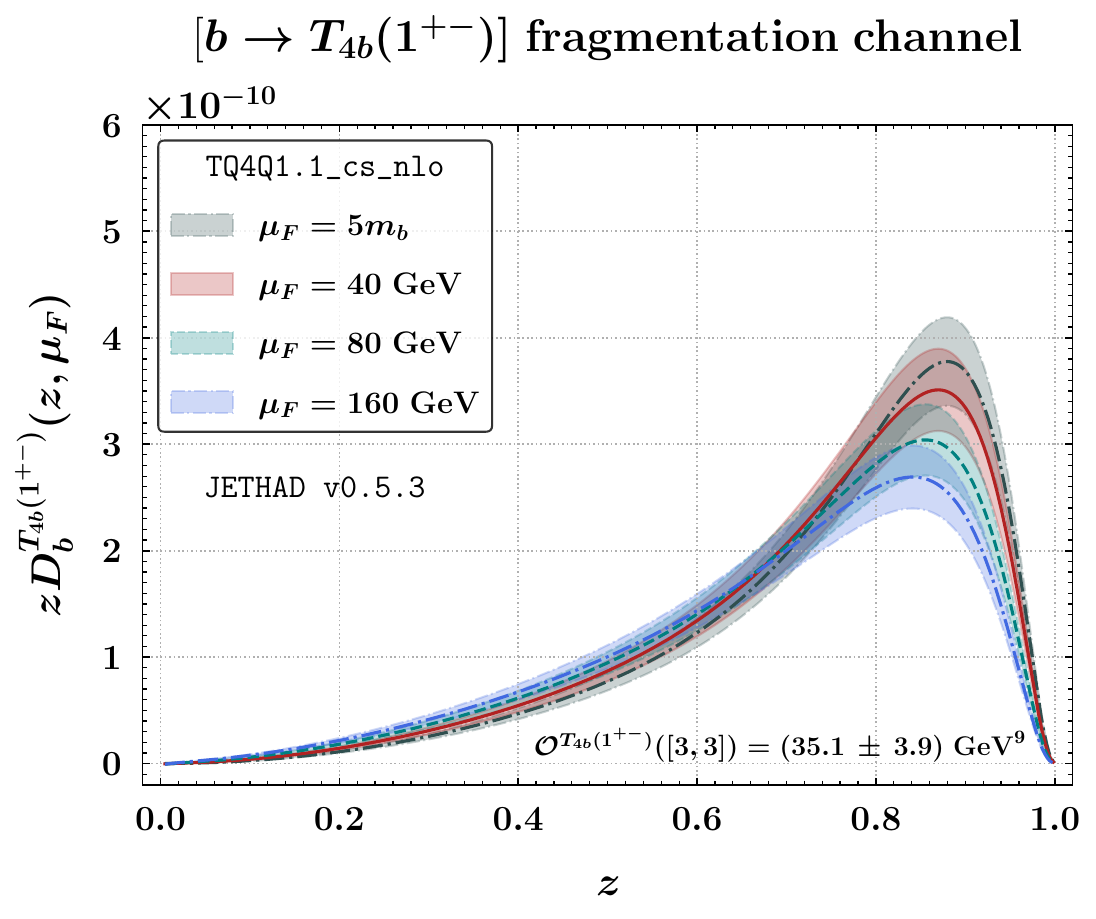}
%   \hspace{0.05cm}

   %\vspace{0.25cm}

   \hspace{-0.00cm}
   \includegraphics[scale=0.360,clip]{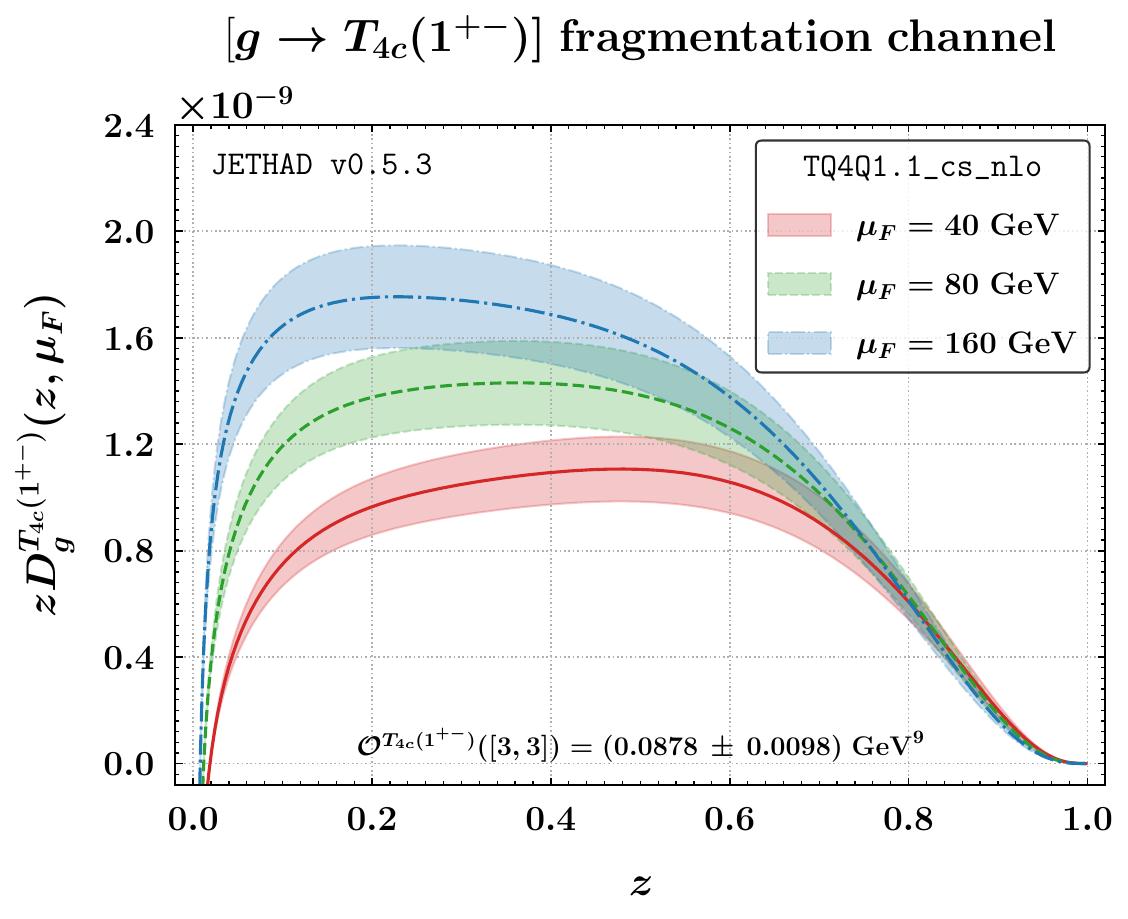}
   \hspace{0.90cm}
   \includegraphics[scale=0.360,clip]{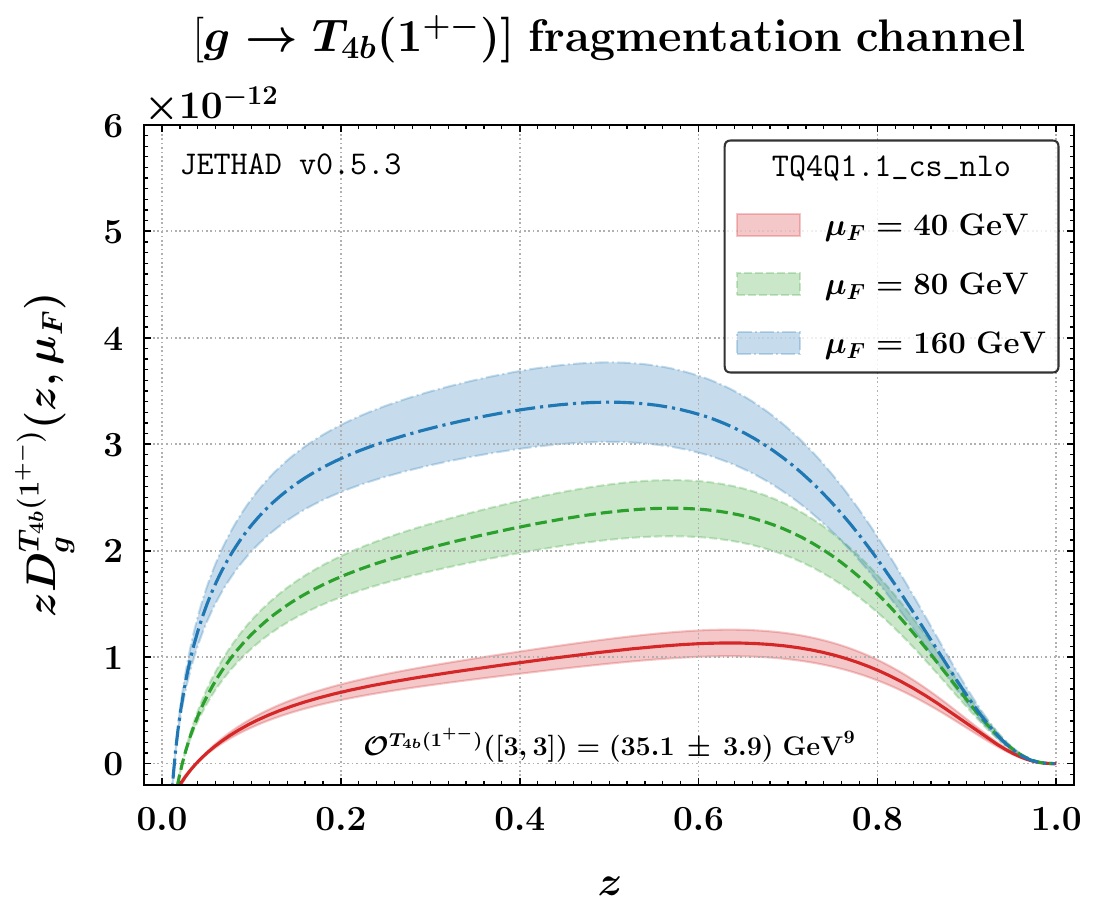}
%   \hspace{0.05cm}

\caption{$z$ dependence of {\tt TQ4Q1.1} FFs for $\TQcOpm$ (left) and $\TQbOpm$ (right) at different $\mu_F$ values.
Upper (lower) plots show the heavy-quark (gluon) channel.
Shaded bands indicate LDME uncertainty.}
\label{fig:FFs-z_TQ1}
\end{figure*}
%%%
\begin{figure}[!t]
\centering

   \hspace{-0.00cm} 
   \includegraphics[scale=0.350,clip]{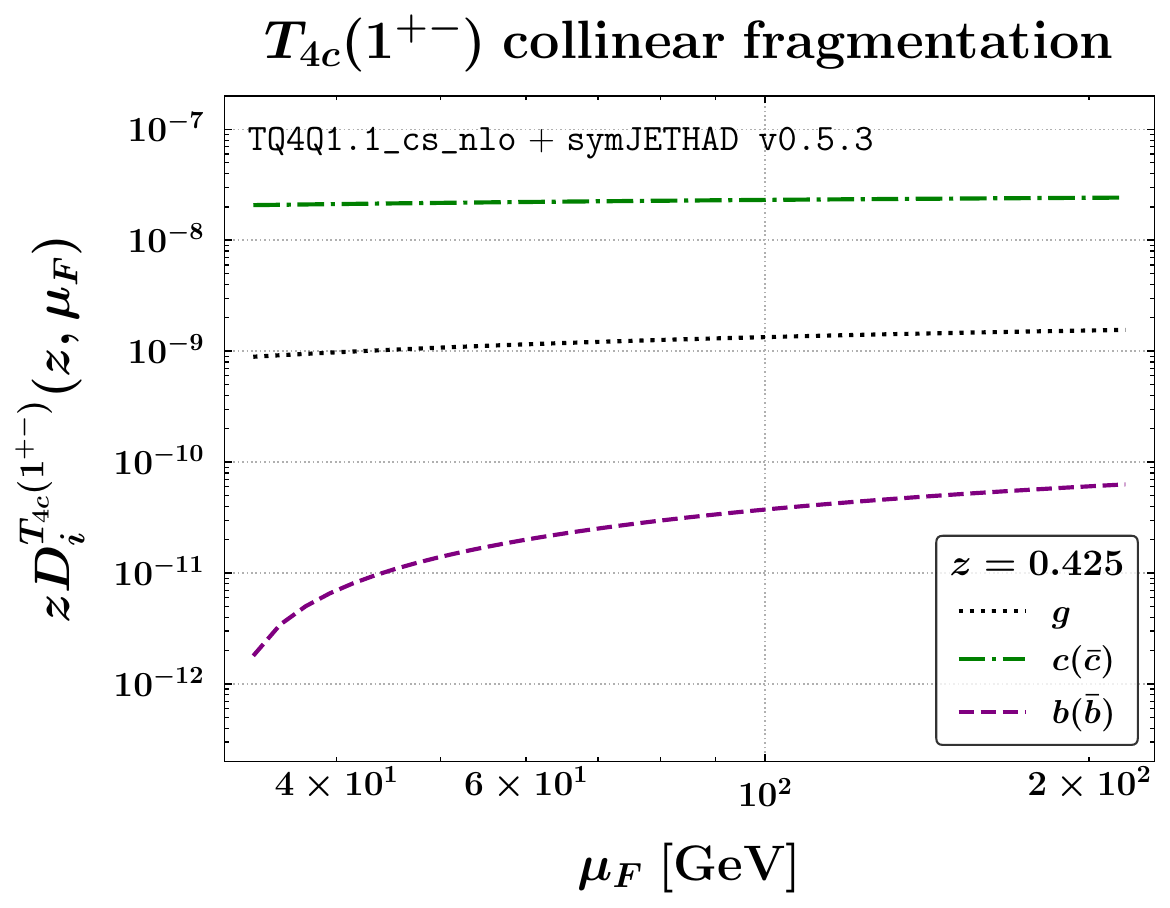}%\vspace{0.35cm}
   
   %\hspace{0.10cm}
   \includegraphics[scale=0.350,clip]{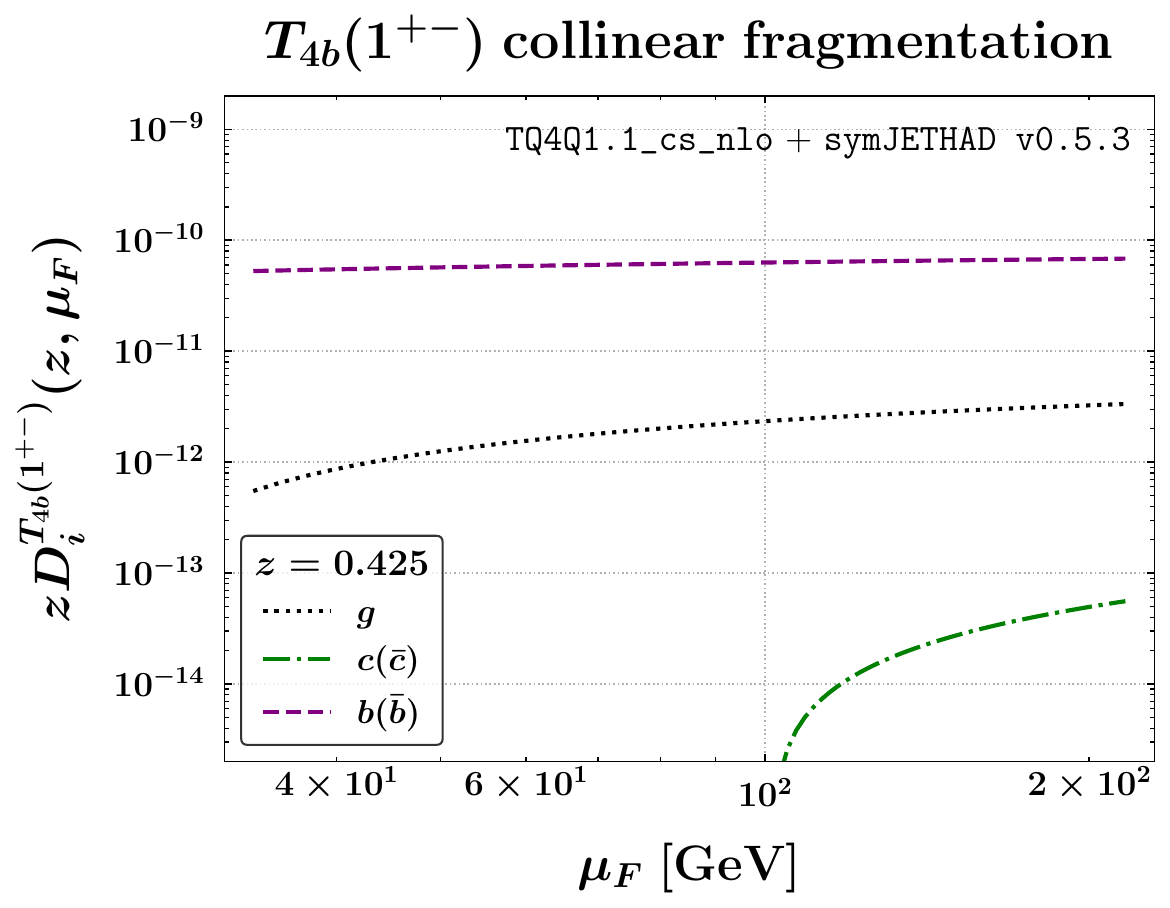}
%   \hspace{0.05cm}

\caption{Energy dependence of {\tt TQ4Q1.1} FFs for the $\TQcOpm$ (upper) and $\TQbOpm$ (lower) states at $z = \langle z \rangle \simeq 0.425$.}
\label{fig:FFs-muF_TQ1}
\end{figure}
%%%
The $[Q \to \TQQ(1^{+-})]$ SDC was first computed in~\cite{Bai:2024ezn} and rederived via {\psymJethad}~\cite{Celiberto:2020wpk,Celiberto:2022rfj,Celiberto:2024mrq}.  
Its full form is given in Eq.~(S1) of~\cite{Celiberto:2025_TQ4Q11_AVT_suppl}.  
LDMEs are based on Cornell-inspired potential models~\cite{Feng:2020riv,Bai:2024ezn}.  
To estimate uncertainty, we average results from~\cite{Lu:2020cns} and~\cite{Yu:2022lak}, following validation in~\cite{Celiberto:2024mab,Celiberto:2024beg}, and use their spread as variation:
\begin{eqnarray}
\begin{aligned}
\label{eq:LDMEs_TQ1}
{\cal O}^{\TQcOpm}([3,3]) &\,=\, (0.0878\,\pm\,0.0098)\mbox{ GeV}^9 \;,
 \\%[0.15cm]
 {\cal O}^{\TQbOpm}([3,3]) &\,\simeq\, (35.1\,\pm\,3.9)\mbox{ GeV}^9 \;. 
\end{aligned}
\end{eqnarray}
More details on LDMEs are given in~\cite{Celiberto:2025_TQ4Q11_AVT_suppl}, \S{1}.
%%%%%%%
The final step in building our {\tt TQ4Q1.1} collinear FFs for $\TQQ(1^{+-})$ states~\cite{Celiberto:2025_TQ4Q11_AVT} is implementing a threshold-consistent DGLAP evolution.
Kinematics sets $\mu_{F,0} = 5 m_Q$ as the minimal invariant mass for the $[Q \to (Q\bar{Q}Q\bar{Q}) + Q]$ splitting in Fig.~\ref{fig:FF_diagram}, which we adopt as the threshold for $Q$ fragmentation.
In the {\HFNRevo} scheme~\cite{Celiberto:2024mex,Celiberto:2024bxu}, DGLAP evolution involves two steps: an expanded and semianalytic decoupled evolution ({\tt EDevo}), which handles channel-dependent thresholds, followed by a numerically implemented all-order evolution ({\tt AOevo}).
However, since only the $[Q \to \TQQ(1^{+-})]$ channel is modeled at $\mu_{F,0}$, we skip {\tt EDevo} and directly apply {\tt AOevo}.
Starting from the $[Q \to \TQQ(1^{+-})]$ input defined at the \emph{evolution-ready} scale $\mu_{F,0}$, we construct the {\tt TQ4Q1.1} sets for axial-vector heavy tetraquarks via NLO DGLAP evolution using {\APFELpp}~\cite{Bertone:2013vaa}, and release them in {\tt LHAPDF} format~\cite{Buckley:2014ana}.
Plots in Fig.~\ref{fig:FFs-z_TQ1} show the $z$ dependence of the $Q$ quark (upper) and gluon (lower) FFs to $\TQcOpm$ (left) and $\TQbOpm$ (right).
We consider representative factorization scales: $\mu_{F,0} \equiv 5m_Q$, 40, 80, and 160~GeV.
The moderate-$z$ peak rises with energy, as expected from DGLAP evolution.
At $\mu_{F,0}$, only the $Q$-quark FF is nonzero, since the gluon FF vanishes at this scale and emerges only above $\mu_{F,0}$.
We use $m_c = 1.5$ GeV and $m_b = 4.9$ GeV.
$Q$-quark FFs peak near $z \simeq 0.85$, consistent with heavy-light trends~\cite{Suzuki:1977km,Bjorken:1977md}.
Gluon FFs show a broad distribution at moderate $z$, with a sharp rise at low $z$ and slow falloff at large $z$.
Both channels exhibit shapes different from scalar and tensor cases~\cite{Celiberto:2025_TQ4Q11_AVT_suppl}, highlighting the dynamics of axial states.
FFs for $1^{+-}$ are suppressed compared to $0^{++}$ and $2^{++}$, particularly for gluons.
This pattern follows from NRQCD expectations based on HQSS, selection rules, and production dynamics.
HQSS enhances symmetric diquark-antidiquark combinations, favoring $0^{++}$ and $2^{++}$ over the antisymmetric $1^{+-}$~\cite{Weng:2020jao}.
Collinear NRQCD selection rules also disfavor $1^{+-}$, often requiring orbital excitation, unlike the favored $S$-wave or spin-aligned channels~\cite{Bodwin:2002cfe,Ma:2015yka,Xu:2021mju}.
Production mechanisms are also less efficient, as $1^{+-}$ states couple more weakly to LO gluon channels~\cite{Wang:2008mw,Feng:2023agq,LHCb:2024ybz}.
A softer fragmentation spectrum further reduces high-$p_T$ production~\cite{Ma:2015yka,Faustov:2022mvs}.
This suppression hierarchy is thus well motivated within NRQCD.
Figure~\ref{fig:FFs-muF_TQ1} shows gluon, charm, and bottom FFs at $z \simeq 0.425$, a typical hadroproduction value~\cite{Celiberto:2020wpk,Celiberto:2021dzy,Celiberto:2021fdp,Celiberto:2022dyf,Celiberto:2022keu,Celiberto:2024omj}.
Charm and bottom FFs dominate, while the gluon FF grows with $\mu_F$.
This behavior stabilizes high-energy observables against NLO corrections and MHOUs.
One might question the absence of initial-scale inputs for light or nonconstituent heavy partons.
However, since these SDCs vanish at LO and no NLO results exist, our NLO-evolved approach is the most complete framework available for $\TQQ(1^{+-})$.
Moreover, the gluon FF vanishes as $z \to 1$, as required for leading-power channels.
Thus, our method overcomes the well-known issue of NRQCD gluon FFs, which often fail to satisfy this condition~\cite{Braaten:1993rw,Artoisenet:2014lpa,Feng:2020riv}.

\begin{figure}[!t]
\centering

   \hspace{0.00cm}
   \includegraphics[scale=0.330,clip]{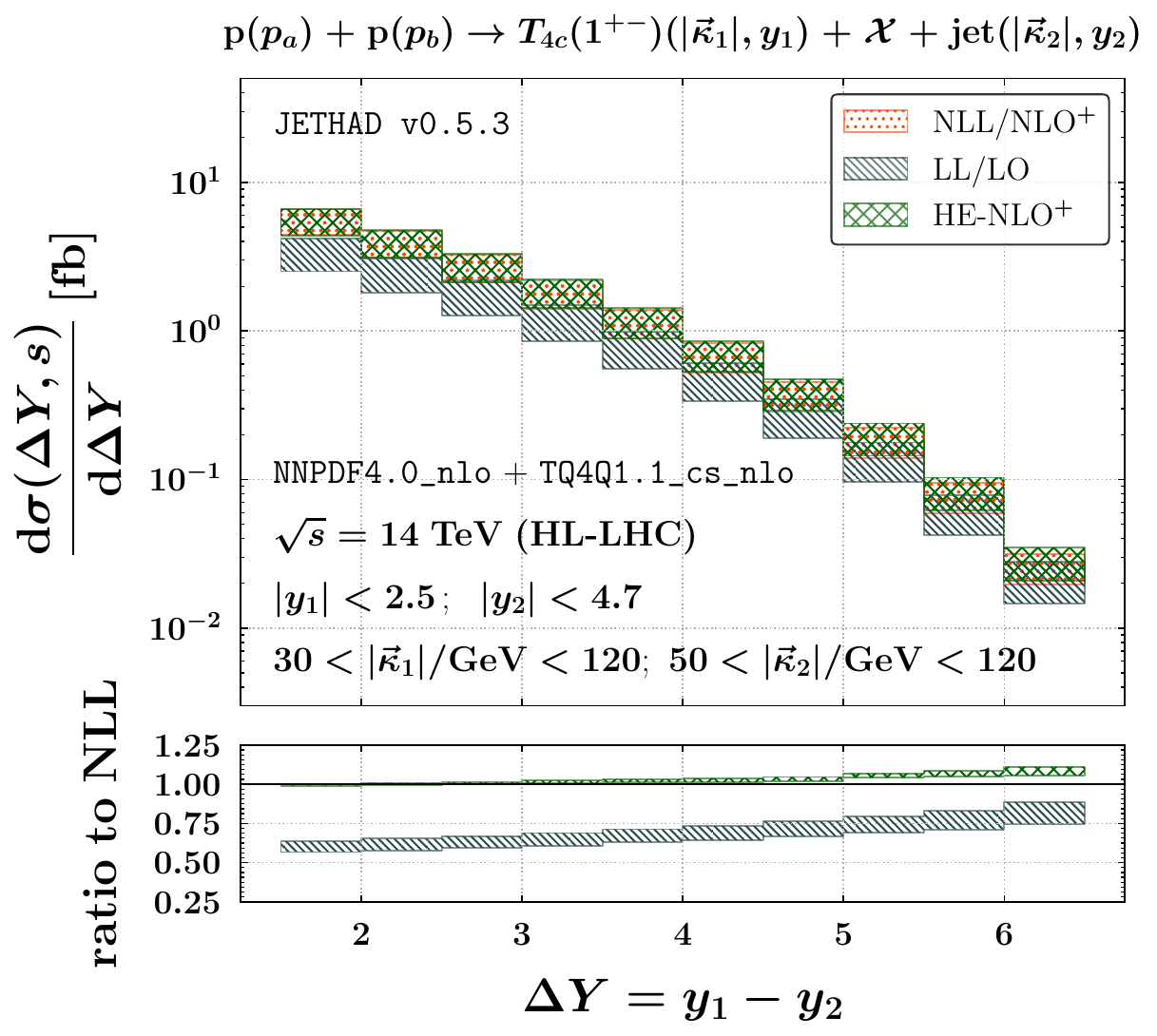}
   %\hspace{0.90cm}
   \includegraphics[scale=0.330,clip]{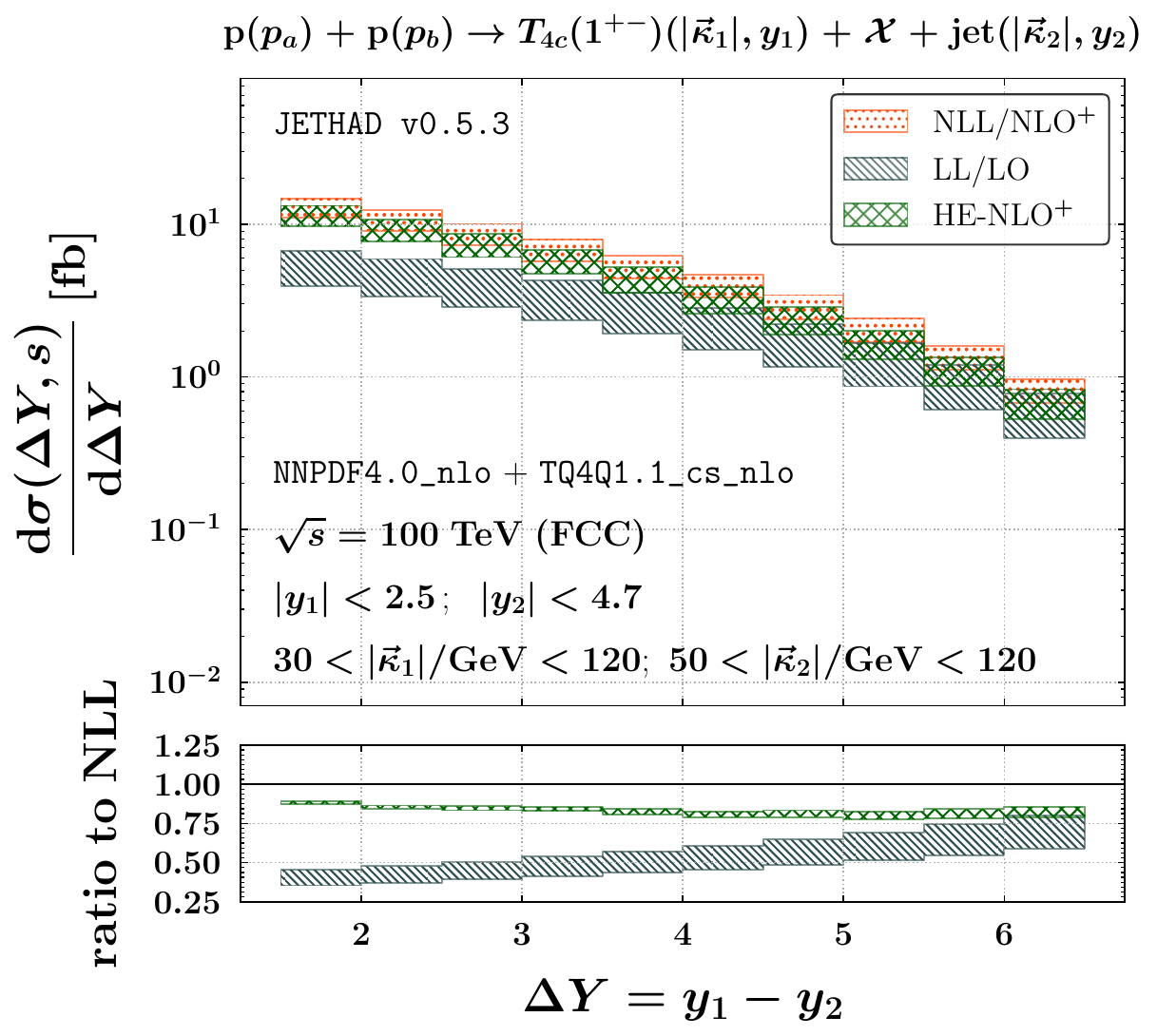}

\caption{$\TQcOpm$ plus jet $\DY$ rates at $\sqrt{s} = 14$ TeV (LHC, left) and $100$ TeV (FCC, right).
Shaded bands in main panels show MHOUs, LDMEs, and phase-space combined uncertainties.
Ancillary panels display $\LL$ or $\HENLOp$ to $\NLLp$ ratios, with bands for MHOUs only.}
\label{fig:rapidity_distributions}
\end{figure}

\vspace{0.10cm}
%==========================
\textbf{\textit{Hadron-collider phenomenology.}}
%==========================

We present predictions for rapidity-interval distributions in the semi-inclusive hadroproduction of a $\TQc(1^{+-})$ plus a light jet, working at $\NLLp$ accuracy in the hybrid factorization framework.  
This approach resums high-energy logarithms~\cite{Fadin:1975cb,Balitsky:1978ic} within the collinear NLO scheme~\cite{Colferai:2015zfa,Boussarie:2017oae,Celiberto:2015yba,Ivanov:2012iv,Bolognino:2021mrc,Celiberto:2021dzy,Celiberto:2021fdp,Celiberto:2018muu,Bolognino:2018rhb,Ball:2017otu,Bacchetta:2020vty,Bacchetta:2024fci,Celiberto:2017ius}.  
Numerical results were obtained with the {\Jethad} interface and {\symJethad} symbolic engine~\cite{Celiberto:2020wpk,Celiberto:2022rfj,Celiberto:2024mrq} at $14$~TeV HL-LHC and $100$~TeV (nominal) FCC.
Figure~\ref{fig:rapidity_distributions} shows the differential cross section in $\DY = y_1 - y_2$, computed using CMS-consistent rapidity cuts~\cite{Khachatryan:2016udy} and asymmetric transverse-momentum bins: $30 < |\vec \kappa_1| < 120$~GeV for the tetraquark and $50 < |\vec \kappa_2| < 120$~GeV for the jet.  
These kinematic selections enhance the discriminating power between the resummed NLL signal and the high-energy NLO$^+$ limit~\cite{Celiberto:2025_TQ4Q11_AVT_suppl}, particularly in forward configurations.
Shaded bands represent the combined effect of MHOUs, LDME variations, and numerical uncertainties.  
Event rates, although reduced by axial-vector FF suppression~\cite{Celiberto:2024beg}, remain promising.  
The $\NLLp$ and $\LL$ bands partially overlap at large $\DY$, indicating good stability.  
This agrees with previous analyses on exotic states~\cite{Celiberto:2023rzw,Celiberto:2024mab,Celiberto:2024beg} and reflects the known $\mu_F$-driven growth of gluon FFs~\cite{Celiberto:2021dzy,Celiberto:2021fdp,Celiberto:2022dyf,Celiberto:2022keu,Celiberto:2024omj}.
While earlier studies~\cite{Celiberto:2022keu,Celiberto:2024mrq} questioned the discriminating power of $\DY$ observables, here we find that the $\NLLp$ to $\LL$ ratio rises from $50\%$ to near unity, confirming the axial-vector channel as a sensitive probe of high-energy dynamics.

\vspace{0.10cm}
%==========================
\textbf{\textit{Conclusions.}}
%==========================

We presented and released new sets of VFNS FFs for axial-vector, fully heavy $\TQcOpm$ and $\TQbOpm$ tetraquarks.
Starting from NRQCD-modeled perturbative inputs for the constituent heavy-quark channel, and evolving them through threshold-matched DGLAP equations, the {\tt TQ4Q1.1} sets offer a realistic tool for collider phenomenology.
We employed them to generate high-energy predictions for tetraquark-jet observables at HL-LHC and FCC, working at NLL/NLO$^+$ accuracy within the {\psymJethad} framework.
Advancing hadron structure studies requires deeper insight into exotic-matter dynamics, supported by next-generation collider data.
To this end, we aim to refine heavy-tetraquark fragmentation by improving MHOU quantification~\cite{Kassabov:2022orn,Harland-Lang:2018bxd,NNPDF:2024dpb} and by including color-octet contributions once available.
Recent evidence for (valence~\cite{NNPDF:2023tyk}) intrinsic charm~\cite{Ball:2022qks,Guzzi:2022rca} may also motivate future studies of intrinsic bottom.
Understanding exotic and ordinary bottom physics remains essential, as it was for charm~\cite{Vogt:2024fky}.
The release of {\tt TQ4Q1.1}~\cite{Celiberto:2025_TQ4Q11_AVT}, now extended to axial-vector states, supports future investigations of heavy-tetraquark production via VFNS collinear fragmentation.
The newly available VFNS, DGLAP-evolving FFs for axial-vector fully heavy tetraquarks provide a novel foundation for high-energy exotic-matter studies and open a new direction in probing the core effects of strong interactions.
The $1^{+-}$ channel, with its distinctive dynamics and suppressed production, offers a clean window into exotic spectroscopy.
Our work timely contributes to the exploration of exotic QCD at present and future colliders.

\vspace{0.10cm}
%==========================
\textbf{\textit{Acknowledgments.}}
%==========================
We acknowledge the use of calculations from~\cite{Bai:2024ezn}, rederived via {\psymJethad}~\cite{Celiberto:2020wpk,Celiberto:2022rfj,Celiberto:2024mrq}, and used as proxies for the initial-scale fragmentation process.
We thank Angelo Esposito, Alessandro Papa, Fulvio Piccinini, and Alessandro Pilloni for insightful discussions on the physics of exotic states.
This work is supported from the Atracción de Talento Grant No. 2022-T1/TIC-24176 of the Comunidad Autónoma de Madrid, Spain.

\vspace{0.10cm}
%==========================
\textbf{\textit{Data availability.}}
%==========================
The {\tt TQ4Q1.1} FFs for $\TQQ(1^{+-})$ states~\cite{Celiberto:2025_TQ4Q11_AVT} can be accessed at: \url{https://github.com/FGCeliberto/Collinear_FFs/}.
For simplicity, only the central value is provided.
Since FFs scale linearly with the LDME (see Eq.~\eqref{eq:TQQ_FF_initial-scale}), there is no need to release additional error sets: Users can directly rescale FFs by varying the LDME within the uncertainty range given in Eq.~\eqref{eq:LDMEs_TQ1}.

%%%%%%%%%%%%%%%%%%

\bibliographystyle{apsrev4-1} %apsrev4-2 gives an error when "journal" entry is missing in a reference
\bibliography{bibliography}

% PDF for the supplemental material
\input{suppl_material}

\end{document}

%% file: suppl_material.tex
\newpage

\onecolumngrid
\newpage

\graphicspath{{./suppl_figures/}}

\newcommand{\sLL}{{\rm LL/LO}}
\newcommand{\sNLL}{{\rm NLL/NLO}}
\newcommand{\sNLLp}{{\rm NLL/NLO^+}}
\newcommand{\sNLLpp}{{\rm NLL/NLO^{(+)}}}
\newcommand{\sHENLOp}{{\rm HE}\mbox{-}{\rm NLO^+}}
\newcommand{\sHENLO}{{\rm HE}\mbox{-}{\rm NLO}}

% uncomment thisto start the supplemental material at page 1 (e.g. for
% PRL submission)
%\setcounter{page}{1}

\appendix

\makeatletter
\renewcommand\@biblabel[1]{[#1S]}
\makeatother

\setcounter{equation}{0}
\setcounter{figure}{0}
\setcounter{table}{0}
\renewcommand{\theequation}{S\arabic{equation}}
\renewcommand{\thefigure}{S\arabic{figure}}
\renewcommand{\thetable}{S\arabic{table}}
%\interfootnotelinepenalty=100000 % force footnotes to be together
%\setcounter{secnumdepth}{2}

%======================================================================

\begin{flushleft}
\large{\bf Fragmentation functions for axial-vector $T_{4c}$ and $T_{4b}$ tetraquarks: A TQ4Q1.1 update}
\end{flushleft}
\vspace{-0.60cm}
%======================================================================
\section*{Supplemental material}
\vspace{0.30cm}
%----------------------------------------------------------------------

The first section of this Supplemental Material provides additional insight into our {\tt TQ4Q1.1} FFs.
The second section presents technical details of the hybrid factorization for tetraquark-jet systems at $\NLLp$ accuracy.

\subsection{\it \textbf{Additional insights into {\tt TQ4Q1.1} functions}}
\label{ssec:supp_FFs}

We start by reporting the analytic expression for the $[Q \to \TQQ(1^{+-})]$ SDC, which was originally calculated by Authors of~\cite{Bai:2024ezn}, and then re-derived by making use of {\psymJethad}~\cite{Celiberto:2020wpk,Celiberto:2022rfj,Celiberto:2024mrq}.
One has

\begin{equation}
\begin{split}
\label{sDQ_FF_SDC_1pm_33}
 \mathcal{S}^{(1^{+-})}_Q&(z,[\tau \equiv [3,3]])
 \;=\;
 \frac{\pi^{2}\left[\alpha_{s}(\mu_R=5m_Q)\right]^4}{279936 (4-3 z)^6 (z-4)^2 z (11 z-12)\left(z^2-16 z+16\right)}
 \\[0.20cm]
 \;&\times\; 
 \bigg\{480 (z-4)
 (11 z-12)\left(z^2-16 z+16\right) \left(4 z^4+115 z^3-316 z^2+112z+64\right) (3 z-4)^5
 \\[0.20cm]
 \;&\times\; \log \left(z^2-16 z+16\right)+6 (11 z-12)\left(z^2-16 z+16\right) (4825 z^5-56232 z^4+378480z^3
 \\[0.20cm]
 \;&-\; 942528 z^2+672768 z-60416) (3 z-4)^5 \log (4-3 z)-3 (11z-12) \left(z^2-16 z+16\right) (5465 z^5
 \\[0.20cm]
 \;&-\; 40392 z^4+254320z^3-722368 z^2+611328 z-101376) (3 z-4)^5 \log\left[\left(\frac{11z}{3}-4\right)(z-4)\right]
 \\[0.20cm]
 \;&+\; 16 (z-1) z(476423 z^{11}+32559240 z^{10}-934590720 z^9+8015251776z^8
 \\[0.20cm]
 \;&-\; 35393754624 z^7+94265413632 z^6-160779010048 z^5+177897046016z^4
 \\[0.20cm]
 \;&-\; 124600254464 z^3 + 51223461888 z^2-10217324544z+490733568)\bigg\} \;,
\end{split}
\end{equation}
where, as previously mentioned, $\tau \equiv [3,3]$ is the only nonvanishing composite quantum number for axial-vector tetraquark states.

Regarding the nonperturbative component of our tetraquark FFs, as noted in our previous release of the {\tt TQ4Q1.0} sets~\cite{Celiberto:2024mab}, an effective approach involves calculating the radial wave functions at the origin using potential models and then connecting them to the LDMEs through the vacuum saturation approximation.
In Section~V of~\cite{Feng:2020riv}, three such models were proposed, all employing a Cornell-like potential and incorporating certain spin-dependent features.
The first and third models are based on nonrelativistic quark fields, while the second one includes relativistic corrections.

\begin{figure*}[!t]
\centering

   \hspace{-0.00cm}
   \includegraphics[scale=0.450,clip]{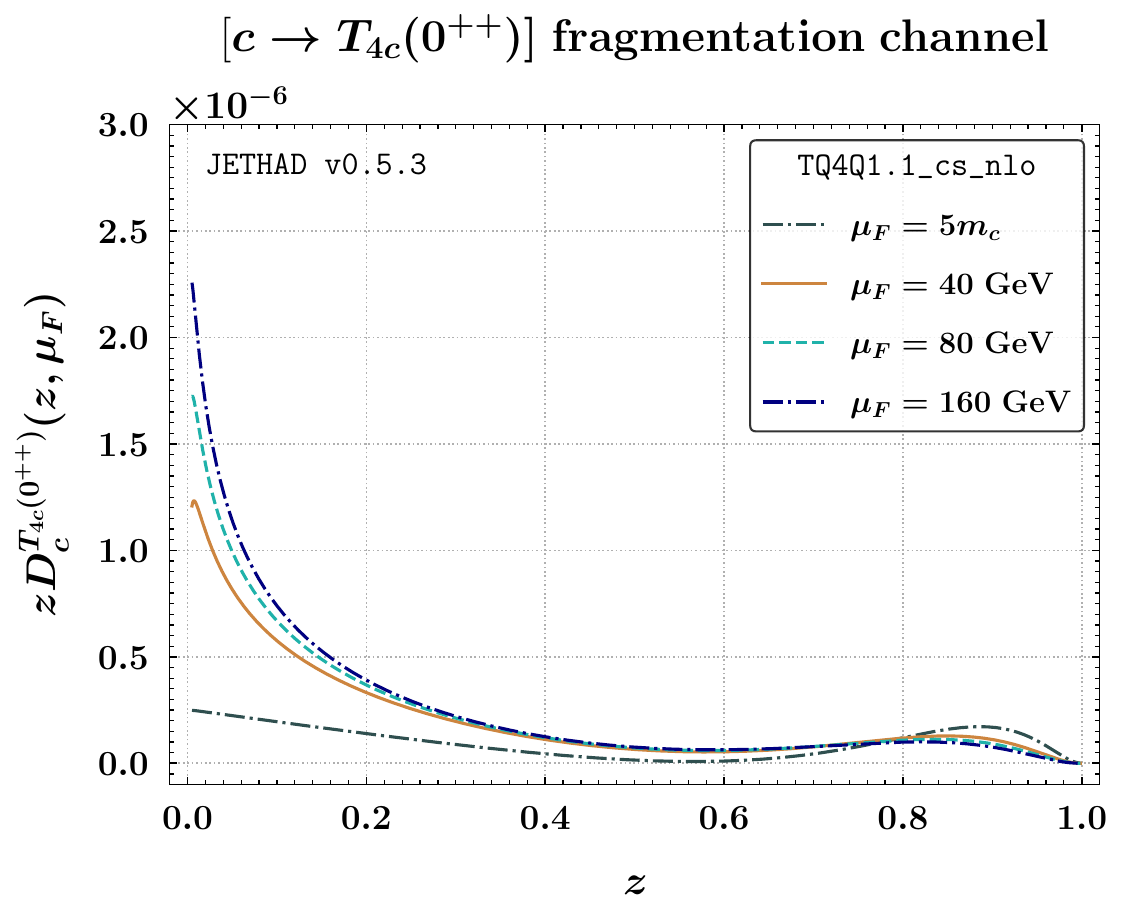}
   \hspace{0.30cm}
   \includegraphics[scale=0.450,clip]{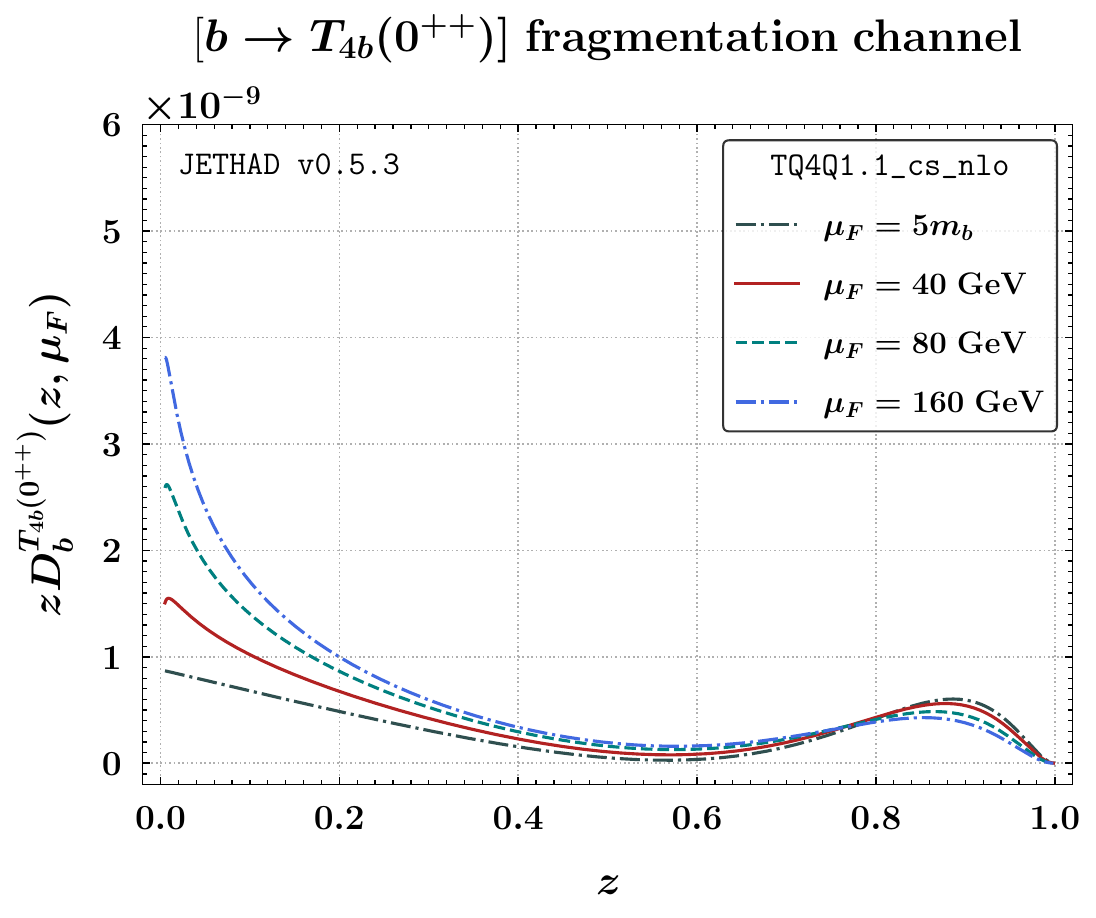}
%   \hspace{0.05cm}

   \vspace{0.25cm}

   \hspace{-0.00cm}
   \includegraphics[scale=0.450,clip]{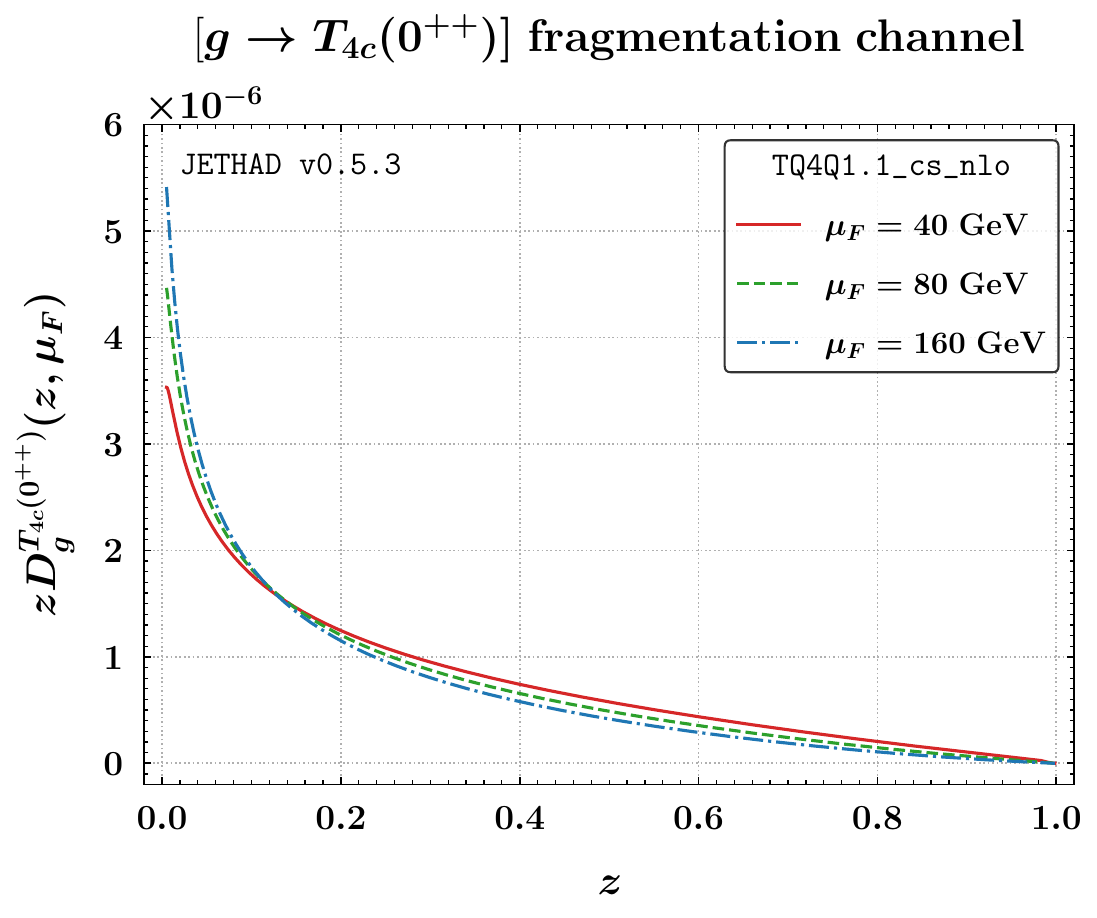}
   \hspace{0.30cm}
   \includegraphics[scale=0.450,clip]{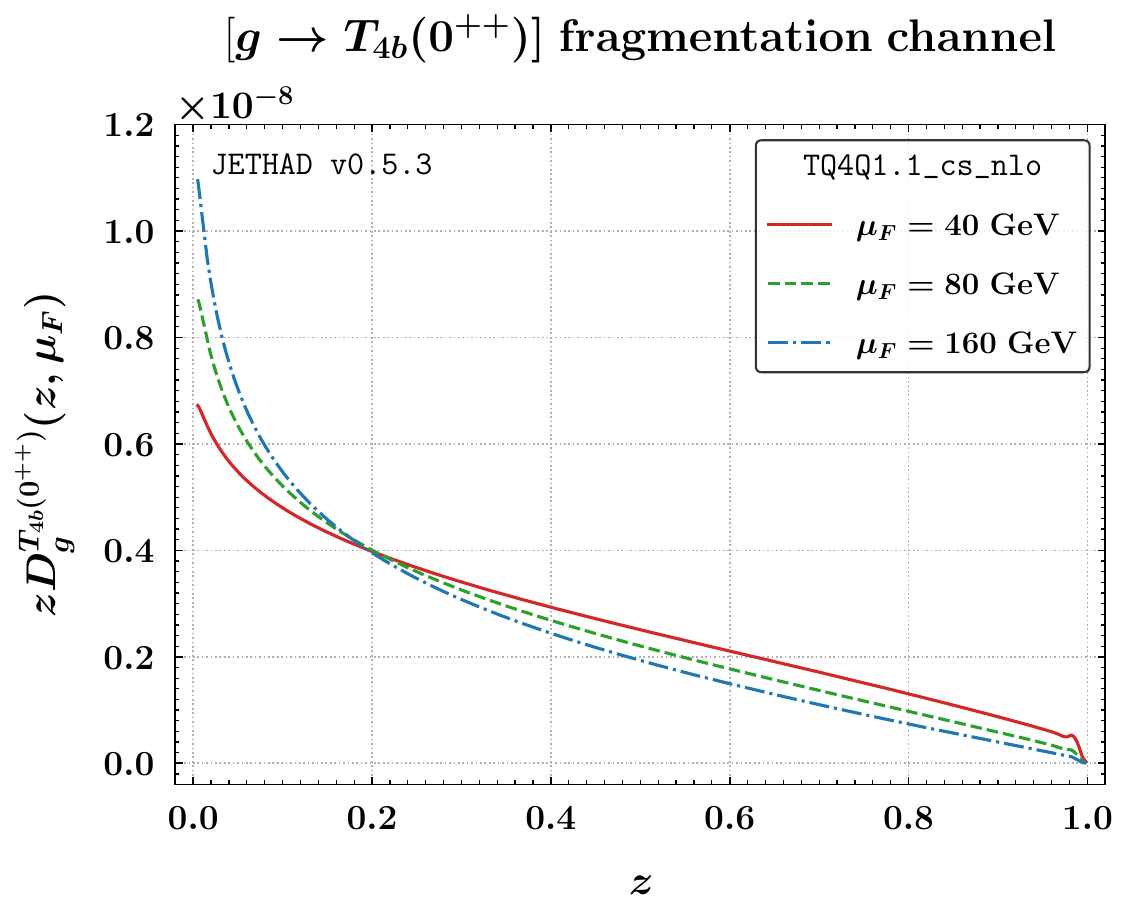}
%   \hspace{0.05cm}

\caption{Dependence on $z$ of {\tt TQ4Q1.1} collinear FFs for the $T_{4c}(0^{++})$ (left) and $T_{4b}(0^{++})$ (right) states for different $\mu_F$ values.
Upper (lower) plots are for the constituent heavy-quark (gluon) channel.}
\label{fig:supp_FFs-z_TQ0}
\end{figure*}

\begin{figure*}[!t]
\centering

   \hspace{-0.00cm}
   \includegraphics[scale=0.450,clip]{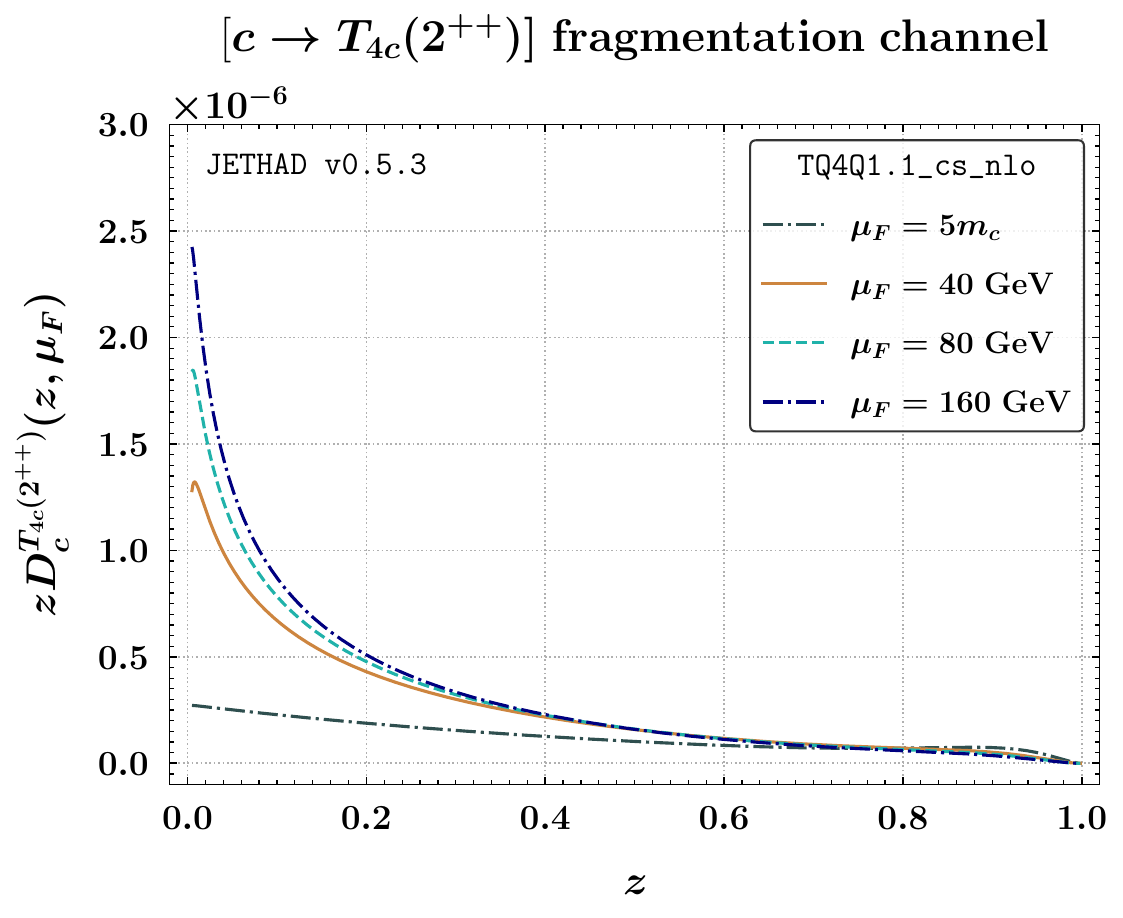}
   \hspace{0.30cm}
   \includegraphics[scale=0.450,clip]{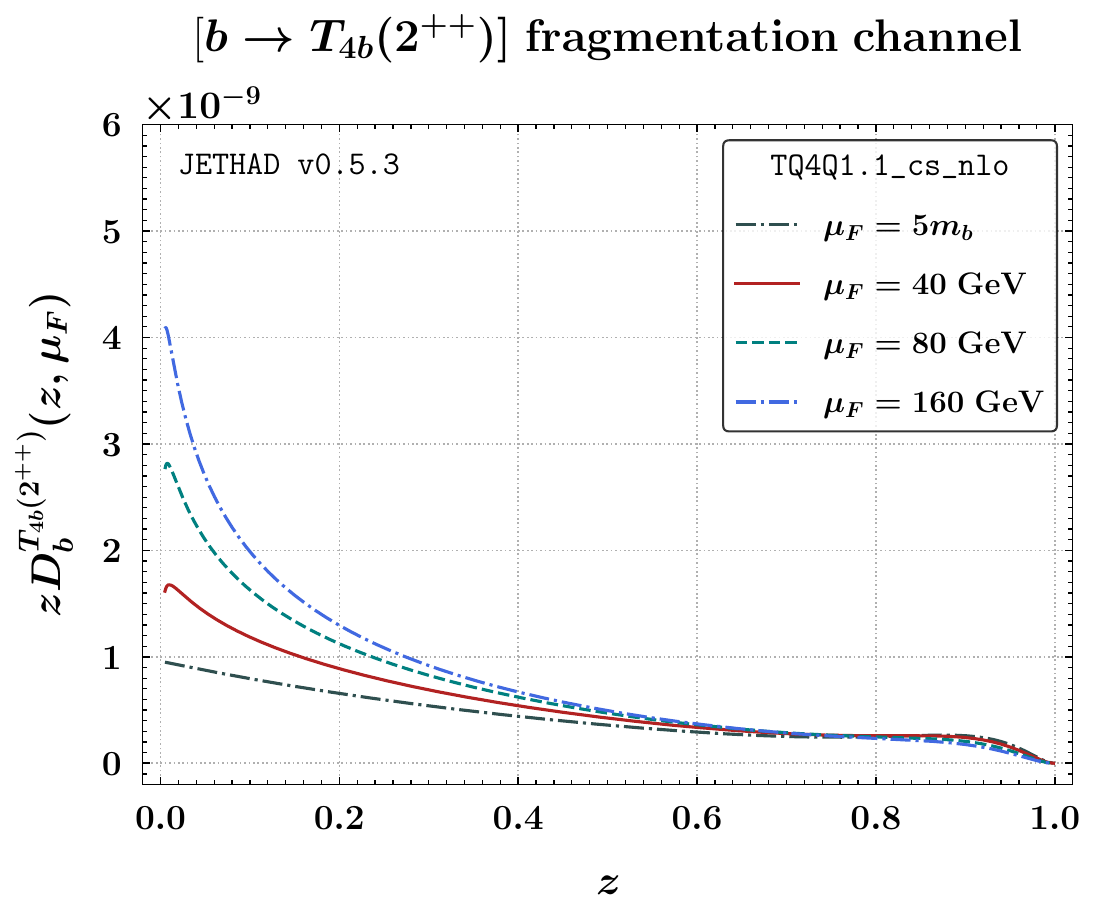}
%   \hspace{0.05cm}

   \vspace{0.25cm}

   \hspace{-0.00cm}
   \includegraphics[scale=0.450,clip]{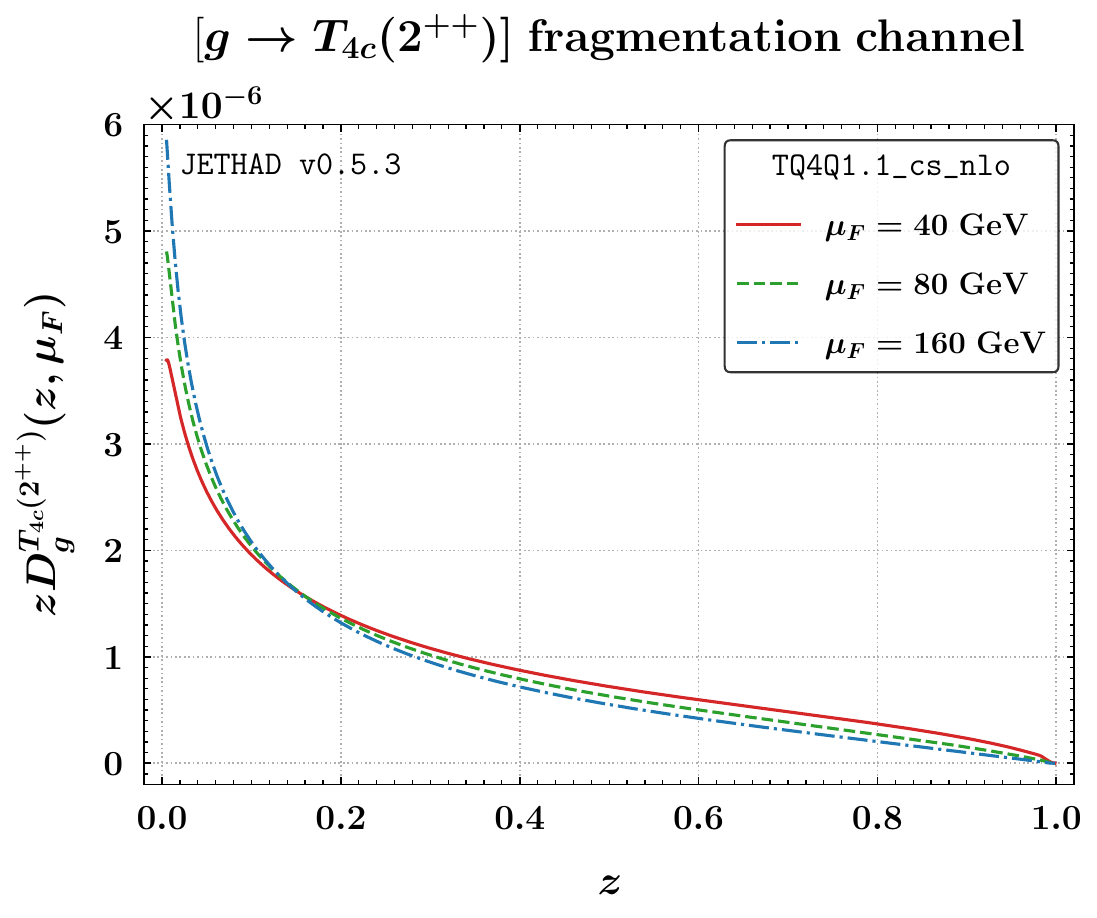}
   \hspace{0.30cm}
   \includegraphics[scale=0.450,clip]{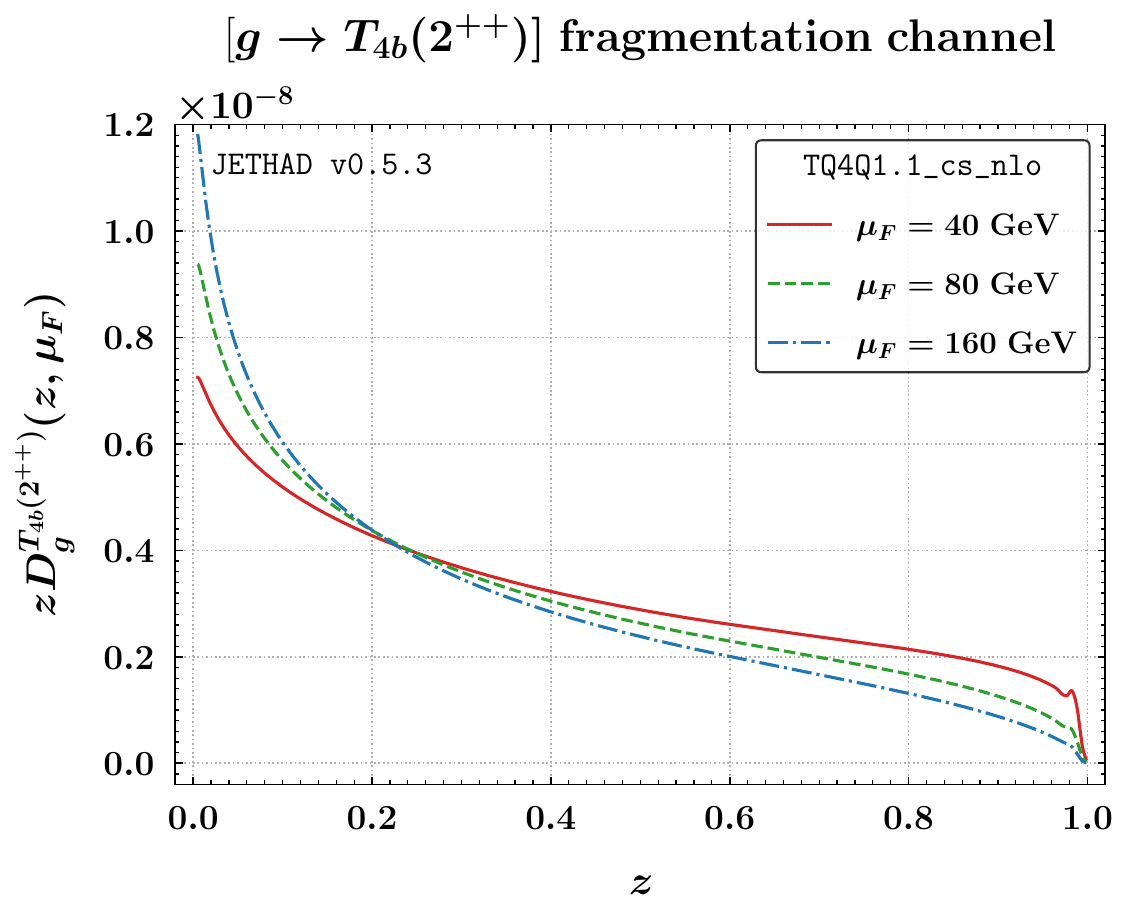}
%   \hspace{0.05cm}

\caption{Dependence on $z$ of {\tt TQ4Q1.1} collinear FFs for the $T_{4c}(2^{++})$ (left) and $T_{4b}(2^{++})$ (right) states for different $\mu_F$ values.
Upper (lower) plots are for the constituent heavy-quark (gluon) channel.}
\label{fig:supp_FFs-z_TQ2}
\end{figure*}

As discussed in~\cite{Feng:2020riv}, the first model significantly overestimates the cross section when compared to data on $\Jpsi$ production at 13 TeV (CMS), which are, in any case, expected to be well above the $\TQc$ rate.
In addition, numerical checks not shown in this work revealed that the FFs constructed using LDMEs from the third model are highly unstable, even under very small variations in their values, at the level of about 0.1 percent.
For these reasons, when building the scalar [$0^{++}$] and tensor [$2^{++}$] channels in both our {\tt TQ4Q1.0} sets~\cite{Celiberto:2024mab} and the {\tt 1.1} update~\cite{Celiberto:2024beg}, we adopted the second model proposed in~\cite{Lu:2020cns}.
The corresponding LDMEs are listed below. A comparison with the values predicted by the other two models can be found in Table I of the published version of~\cite{Feng:2020riv}:
\begin{equation}
\begin{split}
\label{sLDMEs_T4c_02}
 {\cal O}^{\TQcZpp}([3,3]) &\,=\, 0.0347\mbox{ GeV}^9 \;, \qquad\quad
 {\cal O}^{\TQcTpp}([3,3]) \,=\, 0.072\mbox{ GeV}^9 \;,
 \\
 {\cal O}^{\TQcZpp}([3,6]) &\,=\, 0.0211\mbox{ GeV}^9 \;, \qquad\quad
 {\cal O}^{\TQcTpp}([3,6]) \,=\, 0 \;,
 \\
 {\cal O}^{\TQcZpp}([6,6]) &\,=\, 0.0128\mbox{ GeV}^9 \;, \qquad\quad
 {\cal O}^{\TQcTpp}([6,6]) \,=\, 0 \;.
\end{split}
\end{equation}

Then, in~\cite{Bai:2024ezn}, the NRQCD calculation of the constituent heavy-quark FF at the initial scale was extended by including two additional LDME models (Models~IV and~V in Table~1 of that work).
From the inspection of the numerical values for the $1S$ channels, we note that Model IV~\cite{Yu:2022lak} yields LDMEs that are close to those in~\cite{Lu:2020cns}, particularly for the axial-vector channel, whereas the other model gives values that are roughly one order of magnitude smaller.
For this reason, in our main analysis of the $T_{4c}(1^{+-})$ case, we used the average of the values from~\cite{Lu:2020cns} and~\cite{Yu:2022lak} as the reference for the corresponding LDME (see the first line of Eq.~\eqref{eq:LDMEs_TQ1}).

Since exact values for the LDMEs of the fully bottomed states have not yet been computed, we can adopt a physically motivated \emph{Ansatz} as a reasonable starting point.
We assume that the $\TQb$ tetraquark forms a compact diquark-antidiquark system, where the binding dynamics are predominantly driven by attractive color-Coulomb forces.
Under this assumption, the ratio between the four-body Schr{\"o}dinger wave functions at the origin for $\TQc$ and $\TQb$ states can be estimated through dimensional analysis.
Following the approach proposed in~\cite{Feng:2023agq} and already used to derive the {\tt TQ4Q1.1} FFs for $0^{++}$ and $2^{++}$ hadrons~\cite{Celiberto:2024beg}, we write
\begin{equation}
\label{sLDMEs_T4b}
 \frac{\langle {\cal O}^{\TQb(J^{PC})}([n]) \rangle}{\langle {\cal O}^{\TQc(J^{PC})}([n]) \rangle} 
 \,=\, 
 \frac{{\langle \cal O}^{\TQb}_{\rm [Coulomb]} \rangle}{{\langle \cal O}^{\TQc}_{\rm [Coulomb]}\rangle} 
 \,\simeq\,
 \left( \frac{m_b \, \as^{[b]}}{m_c \, \as^{[c]}} \right)^9 
 \,\simeq\, 
 400
 \,.
\end{equation}

Here, $\as^{[Q=c,b]}$ denotes the strong coupling evaluated at the scale $m_Q \upsilon_Q$, with $\upsilon_Q$ being the relative velocity between the two constituent heavy quarks.
The subscript `${\rm [Coulomb]}$' indicates that the LDME is computed within a Coulomb-like potential model for the diquark system.
The numerical value of the ${\cal O}^{\TQcOpm}([3,3])$ LDME used in this study (second line of Eq.~\eqref{eq:LDMEs_TQ1}) was obtained by combining the first line of Eq.~\eqref{eq:LDMEs_TQ1} with Eq.~\eqref{sLDMEs_T4b}.
In the same way, the ${\cal O}^{\TQcZpp}$ and ${\cal O}^{\TQcTpp}$ LDMEs can be derived by combining Eq.~\eqref{sLDMEs_T4c_02} with Eq.~\eqref{sLDMEs_T4b}.

For illustration purposes, Figs.~\ref{fig:supp_FFs-z_TQ0} and~\ref{fig:supp_FFs-z_TQ2} display the $z$-dependence of the {\tt TQ4Q1.1} functions for the $0^{++}$ and $2^{++}$ tetraquark states.  
The left panels correspond to charmed states, while the right panels refer to bottomed ones.  
The upper plots show the constituent $Q$-quark FFs, and the lower plots show the gluon FFs.  
As in Fig.~\ref{fig:FFs-z_TQ1} of the main text, four representative values for $\mu_F$ are considered: $\mu_{F,0} \equiv 5m_Q$, as well as 40, 80, and 160~GeV. At the input scale $\mu_{F,0}$, the gluon fragmentation function is identically zero, as it is entirely generated by the DGLAP evolution above this scale. 
Therefore, only the $Q$-quark channel contributes at $\mu_{F,0}$ and is shown in the corresponding plots.
Unlike the axial-vector case, the scalar and tensor FFs are not accompanied by uncertainty bands, as these were not included in our earlier dedicated analysis~\cite{Celiberto:2024beg}.
The results shown in Figs.~\ref{fig:supp_FFs-z_TQ0} and~\ref{fig:supp_FFs-z_TQ2} not only allow for a direct comparison among the $0^{++}$, $1^{+-}$, and $2^{++}$ states, but also complement the findings of~\cite{Celiberto:2024beg}, where only the~$\mu_F$-dependence was discussed, while the~$z$-dependence was not addressed.

%----------------------------------------------------------------------
\subsection{\it \textbf{$\boldsymbol{\NLLp}$ hybrid factorization for tetraquark-jet systems}}
\label{ssec:supp_HyF}

\begin{figure*}[!t]
\includegraphics[width=0.575\textwidth]{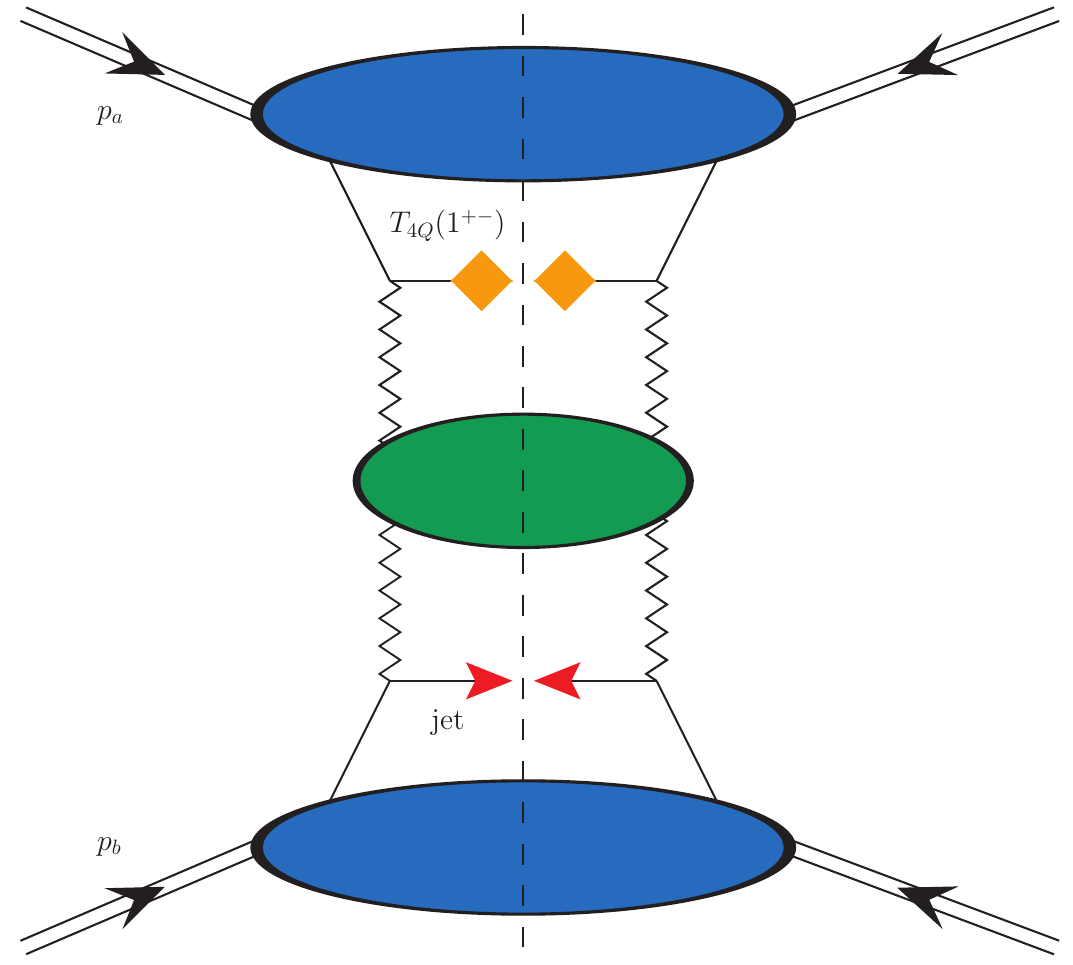}

\caption{Schematic representation of the semi-inclusive hadroproduction of tetraquark-jet systems within the hybrid collinear and high-energy factorization framework.
Orange rhombi denote the collinear FFs of the $\TQQ(1^{+-})$ tetraquark.
Red arrows illustrate light-flavored jets, while blue ovals correspond to proton collinear PDFs.
The exponentiated resummation kernel (green blob) is linked to the two off-shell emission functions via waggle Reggeon lines.}
\label{fig:supp_process}
\end{figure*}

The process under analysis is (see Fig.~\ref{fig:supp_process})
\begin{eqnarray}
\label{sprocess}
 {\rm p}(p_a) + {\rm p}(p_b) \to T_{4Q}(\kappa_1, y_1) + {\cal X} + {\rm jet}(\kappa_2, y_2) \;,
\end{eqnarray}
where an outgoing heavy axial-vector fully heavy tetraquark, $T_{4c}(1^{+-})$ or $T_{4b}(1^{+-})$, is produced in association with a light-flavored jet, and together with an undetected gluon system, ${\cal X}$.
The final-state objects possess transverse momenta satisfying $|\vec \kappa_{1,2}| \gg \Lambda_{\rm QCD}$, with $\Lambda_{\rm QCD}$ the QCD hadronization scale, and they are separated by a large rapidity interval, $\DY = y_1 - y_2$.
We perform a Sudakov decomposition of the $\kappa_{1,2}$ four-momenta in terms of the colliding-proton momenta, $p_{a,b}$, obtaining
\begin{eqnarray}
\label{ssudakov}
\kappa_{1,2} = x_{1,2} p_{a,b} + \frac{\vec \kappa_{1,2}^{\,2}}{x_{1,2} s}p_{b,a} + \kappa_{1,2\perp} \, \qquad
\kappa_{1,2\perp}^2 = - \vec \kappa_{1,2}^{\,2} \;,
\end{eqnarray}
In the center-of-mass frame, the rapidities are given by
\begin{eqnarray}
\label{srapidities}
y_{1,2} = \pm \ln  \frac{x_{1,2} \sqrt{s}}{|\vec \kappa_{1,2}| \;,
}
\end{eqnarray}

One can recast the $\DY$- and $\varphi$-differential cross section, where $\varphi = \phi_1 - \phi_2 - \pi$ and $\phi_{1,2}$ denote the azimuthal angles of the two identified final-state objects, as a Fourier series of azimuthal coefficients, $C_{n \ge 0}$.
We write
\begin{eqnarray}
 \label{sdsigmaFourier}
 \frac{\drv \sigma}{\drv \DY \, \drv \varphi \, \drv |\kappa_1| \, \drv |\vec \kappa_2|} 
 =
 \frac{1}{\pi} \left[ \frac{1}{2} C_0 + \sum_{n=1}^\infty \cos (n \varphi)\,
 C_n \right]\;.
\end{eqnarray}
Within the hybrid high-energy and collinear factorization framework and adopting the $\MSb$ renormalization scheme, we derive a master expression for the $C_n$ coefficients.
This formulation holds within the NLO perturbative expansion and incorporates the NLL resummation of high-energy logarithms.
Explicitly, one obtains
\begin{equation}
\begin{split}
\label{sCnNLL}
 C_n^{\sNLLp} \;&=\; 
 \int_{\kappa_1^{\rm min}}^{\kappa_1^{\rm max}} \drv |\vec \kappa_1|
 \int_{\kappa_2^{\rm min}}^{\kappa_2^{\rm max}} \drv |\vec \kappa_2|
 \int_{y_1^{\rm min}}^{y_1^{\rm max}}
 \drv y_1
 \int_{y_2^{\rm min}}^{y_2^{\rm max}} 
 \drv y_2
 \; \delta(\DY - y_1 + y_2)
 \int_{-\infty}^{+\infty} \drv \nu \, e^{\bar \alpha_s \DY \chi^{\rm NLL}(n,\nu)}
\\[0.20cm]
 &\times \,
 \frac{e^{\DY}}{s}
 \alpha_s^2(\mu_R)
 \left\{ 
 \Phi_1^{\rm NLO}(n,\nu,|\vec \kappa_1|, x_1)[\Phi_2^{\rm NLO}(n,\nu,|\vec \kappa_2|,x_2)]^*
 + \bar \alpha_s^2 \frac{\beta_0 \DY}{4 N_c}\chi(n,\nu)\upsilon(\nu)
 \right\} \;,
\end{split}
\end{equation}
where $\bar \alpha_s(\mu_R) = \alpha_s(\mu_R) N_c/\pi$, $N_c$ is the color number, and $\beta_0 = 11N_c/3 - 2 n_f/3$ the QCD $\beta$-function leading coefficient.
We adopt a two-loop running-coupling setup with $\alpha_s\left(M_Z\right)=0.11707$ and a dynamic number of flavors, $n_f$.
The $\chi(n,\nu)$ function in the exponent of~\eqref{sCnNLL} represents the Balitsky-Fadin-Kuraev-Lipatov (BFKL) kernel~\cite{Fadin:1975cb,Balitsky:1978ic}, which resums NLL energy logarithms.
\begin{eqnarray}
 \label{schi}
 \chi^{\rm NLL}(n,\nu) = \chi(n,\nu) + \bar\alpha_s \hat \chi(n,\nu) \;,
\end{eqnarray}
with $\chi(n,\nu)$ the LO BFKL eigenvalues
\begin{eqnarray}
\chi\left(n,\nu\right) = -2\left\{\gamma_{\rm E}+{\rm Re} \left[\psi\left( (n + 1)/2 + i \nu \right)\right] \right\} \;.
\label{schiLO}
\end{eqnarray}
Here, $\psi(z) = \Gamma^\prime(z)/\Gamma(z)$ is the logarithmic derivative of the Gamma function, and $\gamma_{\rm E}$ denotes the Euler-Mascheroni constant.
The $\hat\chi(n,\nu)$ expression is the NLO kernel correction
\begin{align}
\label{schiNLO}
\hat \chi\left(n,\nu\right) \;&=\; \bar\chi(n,\nu)+\frac{\beta_0}{8 N_c}\chi(n,\nu)
%\\ \nonumber &\times \,
\left\{-\chi(n,\nu)+2\ln\left(\mu_R^2/\hat{\mu}^2\right)+\frac{10}{3}\right\} \;,
\end{align}
with $\hat{\mu} = \sqrt{|\vec \kappa_1| |\vec \kappa_2|}$.
The analytic expression of the characteristic $\bar\chi(n,\nu)$ function can be found, \emph{e.g.} in Sec.~2.1.1 of~\cite{Celiberto:2020wpk}.
The two terms
\begin{eqnarray}
\label{sEFs}
\Phi_{1,2}^{\rm NLO}(n,\nu,|\vec \kappa|,x) =
\Phi_{1,2}(n,\nu,|\vec \kappa|,x) +
\alpha_s(\mu_R) \, \hat \Phi_{1,2}(n,\nu,|\vec \kappa|,x)
\end{eqnarray}
represent the NLO singly off-shell, transverse-momentum-dependent emissions functions NLO emission functions, also known in the BFKL jargon as forward-production impact factors.
Tetraquark emissions are described by the NLO forward-hadron impact factor~\cite{Ivanov:2012iv}.
Although originally formulated for light hadrons, its applicability extends to our VFNS approach~\cite{Mele:1990cw,Cacciari:1993mq}, as long as the considered transverse-momentum ranges remain well above the DGLAP-evolution thresholds for heavy quarks.
At LO, one has
\begin{equation}
\begin{split}
\label{sLOHEF}
\Phi_{T_{4Q}}(n,\nu,|\vec \kappa|,x) \;&=\; 2 \sqrt{\frac{C_F}{C_A}} \; |\vec \kappa|^{2i\nu-1}\int_{x}^1 \frac{\drv \zeta}{\zeta} \left( \frac{x}{\zeta} \right)^{1-2i\nu} 
%\nonumber 
\\
 &\times \, 
 \left[ \frac{C_A}{C_F} f_g(\zeta, \mu_F)D_g^{T_{4Q}}\left( \frac{x}{\zeta}, \mu_F \right)
 +\sum_{i=q,\bar q}f_i(\zeta, \mu_F)D_i^{T_{4Q}}\left( \frac{x}{\zeta}, \mu_F \right) \right] \;,
\end{split}
\end{equation}
with $C_F = (N_c^2-1)/(2N_c)$ and $C_A \equiv N_c$ the Casimir constants connected to gluon emissions from a quark and a gluon, respectively. 
Here, $f_i\left(x, \mu_F \right)$ denotes the PDF of parton $i$ within the parent proton, while $D_i^{T_{4Q}}\left(x/\zeta, \mu_F \right)$ represents the FF describing the fragmentation of parton $i$ into the identified tetraquark, $T_{4Q}$.
The NLO correction can be found in~\cite{Ivanov:2012iv}.
The LO light-jet emission function reads
\begin{eqnarray}
 \label{sLOJEF}
 \hspace{-0.09cm}
 \Phi_J(n,\nu,|\vec \kappa|,x) = 2 \sqrt{\frac{C_F}{C_A}} \;
 |\vec \kappa|^{2i\nu-1}\,\hspace{-0.05cm} \left[ \frac{C_A}{C_F} f_g(x, \mu_F)
 +\hspace{-0.15cm}\sum_{j=q,\bar q}\hspace{-0.10cm}f_j(x, \mu_F) \right] \;.
\end{eqnarray}
while its NLO correction is derived from~\cite{Colferai:2015zfa}.
It employs small-cone selection functions with the jet-cone radius set to $R_J = 0.5$, in accordance with recent analyses at CMS~\cite{Khachatryan:2016udy}.
The last ingredient of~\eqref{sCnNLL} is the  $\upsilon(\nu)$ function
\begin{eqnarray}
 \upsilon(\nu) = \frac{1}{2} \left[ 4 \ln \hat{\mu} + i \frac{\drv}{\drv \nu} \ln\frac{\Phi_1(n,\nu,|\vec \kappa_1|, x_1)}{\Phi_2[(n,\nu,|\vec \kappa_1|, x_1)]^*} \right] \;.
\label{sfnu}
\end{eqnarray}
Equations~\eqref{sCnNLL} to~\eqref{sLOJEF} provide insight into the structure of our hybrid-factorization setup.
In accordance with BFKL, the cross section undergoes high-energy factorization, expressed as a convolution between the Green's function and two singly off-shell emission functions.
These emission functions encapsulate collinear inputs, specifically the collinear convolutions of the incoming protons' PDFs with the outgoing hadrons' FFs.
The label $\NLLp$ signifies a fully resummed NLL treatment of energy logarithms within the NLO perturbative framework. The `$+$' superscript indicates the inclusion of contributions beyond the NLL level, originating from the cross product of NLO emission-function corrections, in our representation of azimuthal coefficients.

For comparison, we also examine the pure LL limit within the $\MSb$ scheme, obtained by discarding NLO corrections in both the resummation kernel (Eq.~\eqref{schi}) and the impact factors (Eq.~\eqref{sEFs}). We obtain
\begin{eqnarray}
\label{sCnLL}%\nonumber
 C_n^{\sLL} \propto 
 \frac{e^{\DY}}{s} 
 \int_{-\infty}^{+\infty} \drv \nu \, 
 e^{\bar \alpha_s \DY \chi(n,\nu)} %\,
 \alpha_s^2(\mu_R) \, \Phi_{\TQQ}(n,\nu,|\vec \kappa_1|, x_1)[\Phi_J(n,\nu,|\vec \kappa_2|,x_2)]^* \;.
\end{eqnarray}
For simplicity, the integration over transverse momenta and rapidities of final-state objects, explicitly shown in the first line of~\eqref{sCnNLL}, is omitted in~\eqref{sCnLL} but remains implicit.

A thorough comparison between high-energy and fixed-order approaches relies on confronting NLL-resummed predictions with purely fixed-order calculations.
However, to the best of our knowledge, no numerical tool is currently available for computing NLO observables sensitive to two-particle hadroproduction.
To establish a fixed-order reference, we truncate the expansion of the $C_n$ coefficients in~(\ref{sCnNLL}) at ${\cal O}(\alpha_s^3)$. This leads to an effective high-energy fixed-order ($\sHENLOp$) formulation suitable for phenomenological studies.
We retain the leading-power asymptotic contributions present in a full NLO calculation while neglecting terms suppressed by inverse powers of the partonic center-of-mass energy.
The $\MSb$ expression for $\sHENLOp$ angular coefficients is given by
\begin{align}
\label{sCnHENLO}%\nonumber
 C_n^{\sHENLOp} &\propto %\frac{x_1 x_2}{|\vec \kappa_1| |\vec \kappa_2|} 
 \frac{e^{\DY}}{s} 
 \int_{-\infty}^{+\infty} \drv \nu \, 
 %e^{{\DY} \bar \alpha_s(\mu_R) \chi^{\rm NLO}(m,\nu)}
 \alpha_s^2(\mu_R) \,
 \left[ 1 + \bar \alpha_s(\mu_R) \DY \chi(n,\nu) \right] \,
% \end{equation}
%\[
 %\\ \nonumber
 %&\times
 \Phi_{\TQQ}^{\rm NLO}(n,\nu,|\vec \kappa_1|, x_1)[\Phi_J^{\rm NLO}(n,\nu,|\vec \kappa_2|,x_2)]^* \;,
% \]
\end{align}
where the exponentiated kernel has been expanded up to ${\cal O}(\alpha_s)$.
Also in this case, for the sake of brevity, the integration over transverse momenta and rapidities of final-state objects is understood.

%% file: manuscript_v2a.bbl
\begin{thebibliography}{141}%
\makeatletter
\providecommand \@ifxundefined [1]{%
 \@ifx{#1\undefined}
}%
\providecommand \@ifnum [1]{%
 \ifnum #1\expandafter \@firstoftwo
 \else \expandafter \@secondoftwo
 \fi
}%
\providecommand \@ifx [1]{%
 \ifx #1\expandafter \@firstoftwo
 \else \expandafter \@secondoftwo
 \fi
}%
\providecommand \natexlab [1]{#1}%
\providecommand \enquote  [1]{``#1''}%
\providecommand \bibnamefont  [1]{#1}%
\providecommand \bibfnamefont [1]{#1}%
\providecommand \citenamefont [1]{#1}%
\providecommand \href@noop [0]{\@secondoftwo}%
\providecommand \href [0]{\begingroup \@sanitize@url \@href}%
\providecommand \@href[1]{\@@startlink{#1}\@@href}%
\providecommand \@@href[1]{\endgroup#1\@@endlink}%
\providecommand \@sanitize@url [0]{\catcode `\\12\catcode `\$12\catcode `\&12\catcode `\#12\catcode `\^12\catcode `\_12\catcode `\%12\relax}%
\providecommand \@@startlink[1]{}%
\providecommand \@@endlink[0]{}%
\providecommand \url  [0]{\begingroup\@sanitize@url \@url }%
\providecommand \@url [1]{\endgroup\@href {#1}{\urlprefix }}%
\providecommand \urlprefix  [0]{URL }%
\providecommand \Eprint [0]{\href }%
\providecommand \doibase [0]{http://dx.doi.org/}%
\providecommand \selectlanguage [0]{\@gobble}%
\providecommand \bibinfo  [0]{\@secondoftwo}%
\providecommand \bibfield  [0]{\@secondoftwo}%
\providecommand \translation [1]{[#1]}%
\providecommand \BibitemOpen [0]{}%
\providecommand \bibitemStop [0]{}%
\providecommand \bibitemNoStop [0]{.\EOS\space}%
\providecommand \EOS [0]{\spacefactor3000\relax}%
\providecommand \BibitemShut  [1]{\csname bibitem#1\endcsname}%
\let\auto@bib@innerbib\@empty
%</preamble>
\bibitem [{\citenamefont {Esposito}\ \emph {et~al.}(2017)\citenamefont {Esposito}, \citenamefont {Pilloni},\ and\ \citenamefont {Polosa}}]{Esposito:2016noz}%
  \BibitemOpen
  \bibfield  {author} {\bibinfo {author} {\bibfnamefont {A.}~\bibnamefont {Esposito}}, \bibinfo {author} {\bibfnamefont {A.}~\bibnamefont {Pilloni}}, \ and\ \bibinfo {author} {\bibfnamefont {A.~D.}\ \bibnamefont {Polosa}},\ }\href {\doibase 10.1016/j.physrep.2016.11.002} {\bibfield  {journal} {\bibinfo  {journal} {Phys. Rept.}\ }\textbf {\bibinfo {volume} {668}},\ \bibinfo {pages} {1} (\bibinfo {year} {2017})},\ \Eprint {http://arxiv.org/abs/1611.07920} {arXiv:1611.07920 [hep-ph]} \BibitemShut {NoStop}%
\bibitem [{\citenamefont {Olsen}\ \emph {et~al.}(2018)\citenamefont {Olsen}, \citenamefont {Skwarnicki},\ and\ \citenamefont {Zieminska}}]{Olsen:2017bmm}%
  \BibitemOpen
  \bibfield  {author} {\bibinfo {author} {\bibfnamefont {S.~L.}\ \bibnamefont {Olsen}}, \bibinfo {author} {\bibfnamefont {T.}~\bibnamefont {Skwarnicki}}, \ and\ \bibinfo {author} {\bibfnamefont {D.}~\bibnamefont {Zieminska}},\ }\href {\doibase 10.1103/RevModPhys.90.015003} {\bibfield  {journal} {\bibinfo  {journal} {Rev. Mod. Phys.}\ }\textbf {\bibinfo {volume} {90}},\ \bibinfo {pages} {015003} (\bibinfo {year} {2018})},\ \Eprint {http://arxiv.org/abs/1708.04012} {arXiv:1708.04012 [hep-ph]} \BibitemShut {NoStop}%
\bibitem [{\citenamefont {Brambilla}\ \emph {et~al.}(2020)\citenamefont {Brambilla}, \citenamefont {Eidelman}, \citenamefont {Hanhart}, \citenamefont {Nefediev}, \citenamefont {Shen}, \citenamefont {Thomas}, \citenamefont {Vairo},\ and\ \citenamefont {Yuan}}]{Brambilla:2019esw}%
  \BibitemOpen
  \bibfield  {author} {\bibinfo {author} {\bibfnamefont {N.}~\bibnamefont {Brambilla}}, \bibinfo {author} {\bibfnamefont {S.}~\bibnamefont {Eidelman}}, \bibinfo {author} {\bibfnamefont {C.}~\bibnamefont {Hanhart}}, \bibinfo {author} {\bibfnamefont {A.}~\bibnamefont {Nefediev}}, \bibinfo {author} {\bibfnamefont {C.-P.}\ \bibnamefont {Shen}}, \bibinfo {author} {\bibfnamefont {C.~E.}\ \bibnamefont {Thomas}}, \bibinfo {author} {\bibfnamefont {A.}~\bibnamefont {Vairo}}, \ and\ \bibinfo {author} {\bibfnamefont {C.-Z.}\ \bibnamefont {Yuan}},\ }\href {\doibase 10.1016/j.physrep.2020.05.001} {\bibfield  {journal} {\bibinfo  {journal} {Phys. Rept.}\ }\textbf {\bibinfo {volume} {873}},\ \bibinfo {pages} {1} (\bibinfo {year} {2020})},\ \Eprint {http://arxiv.org/abs/1907.07583} {arXiv:1907.07583 [hep-ex]} \BibitemShut {NoStop}%
\bibitem [{\citenamefont {Karliner}\ \emph {et~al.}(2018)\citenamefont {Karliner}, \citenamefont {Rosner},\ and\ \citenamefont {Skwarnicki}}]{Karliner:2017qhf}%
  \BibitemOpen
  \bibfield  {author} {\bibinfo {author} {\bibfnamefont {M.}~\bibnamefont {Karliner}}, \bibinfo {author} {\bibfnamefont {J.~L.}\ \bibnamefont {Rosner}}, \ and\ \bibinfo {author} {\bibfnamefont {T.}~\bibnamefont {Skwarnicki}},\ }\href {\doibase 10.1146/annurev-nucl-101917-020902} {\bibfield  {journal} {\bibinfo  {journal} {Ann. Rev. Nucl. Part. Sci.}\ }\textbf {\bibinfo {volume} {68}},\ \bibinfo {pages} {17} (\bibinfo {year} {2018})},\ \Eprint {http://arxiv.org/abs/1711.10626} {arXiv:1711.10626 [hep-ph]} \BibitemShut {NoStop}%
\bibitem [{\citenamefont {Lebed}\ \emph {et~al.}(2017)\citenamefont {Lebed}, \citenamefont {Mitchell},\ and\ \citenamefont {Swanson}}]{Lebed:2016hpi}%
  \BibitemOpen
  \bibfield  {author} {\bibinfo {author} {\bibfnamefont {R.~F.}\ \bibnamefont {Lebed}}, \bibinfo {author} {\bibfnamefont {R.~E.}\ \bibnamefont {Mitchell}}, \ and\ \bibinfo {author} {\bibfnamefont {E.~S.}\ \bibnamefont {Swanson}},\ }\href {\doibase 10.1016/j.ppnp.2016.11.003} {\bibfield  {journal} {\bibinfo  {journal} {Prog. Part. Nucl. Phys.}\ }\textbf {\bibinfo {volume} {93}},\ \bibinfo {pages} {143} (\bibinfo {year} {2017})},\ \Eprint {http://arxiv.org/abs/1610.04528} {arXiv:1610.04528 [hep-ph]} \BibitemShut {NoStop}%
\bibitem [{\citenamefont {Kou}\ and\ \citenamefont {Pene}(2005)}]{Kou:2005gt}%
  \BibitemOpen
  \bibfield  {author} {\bibinfo {author} {\bibfnamefont {E.}~\bibnamefont {Kou}}\ and\ \bibinfo {author} {\bibfnamefont {O.}~\bibnamefont {Pene}},\ }\href {\doibase 10.1016/j.physletb.2005.09.013} {\bibfield  {journal} {\bibinfo  {journal} {Phys. Lett. B}\ }\textbf {\bibinfo {volume} {631}},\ \bibinfo {pages} {164} (\bibinfo {year} {2005})},\ \Eprint {http://arxiv.org/abs/hep-ph/0507119} {arXiv:hep-ph/0507119} \BibitemShut {NoStop}%
\bibitem [{\citenamefont {Braaten}(2013)}]{Braaten:2013boa}%
  \BibitemOpen
  \bibfield  {author} {\bibinfo {author} {\bibfnamefont {E.}~\bibnamefont {Braaten}},\ }\href {\doibase 10.1103/PhysRevLett.111.162003} {\bibfield  {journal} {\bibinfo  {journal} {Phys. Rev. Lett.}\ }\textbf {\bibinfo {volume} {111}},\ \bibinfo {pages} {162003} (\bibinfo {year} {2013})},\ \Eprint {http://arxiv.org/abs/1305.6905} {arXiv:1305.6905 [hep-ph]} \BibitemShut {NoStop}%
\bibitem [{\citenamefont {Berwein}\ \emph {et~al.}(2015)\citenamefont {Berwein}, \citenamefont {Brambilla}, \citenamefont {Tarr\'us~Castell\`a},\ and\ \citenamefont {Vairo}}]{Berwein:2015vca}%
  \BibitemOpen
  \bibfield  {author} {\bibinfo {author} {\bibfnamefont {M.}~\bibnamefont {Berwein}}, \bibinfo {author} {\bibfnamefont {N.}~\bibnamefont {Brambilla}}, \bibinfo {author} {\bibfnamefont {J.}~\bibnamefont {Tarr\'us~Castell\`a}}, \ and\ \bibinfo {author} {\bibfnamefont {A.}~\bibnamefont {Vairo}},\ }\href {\doibase 10.1103/PhysRevD.92.114019} {\bibfield  {journal} {\bibinfo  {journal} {Phys. Rev. D}\ }\textbf {\bibinfo {volume} {92}},\ \bibinfo {pages} {114019} (\bibinfo {year} {2015})},\ \Eprint {http://arxiv.org/abs/1510.04299} {arXiv:1510.04299 [hep-ph]} \BibitemShut {NoStop}%
\bibitem [{\citenamefont {Minkowski}\ and\ \citenamefont {Ochs}(1999)}]{Minkowski:1998mf}%
  \BibitemOpen
  \bibfield  {author} {\bibinfo {author} {\bibfnamefont {P.}~\bibnamefont {Minkowski}}\ and\ \bibinfo {author} {\bibfnamefont {W.}~\bibnamefont {Ochs}},\ }\href {\doibase 10.1007/s100520050533} {\bibfield  {journal} {\bibinfo  {journal} {Eur. Phys. J. C}\ }\textbf {\bibinfo {volume} {9}},\ \bibinfo {pages} {283} (\bibinfo {year} {1999})},\ \Eprint {http://arxiv.org/abs/hep-ph/9811518} {arXiv:hep-ph/9811518} \BibitemShut {NoStop}%
\bibitem [{\citenamefont {Mathieu}\ \emph {et~al.}(2009)\citenamefont {Mathieu}, \citenamefont {Kochelev},\ and\ \citenamefont {Vento}}]{Mathieu:2008me}%
  \BibitemOpen
  \bibfield  {author} {\bibinfo {author} {\bibfnamefont {V.}~\bibnamefont {Mathieu}}, \bibinfo {author} {\bibfnamefont {N.}~\bibnamefont {Kochelev}}, \ and\ \bibinfo {author} {\bibfnamefont {V.}~\bibnamefont {Vento}},\ }\href {\doibase 10.1142/S0218301309012124} {\bibfield  {journal} {\bibinfo  {journal} {Int. J. Mod. Phys. E}\ }\textbf {\bibinfo {volume} {18}},\ \bibinfo {pages} {1} (\bibinfo {year} {2009})},\ \Eprint {http://arxiv.org/abs/0810.4453} {arXiv:0810.4453 [hep-ph]} \BibitemShut {NoStop}%
\bibitem [{\citenamefont {Chen}\ \emph {et~al.}(2021)\citenamefont {Chen}, \citenamefont {Chen},\ and\ \citenamefont {Zhu}}]{Chen:2021cjr}%
  \BibitemOpen
  \bibfield  {author} {\bibinfo {author} {\bibfnamefont {H.-X.}\ \bibnamefont {Chen}}, \bibinfo {author} {\bibfnamefont {W.}~\bibnamefont {Chen}}, \ and\ \bibinfo {author} {\bibfnamefont {S.-L.}\ \bibnamefont {Zhu}},\ }\href {\doibase 10.1103/PhysRevD.103.L091503} {\bibfield  {journal} {\bibinfo  {journal} {Phys. Rev. D}\ }\textbf {\bibinfo {volume} {103}},\ \bibinfo {pages} {L091503} (\bibinfo {year} {2021})},\ \Eprint {http://arxiv.org/abs/2103.17201} {arXiv:2103.17201 [hep-ph]} \BibitemShut {NoStop}%
\bibitem [{\citenamefont {Abazov}\ \emph {et~al.}(2021)\citenamefont {Abazov} \emph {et~al.}}]{D0:2020tig}%
  \BibitemOpen
  \bibfield  {author} {\bibinfo {author} {\bibfnamefont {V.~M.}\ \bibnamefont {Abazov}} \emph {et~al.} (\bibinfo {collaboration} {D0, TOTEM}),\ }\href {\doibase 10.1103/PhysRevLett.127.062003} {\bibfield  {journal} {\bibinfo  {journal} {Phys. Rev. Lett.}\ }\textbf {\bibinfo {volume} {127}},\ \bibinfo {pages} {062003} (\bibinfo {year} {2021})},\ \Eprint {http://arxiv.org/abs/2012.03981} {arXiv:2012.03981 [hep-ex]} \BibitemShut {NoStop}%
\bibitem [{\citenamefont {Gell-Mann}(1964)}]{Gell-Mann:1964ewy}%
  \BibitemOpen
  \bibfield  {author} {\bibinfo {author} {\bibfnamefont {M.}~\bibnamefont {Gell-Mann}},\ }\href {\doibase 10.1016/S0031-9163(64)92001-3} {\bibfield  {journal} {\bibinfo  {journal} {Phys. Lett.}\ }\textbf {\bibinfo {volume} {8}},\ \bibinfo {pages} {214} (\bibinfo {year} {1964})}\BibitemShut {NoStop}%
\bibitem [{\citenamefont {Jaffe}(1977)}]{Jaffe:1976ig}%
  \BibitemOpen
  \bibfield  {author} {\bibinfo {author} {\bibfnamefont {R.~L.}\ \bibnamefont {Jaffe}},\ }\href {\doibase 10.1103/PhysRevD.15.267} {\bibfield  {journal} {\bibinfo  {journal} {Phys. Rev. D}\ }\textbf {\bibinfo {volume} {15}},\ \bibinfo {pages} {267} (\bibinfo {year} {1977})}\BibitemShut {NoStop}%
\bibitem [{\citenamefont {Ader}\ \emph {et~al.}(1982)\citenamefont {Ader}, \citenamefont {Richard},\ and\ \citenamefont {Taxil}}]{Ader:1981db}%
  \BibitemOpen
  \bibfield  {author} {\bibinfo {author} {\bibfnamefont {J.~P.}\ \bibnamefont {Ader}}, \bibinfo {author} {\bibfnamefont {J.~M.}\ \bibnamefont {Richard}}, \ and\ \bibinfo {author} {\bibfnamefont {P.}~\bibnamefont {Taxil}},\ }\href {\doibase 10.1103/PhysRevD.25.2370} {\bibfield  {journal} {\bibinfo  {journal} {Phys. Rev. D}\ }\textbf {\bibinfo {volume} {25}},\ \bibinfo {pages} {2370} (\bibinfo {year} {1982})}\BibitemShut {NoStop}%
\bibitem [{\citenamefont {Maiani}\ \emph {et~al.}(2015)\citenamefont {Maiani}, \citenamefont {Polosa},\ and\ \citenamefont {Riquer}}]{Maiani:2015vwa}%
  \BibitemOpen
  \bibfield  {author} {\bibinfo {author} {\bibfnamefont {L.}~\bibnamefont {Maiani}}, \bibinfo {author} {\bibfnamefont {A.~D.}\ \bibnamefont {Polosa}}, \ and\ \bibinfo {author} {\bibfnamefont {V.}~\bibnamefont {Riquer}},\ }\href {\doibase 10.1016/j.physletb.2015.08.008} {\bibfield  {journal} {\bibinfo  {journal} {Phys. Lett. B}\ }\textbf {\bibinfo {volume} {749}},\ \bibinfo {pages} {289} (\bibinfo {year} {2015})}\BibitemShut {NoStop}%
\bibitem [{\citenamefont {Choi}\ \emph {et~al.}(2003)\citenamefont {Choi} \emph {et~al.}}]{Belle:2003nnu}%
  \BibitemOpen
  \bibfield  {author} {\bibinfo {author} {\bibfnamefont {S.~K.}\ \bibnamefont {Choi}} \emph {et~al.} (\bibinfo {collaboration} {Belle}),\ }\href {\doibase 10.1103/PhysRevLett.91.262001} {\bibfield  {journal} {\bibinfo  {journal} {Phys. Rev. Lett.}\ }\textbf {\bibinfo {volume} {91}},\ \bibinfo {pages} {262001} (\bibinfo {year} {2003})},\ \Eprint {http://arxiv.org/abs/hep-ex/0309032} {arXiv:hep-ex/0309032} \BibitemShut {NoStop}%
\bibitem [{\citenamefont {Chen}\ \emph {et~al.}(2016)\citenamefont {Chen}, \citenamefont {Chen}, \citenamefont {Liu},\ and\ \citenamefont {Zhu}}]{Chen:2016qju}%
  \BibitemOpen
  \bibfield  {author} {\bibinfo {author} {\bibfnamefont {H.-X.}\ \bibnamefont {Chen}}, \bibinfo {author} {\bibfnamefont {W.}~\bibnamefont {Chen}}, \bibinfo {author} {\bibfnamefont {X.}~\bibnamefont {Liu}}, \ and\ \bibinfo {author} {\bibfnamefont {S.-L.}\ \bibnamefont {Zhu}},\ }\href {\doibase 10.1016/j.physrep.2016.05.004} {\bibfield  {journal} {\bibinfo  {journal} {Phys. Rept.}\ }\textbf {\bibinfo {volume} {639}},\ \bibinfo {pages} {1} (\bibinfo {year} {2016})},\ \Eprint {http://arxiv.org/abs/1601.02092} {arXiv:1601.02092 [hep-ph]} \BibitemShut {NoStop}%
\bibitem [{\citenamefont {Liu}\ \emph {et~al.}(2019)\citenamefont {Liu}, \citenamefont {Chen}, \citenamefont {Chen}, \citenamefont {Liu},\ and\ \citenamefont {Zhu}}]{Liu:2019zoy}%
  \BibitemOpen
  \bibfield  {author} {\bibinfo {author} {\bibfnamefont {Y.-R.}\ \bibnamefont {Liu}}, \bibinfo {author} {\bibfnamefont {H.-X.}\ \bibnamefont {Chen}}, \bibinfo {author} {\bibfnamefont {W.}~\bibnamefont {Chen}}, \bibinfo {author} {\bibfnamefont {X.}~\bibnamefont {Liu}}, \ and\ \bibinfo {author} {\bibfnamefont {S.-L.}\ \bibnamefont {Zhu}},\ }\href {\doibase 10.1016/j.ppnp.2019.04.003} {\bibfield  {journal} {\bibinfo  {journal} {Prog. Part. Nucl. Phys.}\ }\textbf {\bibinfo {volume} {107}},\ \bibinfo {pages} {237} (\bibinfo {year} {2019})},\ \Eprint {http://arxiv.org/abs/1903.11976} {arXiv:1903.11976 [hep-ph]} \BibitemShut {NoStop}%
\bibitem [{\citenamefont {Esposito}\ \emph {et~al.}(2021)\citenamefont {Esposito}, \citenamefont {Ferreiro}, \citenamefont {Pilloni}, \citenamefont {Polosa},\ and\ \citenamefont {Salgado}}]{Esposito:2020ywk}%
  \BibitemOpen
  \bibfield  {author} {\bibinfo {author} {\bibfnamefont {A.}~\bibnamefont {Esposito}}, \bibinfo {author} {\bibfnamefont {E.~G.}\ \bibnamefont {Ferreiro}}, \bibinfo {author} {\bibfnamefont {A.}~\bibnamefont {Pilloni}}, \bibinfo {author} {\bibfnamefont {A.~D.}\ \bibnamefont {Polosa}}, \ and\ \bibinfo {author} {\bibfnamefont {C.~A.}\ \bibnamefont {Salgado}},\ }\href {\doibase 10.1140/epjc/s10052-021-09425-w} {\bibfield  {journal} {\bibinfo  {journal} {Eur. Phys. J. C}\ }\textbf {\bibinfo {volume} {81}},\ \bibinfo {pages} {669} (\bibinfo {year} {2021})},\ \Eprint {http://arxiv.org/abs/2006.15044} {arXiv:2006.15044 [hep-ph]} \BibitemShut {NoStop}%
\bibitem [{\citenamefont {Esposito}\ \emph {et~al.}(2025)\citenamefont {Esposito}, \citenamefont {Glioti}, \citenamefont {Germani},\ and\ \citenamefont {Polosa}}]{Esposito:2025hlp}%
  \BibitemOpen
  \bibfield  {author} {\bibinfo {author} {\bibfnamefont {A.}~\bibnamefont {Esposito}}, \bibinfo {author} {\bibfnamefont {A.}~\bibnamefont {Glioti}}, \bibinfo {author} {\bibfnamefont {D.}~\bibnamefont {Germani}}, \ and\ \bibinfo {author} {\bibfnamefont {A.~D.}\ \bibnamefont {Polosa}},\ }\href@noop {} {\  (\bibinfo {year} {2025})},\ \Eprint {http://arxiv.org/abs/2502.02505} {arXiv:2502.02505 [hep-ph]} \BibitemShut {NoStop}%
\bibitem [{\citenamefont {Tornqvist}(1994)}]{Tornqvist:1993ng}%
  \BibitemOpen
  \bibfield  {author} {\bibinfo {author} {\bibfnamefont {N.~A.}\ \bibnamefont {Tornqvist}},\ }\href {\doibase 10.1007/BF01413192} {\bibfield  {journal} {\bibinfo  {journal} {Z. Phys. C}\ }\textbf {\bibinfo {volume} {61}},\ \bibinfo {pages} {525} (\bibinfo {year} {1994})},\ \Eprint {http://arxiv.org/abs/hep-ph/9310247} {arXiv:hep-ph/9310247} \BibitemShut {NoStop}%
\bibitem [{\citenamefont {Braaten}\ and\ \citenamefont {Kusunoki}(2004)}]{Braaten:2003he}%
  \BibitemOpen
  \bibfield  {author} {\bibinfo {author} {\bibfnamefont {E.}~\bibnamefont {Braaten}}\ and\ \bibinfo {author} {\bibfnamefont {M.}~\bibnamefont {Kusunoki}},\ }\href {\doibase 10.1103/PhysRevD.69.074005} {\bibfield  {journal} {\bibinfo  {journal} {Phys. Rev. D}\ }\textbf {\bibinfo {volume} {69}},\ \bibinfo {pages} {074005} (\bibinfo {year} {2004})},\ \Eprint {http://arxiv.org/abs/hep-ph/0311147} {arXiv:hep-ph/0311147} \BibitemShut {NoStop}%
\bibitem [{\citenamefont {Guo}\ \emph {et~al.}(2013)\citenamefont {Guo}, \citenamefont {Hidalgo-Duque}, \citenamefont {Nieves},\ and\ \citenamefont {Valderrama}}]{Guo:2013sya}%
  \BibitemOpen
  \bibfield  {author} {\bibinfo {author} {\bibfnamefont {F.-K.}\ \bibnamefont {Guo}}, \bibinfo {author} {\bibfnamefont {C.}~\bibnamefont {Hidalgo-Duque}}, \bibinfo {author} {\bibfnamefont {J.}~\bibnamefont {Nieves}}, \ and\ \bibinfo {author} {\bibfnamefont {M.~P.}\ \bibnamefont {Valderrama}},\ }\href {\doibase 10.1103/PhysRevD.88.054007} {\bibfield  {journal} {\bibinfo  {journal} {Phys. Rev. D}\ }\textbf {\bibinfo {volume} {88}},\ \bibinfo {pages} {054007} (\bibinfo {year} {2013})},\ \Eprint {http://arxiv.org/abs/1303.6608} {arXiv:1303.6608 [hep-ph]} \BibitemShut {NoStop}%
\bibitem [{\citenamefont {Wang}(2014{\natexlab{a}})}]{Wang:2014gwa}%
  \BibitemOpen
  \bibfield  {author} {\bibinfo {author} {\bibfnamefont {Z.-G.}\ \bibnamefont {Wang}},\ }\href {\doibase 10.1140/epjc/s10052-014-2963-7} {\bibfield  {journal} {\bibinfo  {journal} {Eur. Phys. J. C}\ }\textbf {\bibinfo {volume} {74}},\ \bibinfo {pages} {2963} (\bibinfo {year} {2014}{\natexlab{a}})},\ \Eprint {http://arxiv.org/abs/1403.0810} {arXiv:1403.0810 [hep-ph]} \BibitemShut {NoStop}%
\bibitem [{\citenamefont {Mutuk}(2022)}]{Mutuk:2022ckn}%
  \BibitemOpen
  \bibfield  {author} {\bibinfo {author} {\bibfnamefont {H.}~\bibnamefont {Mutuk}},\ }\href {\doibase 10.1140/epjc/s10052-022-11120-3} {\bibfield  {journal} {\bibinfo  {journal} {Eur. Phys. J. C}\ }\textbf {\bibinfo {volume} {82}},\ \bibinfo {pages} {1142} (\bibinfo {year} {2022})},\ \Eprint {http://arxiv.org/abs/2211.14836} {arXiv:2211.14836 [hep-ph]} \BibitemShut {NoStop}%
\bibitem [{\citenamefont {Esposito}\ \emph {et~al.}(2023)\citenamefont {Esposito}, \citenamefont {Germani}, \citenamefont {Glioti}, \citenamefont {Polosa}, \citenamefont {Rattazzi},\ and\ \citenamefont {Tarquini}}]{Esposito:2023mxw}%
  \BibitemOpen
  \bibfield  {author} {\bibinfo {author} {\bibfnamefont {A.}~\bibnamefont {Esposito}}, \bibinfo {author} {\bibfnamefont {D.}~\bibnamefont {Germani}}, \bibinfo {author} {\bibfnamefont {A.}~\bibnamefont {Glioti}}, \bibinfo {author} {\bibfnamefont {A.~D.}\ \bibnamefont {Polosa}}, \bibinfo {author} {\bibfnamefont {R.}~\bibnamefont {Rattazzi}}, \ and\ \bibinfo {author} {\bibfnamefont {M.}~\bibnamefont {Tarquini}},\ }\href {\doibase 10.1016/j.physletb.2023.138285} {\bibfield  {journal} {\bibinfo  {journal} {Phys. Lett. B}\ }\textbf {\bibinfo {volume} {847}},\ \bibinfo {pages} {138285} (\bibinfo {year} {2023})},\ \Eprint {http://arxiv.org/abs/2307.11400} {arXiv:2307.11400 [hep-ph]} \BibitemShut {NoStop}%
\bibitem [{\citenamefont {Maiani}\ \emph {et~al.}(2005)\citenamefont {Maiani}, \citenamefont {Piccinini}, \citenamefont {Polosa},\ and\ \citenamefont {Riquer}}]{Maiani:2004vq}%
  \BibitemOpen
  \bibfield  {author} {\bibinfo {author} {\bibfnamefont {L.}~\bibnamefont {Maiani}}, \bibinfo {author} {\bibfnamefont {F.}~\bibnamefont {Piccinini}}, \bibinfo {author} {\bibfnamefont {A.~D.}\ \bibnamefont {Polosa}}, \ and\ \bibinfo {author} {\bibfnamefont {V.}~\bibnamefont {Riquer}},\ }\href {\doibase 10.1103/PhysRevD.71.014028} {\bibfield  {journal} {\bibinfo  {journal} {Phys. Rev. D}\ }\textbf {\bibinfo {volume} {71}},\ \bibinfo {pages} {014028} (\bibinfo {year} {2005})},\ \Eprint {http://arxiv.org/abs/hep-ph/0412098} {arXiv:hep-ph/0412098} \BibitemShut {NoStop}%
\bibitem [{\citenamefont {'t~Hooft}\ \emph {et~al.}(2008)\citenamefont {'t~Hooft}, \citenamefont {Isidori}, \citenamefont {Maiani}, \citenamefont {Polosa},\ and\ \citenamefont {Riquer}}]{tHooft:2008rus}%
  \BibitemOpen
  \bibfield  {author} {\bibinfo {author} {\bibfnamefont {G.}~\bibnamefont {'t~Hooft}}, \bibinfo {author} {\bibfnamefont {G.}~\bibnamefont {Isidori}}, \bibinfo {author} {\bibfnamefont {L.}~\bibnamefont {Maiani}}, \bibinfo {author} {\bibfnamefont {A.~D.}\ \bibnamefont {Polosa}}, \ and\ \bibinfo {author} {\bibfnamefont {V.}~\bibnamefont {Riquer}},\ }\href {\doibase 10.1016/j.physletb.2008.03.036} {\bibfield  {journal} {\bibinfo  {journal} {Phys. Lett. B}\ }\textbf {\bibinfo {volume} {662}},\ \bibinfo {pages} {424} (\bibinfo {year} {2008})},\ \Eprint {http://arxiv.org/abs/0801.2288} {arXiv:0801.2288 [hep-ph]} \BibitemShut {NoStop}%
\bibitem [{\citenamefont {Maiani}\ \emph {et~al.}(2018)\citenamefont {Maiani}, \citenamefont {Polosa},\ and\ \citenamefont {Riquer}}]{Maiani:2017kyi}%
  \BibitemOpen
  \bibfield  {author} {\bibinfo {author} {\bibfnamefont {L.}~\bibnamefont {Maiani}}, \bibinfo {author} {\bibfnamefont {A.~D.}\ \bibnamefont {Polosa}}, \ and\ \bibinfo {author} {\bibfnamefont {V.}~\bibnamefont {Riquer}},\ }\href {\doibase 10.1016/j.physletb.2018.01.039} {\bibfield  {journal} {\bibinfo  {journal} {Phys. Lett. B}\ }\textbf {\bibinfo {volume} {778}},\ \bibinfo {pages} {247} (\bibinfo {year} {2018})},\ \Eprint {http://arxiv.org/abs/1712.05296} {arXiv:1712.05296 [hep-ph]} \BibitemShut {NoStop}%
\bibitem [{\citenamefont {Wang}(2014{\natexlab{b}})}]{Wang:2013exa}%
  \BibitemOpen
  \bibfield  {author} {\bibinfo {author} {\bibfnamefont {Z.-G.}\ \bibnamefont {Wang}},\ }\href {\doibase 10.1140/epjc/s10052-014-2874-7} {\bibfield  {journal} {\bibinfo  {journal} {Eur. Phys. J. C}\ }\textbf {\bibinfo {volume} {74}},\ \bibinfo {pages} {2874} (\bibinfo {year} {2014}{\natexlab{b}})},\ \Eprint {http://arxiv.org/abs/1311.1046} {arXiv:1311.1046 [hep-ph]} \BibitemShut {NoStop}%
\bibitem [{\citenamefont {Grinstein}\ \emph {et~al.}(2024)\citenamefont {Grinstein}, \citenamefont {Maiani},\ and\ \citenamefont {Polosa}}]{Grinstein:2024rcu}%
  \BibitemOpen
  \bibfield  {author} {\bibinfo {author} {\bibfnamefont {B.}~\bibnamefont {Grinstein}}, \bibinfo {author} {\bibfnamefont {L.}~\bibnamefont {Maiani}}, \ and\ \bibinfo {author} {\bibfnamefont {A.~D.}\ \bibnamefont {Polosa}},\ }\href {\doibase 10.1103/PhysRevD.109.074009} {\bibfield  {journal} {\bibinfo  {journal} {Phys. Rev. D}\ }\textbf {\bibinfo {volume} {109}},\ \bibinfo {pages} {074009} (\bibinfo {year} {2024})},\ \Eprint {http://arxiv.org/abs/2401.11623} {arXiv:2401.11623 [hep-ph]} \BibitemShut {NoStop}%
\bibitem [{\citenamefont {Dubynskiy}\ and\ \citenamefont {Voloshin}(2008)}]{Dubynskiy:2008mq}%
  \BibitemOpen
  \bibfield  {author} {\bibinfo {author} {\bibfnamefont {S.}~\bibnamefont {Dubynskiy}}\ and\ \bibinfo {author} {\bibfnamefont {M.~B.}\ \bibnamefont {Voloshin}},\ }\href {\doibase 10.1016/j.physletb.2008.07.086} {\bibfield  {journal} {\bibinfo  {journal} {Phys. Lett. B}\ }\textbf {\bibinfo {volume} {666}},\ \bibinfo {pages} {344} (\bibinfo {year} {2008})},\ \Eprint {http://arxiv.org/abs/0803.2224} {arXiv:0803.2224 [hep-ph]} \BibitemShut {NoStop}%
\bibitem [{\citenamefont {Voloshin}(2013)}]{Voloshin:2013dpa}%
  \BibitemOpen
  \bibfield  {author} {\bibinfo {author} {\bibfnamefont {M.~B.}\ \bibnamefont {Voloshin}},\ }\href {\doibase 10.1103/PhysRevD.87.091501} {\bibfield  {journal} {\bibinfo  {journal} {Phys. Rev. D}\ }\textbf {\bibinfo {volume} {87}},\ \bibinfo {pages} {091501} (\bibinfo {year} {2013})},\ \Eprint {http://arxiv.org/abs/1304.0380} {arXiv:1304.0380 [hep-ph]} \BibitemShut {NoStop}%
\bibitem [{\citenamefont {Guo}\ \emph {et~al.}(2018)\citenamefont {Guo}, \citenamefont {Hanhart}, \citenamefont {Mei\ss{}ner}, \citenamefont {Wang}, \citenamefont {Zhao},\ and\ \citenamefont {Zou}}]{Guo:2017jvc}%
  \BibitemOpen
  \bibfield  {author} {\bibinfo {author} {\bibfnamefont {F.-K.}\ \bibnamefont {Guo}}, \bibinfo {author} {\bibfnamefont {C.}~\bibnamefont {Hanhart}}, \bibinfo {author} {\bibfnamefont {U.-G.}\ \bibnamefont {Mei\ss{}ner}}, \bibinfo {author} {\bibfnamefont {Q.}~\bibnamefont {Wang}}, \bibinfo {author} {\bibfnamefont {Q.}~\bibnamefont {Zhao}}, \ and\ \bibinfo {author} {\bibfnamefont {B.-S.}\ \bibnamefont {Zou}},\ }\href {\doibase 10.1103/RevModPhys.90.015004} {\bibfield  {journal} {\bibinfo  {journal} {Rev. Mod. Phys.}\ }\textbf {\bibinfo {volume} {90}},\ \bibinfo {pages} {015004} (\bibinfo {year} {2018})},\ \bibinfo {note} {[Erratum: Rev.Mod.Phys. 94, 029901 (2022)]},\ \Eprint {http://arxiv.org/abs/1705.00141} {arXiv:1705.00141 [hep-ph]} \BibitemShut {NoStop}%
\bibitem [{\citenamefont {Ferretti}\ \emph {et~al.}(2019)\citenamefont {Ferretti}, \citenamefont {Santopinto}, \citenamefont {Naeem~Anwar},\ and\ \citenamefont {Bedolla}}]{Ferretti:2018ojb}%
  \BibitemOpen
  \bibfield  {author} {\bibinfo {author} {\bibfnamefont {J.}~\bibnamefont {Ferretti}}, \bibinfo {author} {\bibfnamefont {E.}~\bibnamefont {Santopinto}}, \bibinfo {author} {\bibfnamefont {M.}~\bibnamefont {Naeem~Anwar}}, \ and\ \bibinfo {author} {\bibfnamefont {M.~A.}\ \bibnamefont {Bedolla}},\ }\href {\doibase 10.1016/j.physletb.2018.09.047} {\bibfield  {journal} {\bibinfo  {journal} {Phys. Lett. B}\ }\textbf {\bibinfo {volume} {789}},\ \bibinfo {pages} {562} (\bibinfo {year} {2019})},\ \Eprint {http://arxiv.org/abs/1807.01207} {arXiv:1807.01207 [hep-ph]} \BibitemShut {NoStop}%
\bibitem [{\citenamefont {Ferretti}\ and\ \citenamefont {Santopinto}(2020)}]{Ferretti:2020ewe}%
  \BibitemOpen
  \bibfield  {author} {\bibinfo {author} {\bibfnamefont {J.}~\bibnamefont {Ferretti}}\ and\ \bibinfo {author} {\bibfnamefont {E.}~\bibnamefont {Santopinto}},\ }\href {\doibase 10.1007/JHEP04(2020)119} {\bibfield  {journal} {\bibinfo  {journal} {JHEP}\ }\textbf {\bibinfo {volume} {04}},\ \bibinfo {pages} {119} (\bibinfo {year} {2020})},\ \Eprint {http://arxiv.org/abs/2001.01067} {arXiv:2001.01067 [hep-ph]} \BibitemShut {NoStop}%
\bibitem [{\citenamefont {Aaij}\ \emph {et~al.}(2022{\natexlab{a}})\citenamefont {Aaij} \emph {et~al.}}]{LHCb:2021vvq}%
  \BibitemOpen
  \bibfield  {author} {\bibinfo {author} {\bibfnamefont {R.}~\bibnamefont {Aaij}} \emph {et~al.} (\bibinfo {collaboration} {LHCb}),\ }\href {\doibase 10.1038/s41567-022-01614-y} {\bibfield  {journal} {\bibinfo  {journal} {Nature Phys.}\ }\textbf {\bibinfo {volume} {18}},\ \bibinfo {pages} {751} (\bibinfo {year} {2022}{\natexlab{a}})},\ \Eprint {http://arxiv.org/abs/2109.01038} {arXiv:2109.01038 [hep-ex]} \BibitemShut {NoStop}%
\bibitem [{\citenamefont {Aaij}\ \emph {et~al.}(2022{\natexlab{b}})\citenamefont {Aaij} \emph {et~al.}}]{LHCb:2021auc}%
  \BibitemOpen
  \bibfield  {author} {\bibinfo {author} {\bibfnamefont {R.}~\bibnamefont {Aaij}} \emph {et~al.} (\bibinfo {collaboration} {LHCb}),\ }\href {\doibase 10.1038/s41467-022-30206-w} {\bibfield  {journal} {\bibinfo  {journal} {Nature Commun.}\ }\textbf {\bibinfo {volume} {13}},\ \bibinfo {pages} {3351} (\bibinfo {year} {2022}{\natexlab{b}})},\ \Eprint {http://arxiv.org/abs/2109.01056} {arXiv:2109.01056 [hep-ex]} \BibitemShut {NoStop}%
\bibitem [{\citenamefont {Aaij}\ \emph {et~al.}(2020)\citenamefont {Aaij} \emph {et~al.}}]{LHCb:2020bwg}%
  \BibitemOpen
  \bibfield  {author} {\bibinfo {author} {\bibfnamefont {R.}~\bibnamefont {Aaij}} \emph {et~al.} (\bibinfo {collaboration} {LHCb}),\ }\href {\doibase 10.1016/j.scib.2020.08.032} {\bibfield  {journal} {\bibinfo  {journal} {Sci. Bull.}\ }\textbf {\bibinfo {volume} {65}},\ \bibinfo {pages} {1983} (\bibinfo {year} {2020})},\ \Eprint {http://arxiv.org/abs/2006.16957} {arXiv:2006.16957 [hep-ex]} \BibitemShut {NoStop}%
\bibitem [{\citenamefont {Chen}\ \emph {et~al.}(2023)\citenamefont {Chen}, \citenamefont {Chen}, \citenamefont {Liu}, \citenamefont {Liu},\ and\ \citenamefont {Zhu}}]{Chen:2022asf}%
  \BibitemOpen
  \bibfield  {author} {\bibinfo {author} {\bibfnamefont {H.-X.}\ \bibnamefont {Chen}}, \bibinfo {author} {\bibfnamefont {W.}~\bibnamefont {Chen}}, \bibinfo {author} {\bibfnamefont {X.}~\bibnamefont {Liu}}, \bibinfo {author} {\bibfnamefont {Y.-R.}\ \bibnamefont {Liu}}, \ and\ \bibinfo {author} {\bibfnamefont {S.-L.}\ \bibnamefont {Zhu}},\ }\href {\doibase 10.1088/1361-6633/aca3b6} {\bibfield  {journal} {\bibinfo  {journal} {Rept. Prog. Phys.}\ }\textbf {\bibinfo {volume} {86}},\ \bibinfo {pages} {026201} (\bibinfo {year} {2023})},\ \Eprint {http://arxiv.org/abs/2204.02649} {arXiv:2204.02649 [hep-ph]} \BibitemShut {NoStop}%
\bibitem [{\citenamefont {Belov}\ \emph {et~al.}(2025)\citenamefont {Belov}, \citenamefont {Giachino},\ and\ \citenamefont {Santopinto}}]{Belov:2024qyi}%
  \BibitemOpen
  \bibfield  {author} {\bibinfo {author} {\bibfnamefont {I.}~\bibnamefont {Belov}}, \bibinfo {author} {\bibfnamefont {A.}~\bibnamefont {Giachino}}, \ and\ \bibinfo {author} {\bibfnamefont {E.}~\bibnamefont {Santopinto}},\ }\href {\doibase 10.1007/JHEP01(2025)093} {\bibfield  {journal} {\bibinfo  {journal} {JHEP}\ }\textbf {\bibinfo {volume} {01}},\ \bibinfo {pages} {093} (\bibinfo {year} {2025})},\ \Eprint {http://arxiv.org/abs/2409.12070} {arXiv:2409.12070 [hep-ph]} \BibitemShut {NoStop}%
\bibitem [{\citenamefont {Pineda}(2012)}]{Pineda:2011dg}%
  \BibitemOpen
  \bibfield  {author} {\bibinfo {author} {\bibfnamefont {A.}~\bibnamefont {Pineda}},\ }\href {\doibase 10.1016/j.ppnp.2012.01.038} {\bibfield  {journal} {\bibinfo  {journal} {Prog. Part. Nucl. Phys.}\ }\textbf {\bibinfo {volume} {67}},\ \bibinfo {pages} {735} (\bibinfo {year} {2012})},\ \Eprint {http://arxiv.org/abs/1111.0165} {arXiv:1111.0165 [hep-ph]} \BibitemShut {NoStop}%
\bibitem [{\citenamefont {Wu}\ \emph {et~al.}(2018)\citenamefont {Wu}, \citenamefont {Liu}, \citenamefont {Chen}, \citenamefont {Liu},\ and\ \citenamefont {Zhu}}]{Wu:2016vtq}%
  \BibitemOpen
  \bibfield  {author} {\bibinfo {author} {\bibfnamefont {J.}~\bibnamefont {Wu}}, \bibinfo {author} {\bibfnamefont {Y.-R.}\ \bibnamefont {Liu}}, \bibinfo {author} {\bibfnamefont {K.}~\bibnamefont {Chen}}, \bibinfo {author} {\bibfnamefont {X.}~\bibnamefont {Liu}}, \ and\ \bibinfo {author} {\bibfnamefont {S.-L.}\ \bibnamefont {Zhu}},\ }\href {\doibase 10.1103/PhysRevD.97.094015} {\bibfield  {journal} {\bibinfo  {journal} {Phys. Rev. D}\ }\textbf {\bibinfo {volume} {97}},\ \bibinfo {pages} {094015} (\bibinfo {year} {2018})},\ \Eprint {http://arxiv.org/abs/1605.01134} {arXiv:1605.01134 [hep-ph]} \BibitemShut {NoStop}%
\bibitem [{\citenamefont {Maciu\l{}a}\ \emph {et~al.}(2021)\citenamefont {Maciu\l{}a}, \citenamefont {Sch\"afer},\ and\ \citenamefont {Szczurek}}]{Maciula:2020wri}%
  \BibitemOpen
  \bibfield  {author} {\bibinfo {author} {\bibfnamefont {R.}~\bibnamefont {Maciu\l{}a}}, \bibinfo {author} {\bibfnamefont {W.}~\bibnamefont {Sch\"afer}}, \ and\ \bibinfo {author} {\bibfnamefont {A.}~\bibnamefont {Szczurek}},\ }\href {\doibase 10.1016/j.physletb.2020.136010} {\bibfield  {journal} {\bibinfo  {journal} {Phys. Lett. B}\ }\textbf {\bibinfo {volume} {812}},\ \bibinfo {pages} {136010} (\bibinfo {year} {2021})},\ \Eprint {http://arxiv.org/abs/2009.02100} {arXiv:2009.02100 [hep-ph]} \BibitemShut {NoStop}%
\bibitem [{\citenamefont {Carvalho}\ \emph {et~al.}(2016)\citenamefont {Carvalho}, \citenamefont {Cazaroto}, \citenamefont {Gon\c{c}alves},\ and\ \citenamefont {Navarra}}]{Carvalho:2015nqf}%
  \BibitemOpen
  \bibfield  {author} {\bibinfo {author} {\bibfnamefont {F.}~\bibnamefont {Carvalho}}, \bibinfo {author} {\bibfnamefont {E.~R.}\ \bibnamefont {Cazaroto}}, \bibinfo {author} {\bibfnamefont {V.~P.}\ \bibnamefont {Gon\c{c}alves}}, \ and\ \bibinfo {author} {\bibfnamefont {F.~S.}\ \bibnamefont {Navarra}},\ }\href {\doibase 10.1103/PhysRevD.93.034004} {\bibfield  {journal} {\bibinfo  {journal} {Phys. Rev. D}\ }\textbf {\bibinfo {volume} {93}},\ \bibinfo {pages} {034004} (\bibinfo {year} {2016})},\ \Eprint {http://arxiv.org/abs/1511.05209} {arXiv:1511.05209 [hep-ph]} \BibitemShut {NoStop}%
\bibitem [{\citenamefont {Abreu}\ \emph {et~al.}(2024)\citenamefont {Abreu}, \citenamefont {Carvalho}, \citenamefont {Oliveira},\ and\ \citenamefont {Gon\c{c}alves}}]{Abreu:2023wwg}%
  \BibitemOpen
  \bibfield  {author} {\bibinfo {author} {\bibfnamefont {L.~M.}\ \bibnamefont {Abreu}}, \bibinfo {author} {\bibfnamefont {F.}~\bibnamefont {Carvalho}}, \bibinfo {author} {\bibfnamefont {J.~V.~C.}\ \bibnamefont {Oliveira}}, \ and\ \bibinfo {author} {\bibfnamefont {V.~P.}\ \bibnamefont {Gon\c{c}alves}},\ }\href {\doibase 10.1140/epjc/s10052-024-12847-x} {\bibfield  {journal} {\bibinfo  {journal} {Eur. Phys. J. C}\ }\textbf {\bibinfo {volume} {84}},\ \bibinfo {pages} {470} (\bibinfo {year} {2024})},\ \Eprint {http://arxiv.org/abs/2306.12731} {arXiv:2306.12731 [hep-ph]} \BibitemShut {NoStop}%
\bibitem [{\citenamefont {Cisek}\ \emph {et~al.}(2022)\citenamefont {Cisek}, \citenamefont {Sch\"afer},\ and\ \citenamefont {Szczurek}}]{Cisek:2022uqx}%
  \BibitemOpen
  \bibfield  {author} {\bibinfo {author} {\bibfnamefont {A.}~\bibnamefont {Cisek}}, \bibinfo {author} {\bibfnamefont {W.}~\bibnamefont {Sch\"afer}}, \ and\ \bibinfo {author} {\bibfnamefont {A.}~\bibnamefont {Szczurek}},\ }\href {\doibase 10.1140/epjc/s10052-022-11029-x} {\bibfield  {journal} {\bibinfo  {journal} {Eur. Phys. J. C}\ }\textbf {\bibinfo {volume} {82}},\ \bibinfo {pages} {1062} (\bibinfo {year} {2022})},\ \Eprint {http://arxiv.org/abs/2203.07827} {arXiv:2203.07827 [hep-ph]} \BibitemShut {NoStop}%
\bibitem [{\citenamefont {Chatrchyan}\ \emph {et~al.}(2013)\citenamefont {Chatrchyan} \emph {et~al.}}]{CMS:2013fpt}%
  \BibitemOpen
  \bibfield  {author} {\bibinfo {author} {\bibfnamefont {S.}~\bibnamefont {Chatrchyan}} \emph {et~al.} (\bibinfo {collaboration} {CMS}),\ }\href {\doibase 10.1007/JHEP04(2013)154} {\bibfield  {journal} {\bibinfo  {journal} {JHEP}\ }\textbf {\bibinfo {volume} {04}},\ \bibinfo {pages} {154} (\bibinfo {year} {2013})},\ \Eprint {http://arxiv.org/abs/1302.3968} {arXiv:1302.3968 [hep-ex]} \BibitemShut {NoStop}%
\bibitem [{\citenamefont {Aaboud}\ \emph {et~al.}(2017)\citenamefont {Aaboud} \emph {et~al.}}]{ATLAS:2016kwu}%
  \BibitemOpen
  \bibfield  {author} {\bibinfo {author} {\bibfnamefont {M.}~\bibnamefont {Aaboud}} \emph {et~al.} (\bibinfo {collaboration} {ATLAS}),\ }\href {\doibase 10.1007/JHEP01(2017)117} {\bibfield  {journal} {\bibinfo  {journal} {JHEP}\ }\textbf {\bibinfo {volume} {01}},\ \bibinfo {pages} {117} (\bibinfo {year} {2017})},\ \Eprint {http://arxiv.org/abs/1610.09303} {arXiv:1610.09303 [hep-ex]} \BibitemShut {NoStop}%
\bibitem [{\citenamefont {Aaij}\ \emph {et~al.}(2022{\natexlab{c}})\citenamefont {Aaij} \emph {et~al.}}]{LHCb:2021ten}%
  \BibitemOpen
  \bibfield  {author} {\bibinfo {author} {\bibfnamefont {R.}~\bibnamefont {Aaij}} \emph {et~al.} (\bibinfo {collaboration} {LHCb}),\ }\href {\doibase 10.1007/JHEP01(2022)131} {\bibfield  {journal} {\bibinfo  {journal} {JHEP}\ }\textbf {\bibinfo {volume} {01}},\ \bibinfo {pages} {131} (\bibinfo {year} {2022}{\natexlab{c}})},\ \Eprint {http://arxiv.org/abs/2109.07360} {arXiv:2109.07360 [hep-ex]} \BibitemShut {NoStop}%
\bibitem [{\citenamefont {Celiberto}\ and\ \citenamefont {Gatto}(2025)}]{Celiberto:2024beg}%
  \BibitemOpen
  \bibfield  {author} {\bibinfo {author} {\bibfnamefont {F.~G.}\ \bibnamefont {Celiberto}}\ and\ \bibinfo {author} {\bibfnamefont {G.}~\bibnamefont {Gatto}},\ }\href {\doibase 10.1103/PhysRevD.111.034037} {\bibfield  {journal} {\bibinfo  {journal} {Phys. Rev. D}\ }\textbf {\bibinfo {volume} {111}},\ \bibinfo {pages} {034037} (\bibinfo {year} {2025})},\ \Eprint {http://arxiv.org/abs/2412.10549} {arXiv:2412.10549 [hep-ph]} \BibitemShut {NoStop}%
\bibitem [{\citenamefont {Mele}\ and\ \citenamefont {Nason}(1991)}]{Mele:1990cw}%
  \BibitemOpen
  \bibfield  {author} {\bibinfo {author} {\bibfnamefont {B.}~\bibnamefont {Mele}}\ and\ \bibinfo {author} {\bibfnamefont {P.}~\bibnamefont {Nason}},\ }\href {\doibase 10.1016/0550-3213(91)90597-Q} {\bibfield  {journal} {\bibinfo  {journal} {Nucl. Phys. B}\ }\textbf {\bibinfo {volume} {361}},\ \bibinfo {pages} {626} (\bibinfo {year} {1991})},\ \bibinfo {note} {[Erratum: Nucl.Phys.B 921, 841--842 (2017)]}\BibitemShut {NoStop}%
\bibitem [{\citenamefont {Cacciari}\ and\ \citenamefont {Greco}(1994{\natexlab{a}})}]{Cacciari:1993mq}%
  \BibitemOpen
  \bibfield  {author} {\bibinfo {author} {\bibfnamefont {M.}~\bibnamefont {Cacciari}}\ and\ \bibinfo {author} {\bibfnamefont {M.}~\bibnamefont {Greco}},\ }\href {\doibase 10.1016/0550-3213(94)90515-0} {\bibfield  {journal} {\bibinfo  {journal} {Nucl. Phys. B}\ }\textbf {\bibinfo {volume} {421}},\ \bibinfo {pages} {530} (\bibinfo {year} {1994}{\natexlab{a}})},\ \Eprint {http://arxiv.org/abs/hep-ph/9311260} {arXiv:hep-ph/9311260} \BibitemShut {NoStop}%
\bibitem [{\citenamefont {Feng}\ \emph {et~al.}(2022)\citenamefont {Feng}, \citenamefont {Huang}, \citenamefont {Jia}, \citenamefont {Sang}, \citenamefont {Xiong},\ and\ \citenamefont {Zhang}}]{Feng:2020riv}%
  \BibitemOpen
  \bibfield  {author} {\bibinfo {author} {\bibfnamefont {F.}~\bibnamefont {Feng}}, \bibinfo {author} {\bibfnamefont {Y.}~\bibnamefont {Huang}}, \bibinfo {author} {\bibfnamefont {Y.}~\bibnamefont {Jia}}, \bibinfo {author} {\bibfnamefont {W.-L.}\ \bibnamefont {Sang}}, \bibinfo {author} {\bibfnamefont {X.}~\bibnamefont {Xiong}}, \ and\ \bibinfo {author} {\bibfnamefont {J.-Y.}\ \bibnamefont {Zhang}},\ }\href {\doibase 10.1103/PhysRevD.106.114029} {\bibfield  {journal} {\bibinfo  {journal} {Phys. Rev. D}\ }\textbf {\bibinfo {volume} {106}},\ \bibinfo {pages} {114029} (\bibinfo {year} {2022})},\ \Eprint {http://arxiv.org/abs/2009.08450} {arXiv:2009.08450 [hep-ph]} \BibitemShut {NoStop}%
\bibitem [{\citenamefont {Bai}\ \emph {et~al.}(2024)\citenamefont {Bai}, \citenamefont {Feng}, \citenamefont {Gan}, \citenamefont {Huang}, \citenamefont {Sang},\ and\ \citenamefont {Zhang}}]{Bai:2024ezn}%
  \BibitemOpen
  \bibfield  {author} {\bibinfo {author} {\bibfnamefont {X.-W.}\ \bibnamefont {Bai}}, \bibinfo {author} {\bibfnamefont {F.}~\bibnamefont {Feng}}, \bibinfo {author} {\bibfnamefont {C.-M.}\ \bibnamefont {Gan}}, \bibinfo {author} {\bibfnamefont {Y.}~\bibnamefont {Huang}}, \bibinfo {author} {\bibfnamefont {W.-L.}\ \bibnamefont {Sang}}, \ and\ \bibinfo {author} {\bibfnamefont {H.-F.}\ \bibnamefont {Zhang}},\ }\href {\doibase 10.1007/JHEP09(2024)002} {\bibfield  {journal} {\bibinfo  {journal} {JHEP}\ }\textbf {\bibinfo {volume} {09}},\ \bibinfo {pages} {002} (\bibinfo {year} {2024})},\ \Eprint {http://arxiv.org/abs/2404.13889} {arXiv:2404.13889 [hep-ph]} \BibitemShut {NoStop}%
\bibitem [{\citenamefont {Caswell}\ and\ \citenamefont {Lepage}(1986)}]{Caswell:1985ui}%
  \BibitemOpen
  \bibfield  {author} {\bibinfo {author} {\bibfnamefont {W.~E.}\ \bibnamefont {Caswell}}\ and\ \bibinfo {author} {\bibfnamefont {G.~P.}\ \bibnamefont {Lepage}},\ }\href {\doibase 10.1016/0370-2693(86)91297-9} {\bibfield  {journal} {\bibinfo  {journal} {Phys. Lett. B}\ }\textbf {\bibinfo {volume} {167}},\ \bibinfo {pages} {437} (\bibinfo {year} {1986})}\BibitemShut {NoStop}%
\bibitem [{\citenamefont {Bodwin}\ \emph {et~al.}(1995)\citenamefont {Bodwin}, \citenamefont {Braaten},\ and\ \citenamefont {Lepage}}]{Bodwin:1994jh}%
  \BibitemOpen
  \bibfield  {author} {\bibinfo {author} {\bibfnamefont {G.~T.}\ \bibnamefont {Bodwin}}, \bibinfo {author} {\bibfnamefont {E.}~\bibnamefont {Braaten}}, \ and\ \bibinfo {author} {\bibfnamefont {G.~P.}\ \bibnamefont {Lepage}},\ }\href {\doibase 10.1103/PhysRevD.55.5853} {\bibfield  {journal} {\bibinfo  {journal} {Phys. Rev. D}\ }\textbf {\bibinfo {volume} {51}},\ \bibinfo {pages} {1125} (\bibinfo {year} {1995})},\ \bibinfo {note} {[Erratum: Phys.Rev.D 55, 5853 (1997)]},\ \Eprint {http://arxiv.org/abs/hep-ph/9407339} {arXiv:hep-ph/9407339} \BibitemShut {NoStop}%
\bibitem [{\citenamefont {Celiberto}(2024{\natexlab{a}})}]{Celiberto:2024mex}%
  \BibitemOpen
  \bibfield  {author} {\bibinfo {author} {\bibfnamefont {F.~G.}\ \bibnamefont {Celiberto}},\ }in\ \href@noop {} {\emph {\bibinfo {booktitle} {{58th Rencontres de Moriond on QCD and High Energy Interactions}}}}\ (\bibinfo {year} {2024})\ \Eprint {http://arxiv.org/abs/2405.08221} {arXiv:2405.08221 [hep-ph]} \BibitemShut {NoStop}%
\bibitem [{\citenamefont {Celiberto}(2025{\natexlab{a}})}]{Celiberto:2024bxu}%
  \BibitemOpen
  \bibfield  {author} {\bibinfo {author} {\bibfnamefont {F.~G.}\ \bibnamefont {Celiberto}},\ }\href {\doibase 10.22323/1.469.0168} {\bibfield  {journal} {\bibinfo  {journal} {PoS}\ }\textbf {\bibinfo {volume} {DIS2024}},\ \bibinfo {pages} {168} (\bibinfo {year} {2025}{\natexlab{a}})},\ \Eprint {http://arxiv.org/abs/2406.10779} {arXiv:2406.10779 [hep-ph]} \BibitemShut {NoStop}%
\bibitem [{\citenamefont {Giacosa}(2007)}]{Giacosa:2006tf}%
  \BibitemOpen
  \bibfield  {author} {\bibinfo {author} {\bibfnamefont {F.}~\bibnamefont {Giacosa}},\ }\href {\doibase 10.1103/PhysRevD.75.054007} {\bibfield  {journal} {\bibinfo  {journal} {Phys. Rev. D}\ }\textbf {\bibinfo {volume} {75}},\ \bibinfo {pages} {054007} (\bibinfo {year} {2007})},\ \Eprint {http://arxiv.org/abs/hep-ph/0611388} {arXiv:hep-ph/0611388} \BibitemShut {NoStop}%
\bibitem [{\citenamefont {Kim}\ \emph {et~al.}(2018)\citenamefont {Kim}, \citenamefont {Kim}, \citenamefont {Cheoun},\ and\ \citenamefont {Oka}}]{Kim:2017yvd}%
  \BibitemOpen
  \bibfield  {author} {\bibinfo {author} {\bibfnamefont {H.}~\bibnamefont {Kim}}, \bibinfo {author} {\bibfnamefont {K.~S.}\ \bibnamefont {Kim}}, \bibinfo {author} {\bibfnamefont {M.-K.}\ \bibnamefont {Cheoun}}, \ and\ \bibinfo {author} {\bibfnamefont {M.}~\bibnamefont {Oka}},\ }\href {\doibase 10.1103/PhysRevD.97.094005} {\bibfield  {journal} {\bibinfo  {journal} {Phys. Rev. D}\ }\textbf {\bibinfo {volume} {97}},\ \bibinfo {pages} {094005} (\bibinfo {year} {2018})},\ \Eprint {http://arxiv.org/abs/1711.08213} {arXiv:1711.08213 [hep-ph]} \BibitemShut {NoStop}%
\bibitem [{\citenamefont {Kim}\ \emph {et~al.}(2019)\citenamefont {Kim}, \citenamefont {Kim}, \citenamefont {Cheoun}, \citenamefont {Jido},\ and\ \citenamefont {Oka}}]{Kim:2018zob}%
  \BibitemOpen
  \bibfield  {author} {\bibinfo {author} {\bibfnamefont {H.}~\bibnamefont {Kim}}, \bibinfo {author} {\bibfnamefont {K.~S.}\ \bibnamefont {Kim}}, \bibinfo {author} {\bibfnamefont {M.-K.}\ \bibnamefont {Cheoun}}, \bibinfo {author} {\bibfnamefont {D.}~\bibnamefont {Jido}}, \ and\ \bibinfo {author} {\bibfnamefont {M.}~\bibnamefont {Oka}},\ }\href {\doibase 10.1103/PhysRevD.99.014005} {\bibfield  {journal} {\bibinfo  {journal} {Phys. Rev. D}\ }\textbf {\bibinfo {volume} {99}},\ \bibinfo {pages} {014005} (\bibinfo {year} {2019})},\ \Eprint {http://arxiv.org/abs/1811.00187} {arXiv:1811.00187 [hep-ph]} \BibitemShut {NoStop}%
\bibitem [{\citenamefont {Zhang}\ and\ \citenamefont {Wang}(2025)}]{Zhang:2025xee}%
  \BibitemOpen
  \bibfield  {author} {\bibinfo {author} {\bibfnamefont {F.-Y.}\ \bibnamefont {Zhang}}\ and\ \bibinfo {author} {\bibfnamefont {L.-M.}\ \bibnamefont {Wang}},\ }\href@noop {} {\  (\bibinfo {year} {2025})},\ \Eprint {http://arxiv.org/abs/2503.01443} {arXiv:2503.01443 [hep-ph]} \BibitemShut {NoStop}%
\bibitem [{\citenamefont {Swanson}(2023)}]{Swanson:2023zlm}%
  \BibitemOpen
  \bibfield  {author} {\bibinfo {author} {\bibfnamefont {E.~S.}\ \bibnamefont {Swanson}},\ }\href {\doibase 10.1103/PhysRevD.107.074028} {\bibfield  {journal} {\bibinfo  {journal} {Phys. Rev. D}\ }\textbf {\bibinfo {volume} {107}},\ \bibinfo {pages} {074028} (\bibinfo {year} {2023})},\ \Eprint {http://arxiv.org/abs/2302.01372} {arXiv:2302.01372 [hep-ph]} \BibitemShut {NoStop}%
\bibitem [{\citenamefont {Wang}\ \emph {et~al.}(2008)\citenamefont {Wang}, \citenamefont {Cotanch},\ and\ \citenamefont {General}}]{Wang:2008mw}%
  \BibitemOpen
  \bibfield  {author} {\bibinfo {author} {\bibfnamefont {P.}~\bibnamefont {Wang}}, \bibinfo {author} {\bibfnamefont {S.~R.}\ \bibnamefont {Cotanch}}, \ and\ \bibinfo {author} {\bibfnamefont {I.~J.}\ \bibnamefont {General}},\ }\href {\doibase 10.1140/epjc/s10052-008-0605-7} {\bibfield  {journal} {\bibinfo  {journal} {Eur. Phys. J. C}\ }\textbf {\bibinfo {volume} {55}},\ \bibinfo {pages} {409} (\bibinfo {year} {2008})},\ \Eprint {http://arxiv.org/abs/0801.4810} {arXiv:0801.4810 [hep-ph]} \BibitemShut {NoStop}%
\bibitem [{\citenamefont {Kim}\ and\ \citenamefont {Kim}(2022)}]{Kim:2022qfj}%
  \BibitemOpen
  \bibfield  {author} {\bibinfo {author} {\bibfnamefont {H.}~\bibnamefont {Kim}}\ and\ \bibinfo {author} {\bibfnamefont {K.~S.}\ \bibnamefont {Kim}},\ }\href {\doibase 10.1140/epjc/s10052-022-11055-9} {\bibfield  {journal} {\bibinfo  {journal} {Eur. Phys. J. C}\ }\textbf {\bibinfo {volume} {82}},\ \bibinfo {pages} {1113} (\bibinfo {year} {2022})},\ \Eprint {http://arxiv.org/abs/2209.02958} {arXiv:2209.02958 [hep-ph]} \BibitemShut {NoStop}%
\bibitem [{\citenamefont {Isgur}\ and\ \citenamefont {Wise}(1991)}]{Isgur:1991wq}%
  \BibitemOpen
  \bibfield  {author} {\bibinfo {author} {\bibfnamefont {N.}~\bibnamefont {Isgur}}\ and\ \bibinfo {author} {\bibfnamefont {M.~B.}\ \bibnamefont {Wise}},\ }\href {\doibase 10.1103/PhysRevLett.66.1130} {\bibfield  {journal} {\bibinfo  {journal} {Phys. Rev. Lett.}\ }\textbf {\bibinfo {volume} {66}},\ \bibinfo {pages} {1130} (\bibinfo {year} {1991})}\BibitemShut {NoStop}%
\bibitem [{\citenamefont {Neubert}(1994)}]{Neubert:1993mb}%
  \BibitemOpen
  \bibfield  {author} {\bibinfo {author} {\bibfnamefont {M.}~\bibnamefont {Neubert}},\ }\href {\doibase 10.1016/0370-1573(94)90091-4} {\bibfield  {journal} {\bibinfo  {journal} {Phys. Rept.}\ }\textbf {\bibinfo {volume} {245}},\ \bibinfo {pages} {259} (\bibinfo {year} {1994})},\ \Eprint {http://arxiv.org/abs/hep-ph/9306320} {arXiv:hep-ph/9306320} \BibitemShut {NoStop}%
\bibitem [{\citenamefont {Weng}\ \emph {et~al.}(2021)\citenamefont {Weng}, \citenamefont {Chen}, \citenamefont {Deng},\ and\ \citenamefont {Zhu}}]{Weng:2020jao}%
  \BibitemOpen
  \bibfield  {author} {\bibinfo {author} {\bibfnamefont {X.-Z.}\ \bibnamefont {Weng}}, \bibinfo {author} {\bibfnamefont {X.-L.}\ \bibnamefont {Chen}}, \bibinfo {author} {\bibfnamefont {W.-Z.}\ \bibnamefont {Deng}}, \ and\ \bibinfo {author} {\bibfnamefont {S.-L.}\ \bibnamefont {Zhu}},\ }\href {\doibase 10.1103/PhysRevD.103.034001} {\bibfield  {journal} {\bibinfo  {journal} {Phys. Rev. D}\ }\textbf {\bibinfo {volume} {103}},\ \bibinfo {pages} {034001} (\bibinfo {year} {2021})},\ \Eprint {http://arxiv.org/abs/2010.05163} {arXiv:2010.05163 [hep-ph]} \BibitemShut {NoStop}%
\bibitem [{\citenamefont {An}\ \emph {et~al.}(2023)\citenamefont {An}, \citenamefont {Luo}, \citenamefont {Liu},\ and\ \citenamefont {Liu}}]{An:2022qpt}%
  \BibitemOpen
  \bibfield  {author} {\bibinfo {author} {\bibfnamefont {H.-T.}\ \bibnamefont {An}}, \bibinfo {author} {\bibfnamefont {S.-Q.}\ \bibnamefont {Luo}}, \bibinfo {author} {\bibfnamefont {Z.-W.}\ \bibnamefont {Liu}}, \ and\ \bibinfo {author} {\bibfnamefont {X.}~\bibnamefont {Liu}},\ }\href {\doibase 10.1140/epjc/s10052-023-11847-7} {\bibfield  {journal} {\bibinfo  {journal} {Eur. Phys. J. C}\ }\textbf {\bibinfo {volume} {83}},\ \bibinfo {pages} {740} (\bibinfo {year} {2023})},\ \Eprint {http://arxiv.org/abs/2208.03899} {arXiv:2208.03899 [hep-ph]} \BibitemShut {NoStop}%
\bibitem [{\citenamefont {Liu}\ \emph {et~al.}(2024)\citenamefont {Liu}, \citenamefont {Liu}, \citenamefont {Zhong},\ and\ \citenamefont {Zhao}}]{liu:2020eha}%
  \BibitemOpen
  \bibfield  {author} {\bibinfo {author} {\bibfnamefont {M.-S.}\ \bibnamefont {Liu}}, \bibinfo {author} {\bibfnamefont {F.-X.}\ \bibnamefont {Liu}}, \bibinfo {author} {\bibfnamefont {X.-H.}\ \bibnamefont {Zhong}}, \ and\ \bibinfo {author} {\bibfnamefont {Q.}~\bibnamefont {Zhao}},\ }\href {\doibase 10.1103/PhysRevD.109.076017} {\bibfield  {journal} {\bibinfo  {journal} {Phys. Rev. D}\ }\textbf {\bibinfo {volume} {109}},\ \bibinfo {pages} {076017} (\bibinfo {year} {2024})},\ \Eprint {http://arxiv.org/abs/2006.11952} {arXiv:2006.11952 [hep-ph]} \BibitemShut {NoStop}%
\bibitem [{\citenamefont {Abdul~Khalek}\ \emph {et~al.}(2022)\citenamefont {Abdul~Khalek} \emph {et~al.}}]{AbdulKhalek:2021gbh}%
  \BibitemOpen
  \bibfield  {author} {\bibinfo {author} {\bibfnamefont {R.}~\bibnamefont {Abdul~Khalek}} \emph {et~al.},\ }\href {\doibase 10.1016/j.nuclphysa.2022.122447} {\bibfield  {journal} {\bibinfo  {journal} {Nucl. Phys. A}\ }\textbf {\bibinfo {volume} {1026}},\ \bibinfo {pages} {122447} (\bibinfo {year} {2022})},\ \Eprint {http://arxiv.org/abs/2103.05419} {arXiv:2103.05419 [physics.ins-det]} \BibitemShut {NoStop}%
\bibitem [{\citenamefont {Feng}\ \emph {et~al.}(2024)\citenamefont {Feng}, \citenamefont {Huang}, \citenamefont {Jia}, \citenamefont {Sang}, \citenamefont {Yang},\ and\ \citenamefont {Zhang}}]{Feng:2023ghc}%
  \BibitemOpen
  \bibfield  {author} {\bibinfo {author} {\bibfnamefont {F.}~\bibnamefont {Feng}}, \bibinfo {author} {\bibfnamefont {Y.}~\bibnamefont {Huang}}, \bibinfo {author} {\bibfnamefont {Y.}~\bibnamefont {Jia}}, \bibinfo {author} {\bibfnamefont {W.-L.}\ \bibnamefont {Sang}}, \bibinfo {author} {\bibfnamefont {D.-S.}\ \bibnamefont {Yang}}, \ and\ \bibinfo {author} {\bibfnamefont {J.-Y.}\ \bibnamefont {Zhang}},\ }\href {\doibase 10.1103/PhysRevD.110.054007} {\bibfield  {journal} {\bibinfo  {journal} {Phys. Rev. D}\ }\textbf {\bibinfo {volume} {110}},\ \bibinfo {pages} {054007} (\bibinfo {year} {2024})},\ \Eprint {http://arxiv.org/abs/2311.08292} {arXiv:2311.08292 [hep-ph]} \BibitemShut {NoStop}%
\bibitem [{\citenamefont {Feng}\ \emph {et~al.}(2023)\citenamefont {Feng}, \citenamefont {Huang}, \citenamefont {Jia}, \citenamefont {Sang}, \citenamefont {Yang},\ and\ \citenamefont {Zhang}}]{Feng:2023agq}%
  \BibitemOpen
  \bibfield  {author} {\bibinfo {author} {\bibfnamefont {F.}~\bibnamefont {Feng}}, \bibinfo {author} {\bibfnamefont {Y.}~\bibnamefont {Huang}}, \bibinfo {author} {\bibfnamefont {Y.}~\bibnamefont {Jia}}, \bibinfo {author} {\bibfnamefont {W.-L.}\ \bibnamefont {Sang}}, \bibinfo {author} {\bibfnamefont {D.-S.}\ \bibnamefont {Yang}}, \ and\ \bibinfo {author} {\bibfnamefont {J.-Y.}\ \bibnamefont {Zhang}},\ }\href {\doibase 10.1103/PhysRevD.108.L051501} {\bibfield  {journal} {\bibinfo  {journal} {Phys. Rev. D}\ }\textbf {\bibinfo {volume} {108}},\ \bibinfo {pages} {L051501} (\bibinfo {year} {2023})},\ \Eprint {http://arxiv.org/abs/2304.11142} {arXiv:2304.11142 [hep-ph]} \BibitemShut {NoStop}%
\bibitem [{\citenamefont {Ma}\ \emph {et~al.}(2025)\citenamefont {Ma}, \citenamefont {Tao},\ and\ \citenamefont {Niu}}]{Ma:2025ryo}%
  \BibitemOpen
  \bibfield  {author} {\bibinfo {author} {\bibfnamefont {H.-H.}\ \bibnamefont {Ma}}, \bibinfo {author} {\bibfnamefont {Z.-K.}\ \bibnamefont {Tao}}, \ and\ \bibinfo {author} {\bibfnamefont {J.-J.}\ \bibnamefont {Niu}},\ }\href {\doibase 10.1140/epjc/s10052-025-14128-7} {\bibfield  {journal} {\bibinfo  {journal} {Eur. Phys. J. C}\ }\textbf {\bibinfo {volume} {85}},\ \bibinfo {pages} {397} (\bibinfo {year} {2025})},\ \Eprint {http://arxiv.org/abs/2502.20891} {arXiv:2502.20891 [hep-ph]} \BibitemShut {NoStop}%
\bibitem [{\citenamefont {Celiberto}(2025{\natexlab{b}})}]{Celiberto:2025_TQ4Q11_AVT}%
  \BibitemOpen
  \bibfield  {author} {\bibinfo {author} {\bibfnamefont {F.~G.}\ \bibnamefont {Celiberto}},\ }\href {{https://github.com/FGCeliberto/Collinear_FFs/}} {\emph {\bibinfo {title} {{TQ4Q1.1: axial-vector TetraQuarks with 4 heavy Quarks VFNS FFs}}}}\ (\bibinfo {year} {2025})\BibitemShut {NoStop}%
\bibitem [{\citenamefont {Chen}\ \emph {et~al.}(2020)\citenamefont {Chen}, \citenamefont {Chen}, \citenamefont {Liu},\ and\ \citenamefont {Zhu}}]{Chen:2020xwe}%
  \BibitemOpen
  \bibfield  {author} {\bibinfo {author} {\bibfnamefont {H.-X.}\ \bibnamefont {Chen}}, \bibinfo {author} {\bibfnamefont {W.}~\bibnamefont {Chen}}, \bibinfo {author} {\bibfnamefont {X.}~\bibnamefont {Liu}}, \ and\ \bibinfo {author} {\bibfnamefont {S.-L.}\ \bibnamefont {Zhu}},\ }\href {\doibase 10.1016/j.scib.2020.08.038} {\bibfield  {journal} {\bibinfo  {journal} {Sci. Bull.}\ }\textbf {\bibinfo {volume} {65}},\ \bibinfo {pages} {1994} (\bibinfo {year} {2020})},\ \Eprint {http://arxiv.org/abs/2006.16027} {arXiv:2006.16027 [hep-ph]} \BibitemShut {NoStop}%
\bibitem [{\citenamefont {Zhu}(2021)}]{Zhu:2020xni}%
  \BibitemOpen
  \bibfield  {author} {\bibinfo {author} {\bibfnamefont {R.}~\bibnamefont {Zhu}},\ }\href {\doibase 10.1016/j.nuclphysb.2021.115393} {\bibfield  {journal} {\bibinfo  {journal} {Nucl. Phys. B}\ }\textbf {\bibinfo {volume} {966}},\ \bibinfo {pages} {115393} (\bibinfo {year} {2021})},\ \Eprint {http://arxiv.org/abs/2010.09082} {arXiv:2010.09082 [hep-ph]} \BibitemShut {NoStop}%
\bibitem [{\citenamefont {Karliner}\ and\ \citenamefont {Rosner}(2020)}]{Karliner:2020dta}%
  \BibitemOpen
  \bibfield  {author} {\bibinfo {author} {\bibfnamefont {M.}~\bibnamefont {Karliner}}\ and\ \bibinfo {author} {\bibfnamefont {J.~L.}\ \bibnamefont {Rosner}},\ }\href {\doibase 10.1103/PhysRevD.102.114039} {\bibfield  {journal} {\bibinfo  {journal} {Phys. Rev. D}\ }\textbf {\bibinfo {volume} {102}},\ \bibinfo {pages} {114039} (\bibinfo {year} {2020})},\ \Eprint {http://arxiv.org/abs/2009.04429} {arXiv:2009.04429 [hep-ph]} \BibitemShut {NoStop}%
\bibitem [{\citenamefont {Becchi}\ \emph {et~al.}(2020)\citenamefont {Becchi}, \citenamefont {Giachino}, \citenamefont {Maiani},\ and\ \citenamefont {Santopinto}}]{Becchi:2020mjz}%
  \BibitemOpen
  \bibfield  {author} {\bibinfo {author} {\bibfnamefont {C.}~\bibnamefont {Becchi}}, \bibinfo {author} {\bibfnamefont {A.}~\bibnamefont {Giachino}}, \bibinfo {author} {\bibfnamefont {L.}~\bibnamefont {Maiani}}, \ and\ \bibinfo {author} {\bibfnamefont {E.}~\bibnamefont {Santopinto}},\ }\href {\doibase 10.1016/j.physletb.2020.135495} {\bibfield  {journal} {\bibinfo  {journal} {Phys. Lett. B}\ }\textbf {\bibinfo {volume} {806}},\ \bibinfo {pages} {135495} (\bibinfo {year} {2020})},\ \Eprint {http://arxiv.org/abs/2002.11077} {arXiv:2002.11077 [hep-ph]} \BibitemShut {NoStop}%
\bibitem [{\citenamefont {Karliner}\ and\ \citenamefont {Rosner}(2017)}]{Karliner:2017qjm}%
  \BibitemOpen
  \bibfield  {author} {\bibinfo {author} {\bibfnamefont {M.}~\bibnamefont {Karliner}}\ and\ \bibinfo {author} {\bibfnamefont {J.~L.}\ \bibnamefont {Rosner}},\ }\href {\doibase 10.1103/PhysRevLett.119.202001} {\bibfield  {journal} {\bibinfo  {journal} {Phys. Rev. Lett.}\ }\textbf {\bibinfo {volume} {119}},\ \bibinfo {pages} {202001} (\bibinfo {year} {2017})},\ \Eprint {http://arxiv.org/abs/1707.07666} {arXiv:1707.07666 [hep-ph]} \BibitemShut {NoStop}%
\bibitem [{\citenamefont {Jaffe}\ and\ \citenamefont {Randall}(1994)}]{Jaffe:1993ie}%
  \BibitemOpen
  \bibfield  {author} {\bibinfo {author} {\bibfnamefont {R.~L.}\ \bibnamefont {Jaffe}}\ and\ \bibinfo {author} {\bibfnamefont {L.}~\bibnamefont {Randall}},\ }\href {\doibase 10.1016/0550-3213(94)90495-2} {\bibfield  {journal} {\bibinfo  {journal} {Nucl. Phys. B}\ }\textbf {\bibinfo {volume} {412}},\ \bibinfo {pages} {79} (\bibinfo {year} {1994})},\ \Eprint {http://arxiv.org/abs/hep-ph/9306201} {arXiv:hep-ph/9306201} \BibitemShut {NoStop}%
\bibitem [{\citenamefont {Helenius}\ and\ \citenamefont {Paukkunen}(2018)}]{Helenius:2018uul}%
  \BibitemOpen
  \bibfield  {author} {\bibinfo {author} {\bibfnamefont {I.}~\bibnamefont {Helenius}}\ and\ \bibinfo {author} {\bibfnamefont {H.}~\bibnamefont {Paukkunen}},\ }\href {\doibase 10.1007/JHEP05(2018)196} {\bibfield  {journal} {\bibinfo  {journal} {JHEP}\ }\textbf {\bibinfo {volume} {05}},\ \bibinfo {pages} {196} (\bibinfo {year} {2018})},\ \Eprint {http://arxiv.org/abs/1804.03557} {arXiv:1804.03557 [hep-ph]} \BibitemShut {NoStop}%
\bibitem [{\citenamefont {Bonino}\ \emph {et~al.}(2024)\citenamefont {Bonino}, \citenamefont {Cacciari},\ and\ \citenamefont {Stagnitto}}]{Bonino:2023icn}%
  \BibitemOpen
  \bibfield  {author} {\bibinfo {author} {\bibfnamefont {L.}~\bibnamefont {Bonino}}, \bibinfo {author} {\bibfnamefont {M.}~\bibnamefont {Cacciari}}, \ and\ \bibinfo {author} {\bibfnamefont {G.}~\bibnamefont {Stagnitto}},\ }\href {\doibase 10.1007/JHEP06(2024)040} {\bibfield  {journal} {\bibinfo  {journal} {JHEP}\ }\textbf {\bibinfo {volume} {06}},\ \bibinfo {pages} {040} (\bibinfo {year} {2024})},\ \Eprint {http://arxiv.org/abs/2312.12519} {arXiv:2312.12519 [hep-ph]} \BibitemShut {NoStop}%
\bibitem [{\citenamefont {Cacciari}\ \emph {et~al.}(2024)\citenamefont {Cacciari}, \citenamefont {Ghira}, \citenamefont {Marzani},\ and\ \citenamefont {Ridolfi}}]{Cacciari:2024kaa}%
  \BibitemOpen
  \bibfield  {author} {\bibinfo {author} {\bibfnamefont {M.}~\bibnamefont {Cacciari}}, \bibinfo {author} {\bibfnamefont {A.}~\bibnamefont {Ghira}}, \bibinfo {author} {\bibfnamefont {S.}~\bibnamefont {Marzani}}, \ and\ \bibinfo {author} {\bibfnamefont {G.}~\bibnamefont {Ridolfi}},\ }\href {\doibase 10.1140/epjc/s10052-024-13245-z} {\bibfield  {journal} {\bibinfo  {journal} {Eur. Phys. J. C}\ }\textbf {\bibinfo {volume} {84}},\ \bibinfo {pages} {889} (\bibinfo {year} {2024})},\ \Eprint {http://arxiv.org/abs/2406.04173} {arXiv:2406.04173 [hep-ph]} \BibitemShut {NoStop}%
\bibitem [{\citenamefont {Czakon}\ \emph {et~al.}(2024)\citenamefont {Czakon}, \citenamefont {Generet}, \citenamefont {Mitov},\ and\ \citenamefont {Poncelet}}]{Czakon:2024tjr}%
  \BibitemOpen
  \bibfield  {author} {\bibinfo {author} {\bibfnamefont {M.}~\bibnamefont {Czakon}}, \bibinfo {author} {\bibfnamefont {T.}~\bibnamefont {Generet}}, \bibinfo {author} {\bibfnamefont {A.}~\bibnamefont {Mitov}}, \ and\ \bibinfo {author} {\bibfnamefont {R.}~\bibnamefont {Poncelet}},\ }\href@noop {} {\  (\bibinfo {year} {2024})},\ \Eprint {http://arxiv.org/abs/2411.09684} {arXiv:2411.09684 [hep-ph]} \BibitemShut {NoStop}%
\bibitem [{\citenamefont {Braaten}\ and\ \citenamefont {Yuan}(1993)}]{Braaten:1993rw}%
  \BibitemOpen
  \bibfield  {author} {\bibinfo {author} {\bibfnamefont {E.}~\bibnamefont {Braaten}}\ and\ \bibinfo {author} {\bibfnamefont {T.~C.}\ \bibnamefont {Yuan}},\ }\href {\doibase 10.1103/PhysRevLett.71.1673} {\bibfield  {journal} {\bibinfo  {journal} {Phys. Rev. Lett.}\ }\textbf {\bibinfo {volume} {71}},\ \bibinfo {pages} {1673} (\bibinfo {year} {1993})},\ \Eprint {http://arxiv.org/abs/hep-ph/9303205} {arXiv:hep-ph/9303205} \BibitemShut {NoStop}%
\bibitem [{\citenamefont {Cacciari}\ and\ \citenamefont {Greco}(1994{\natexlab{b}})}]{Cacciari:1994dr}%
  \BibitemOpen
  \bibfield  {author} {\bibinfo {author} {\bibfnamefont {M.}~\bibnamefont {Cacciari}}\ and\ \bibinfo {author} {\bibfnamefont {M.}~\bibnamefont {Greco}},\ }\href {\doibase 10.1103/PhysRevLett.73.1586} {\bibfield  {journal} {\bibinfo  {journal} {Phys. Rev. Lett.}\ }\textbf {\bibinfo {volume} {73}},\ \bibinfo {pages} {1586} (\bibinfo {year} {1994}{\natexlab{b}})},\ \Eprint {http://arxiv.org/abs/hep-ph/9405241} {arXiv:hep-ph/9405241} \BibitemShut {NoStop}%
\bibitem [{\citenamefont {Zheng}\ \emph {et~al.}(2019)\citenamefont {Zheng}, \citenamefont {Chang},\ and\ \citenamefont {Wu}}]{Zheng:2019dfk}%
  \BibitemOpen
  \bibfield  {author} {\bibinfo {author} {\bibfnamefont {X.-C.}\ \bibnamefont {Zheng}}, \bibinfo {author} {\bibfnamefont {C.-H.}\ \bibnamefont {Chang}}, \ and\ \bibinfo {author} {\bibfnamefont {X.-G.}\ \bibnamefont {Wu}},\ }\href {\doibase 10.1103/PhysRevD.100.014005} {\bibfield  {journal} {\bibinfo  {journal} {Phys. Rev. D}\ }\textbf {\bibinfo {volume} {100}},\ \bibinfo {pages} {014005} (\bibinfo {year} {2019})},\ \Eprint {http://arxiv.org/abs/1905.09171} {arXiv:1905.09171 [hep-ph]} \BibitemShut {NoStop}%
\bibitem [{\citenamefont {Zhang}\ \emph {et~al.}(2019)\citenamefont {Zhang}, \citenamefont {Wang}, \citenamefont {Liu}, \citenamefont {Ma}, \citenamefont {Meng},\ and\ \citenamefont {Chao}}]{Zhang:2018mlo}%
  \BibitemOpen
  \bibfield  {author} {\bibinfo {author} {\bibfnamefont {P.}~\bibnamefont {Zhang}}, \bibinfo {author} {\bibfnamefont {C.-Y.}\ \bibnamefont {Wang}}, \bibinfo {author} {\bibfnamefont {X.}~\bibnamefont {Liu}}, \bibinfo {author} {\bibfnamefont {Y.-Q.}\ \bibnamefont {Ma}}, \bibinfo {author} {\bibfnamefont {C.}~\bibnamefont {Meng}}, \ and\ \bibinfo {author} {\bibfnamefont {K.-T.}\ \bibnamefont {Chao}},\ }\href {\doibase 10.1007/JHEP04(2019)116} {\bibfield  {journal} {\bibinfo  {journal} {JHEP}\ }\textbf {\bibinfo {volume} {04}},\ \bibinfo {pages} {116} (\bibinfo {year} {2019})},\ \Eprint {http://arxiv.org/abs/1810.07656} {arXiv:1810.07656 [hep-ph]} \BibitemShut {NoStop}%
\bibitem [{\citenamefont {Artoisenet}\ and\ \citenamefont {Braaten}(2015)}]{Artoisenet:2014lpa}%
  \BibitemOpen
  \bibfield  {author} {\bibinfo {author} {\bibfnamefont {P.}~\bibnamefont {Artoisenet}}\ and\ \bibinfo {author} {\bibfnamefont {E.}~\bibnamefont {Braaten}},\ }\href {\doibase 10.1007/JHEP04(2015)121} {\bibfield  {journal} {\bibinfo  {journal} {JHEP}\ }\textbf {\bibinfo {volume} {04}},\ \bibinfo {pages} {121} (\bibinfo {year} {2015})},\ \Eprint {http://arxiv.org/abs/1412.3834} {arXiv:1412.3834 [hep-ph]} \BibitemShut {NoStop}%
\bibitem [{\citenamefont {Fleming}\ \emph {et~al.}(2012)\citenamefont {Fleming}, \citenamefont {Leibovich}, \citenamefont {Mehen},\ and\ \citenamefont {Rothstein}}]{Fleming:2012wy}%
  \BibitemOpen
  \bibfield  {author} {\bibinfo {author} {\bibfnamefont {S.}~\bibnamefont {Fleming}}, \bibinfo {author} {\bibfnamefont {A.~K.}\ \bibnamefont {Leibovich}}, \bibinfo {author} {\bibfnamefont {T.}~\bibnamefont {Mehen}}, \ and\ \bibinfo {author} {\bibfnamefont {I.~Z.}\ \bibnamefont {Rothstein}},\ }\href {\doibase 10.1103/PhysRevD.86.094012} {\bibfield  {journal} {\bibinfo  {journal} {Phys. Rev. D}\ }\textbf {\bibinfo {volume} {86}},\ \bibinfo {pages} {094012} (\bibinfo {year} {2012})},\ \Eprint {http://arxiv.org/abs/1207.2578} {arXiv:1207.2578 [hep-ph]} \BibitemShut {NoStop}%
\bibitem [{\citenamefont {Kang}\ \emph {et~al.}(2014)\citenamefont {Kang}, \citenamefont {Ma}, \citenamefont {Qiu},\ and\ \citenamefont {Sterman}}]{Kang:2014tta}%
  \BibitemOpen
  \bibfield  {author} {\bibinfo {author} {\bibfnamefont {Z.-B.}\ \bibnamefont {Kang}}, \bibinfo {author} {\bibfnamefont {Y.-Q.}\ \bibnamefont {Ma}}, \bibinfo {author} {\bibfnamefont {J.-W.}\ \bibnamefont {Qiu}}, \ and\ \bibinfo {author} {\bibfnamefont {G.}~\bibnamefont {Sterman}},\ }\href {\doibase 10.1103/PhysRevD.90.034006} {\bibfield  {journal} {\bibinfo  {journal} {Phys. Rev. D}\ }\textbf {\bibinfo {volume} {90}},\ \bibinfo {pages} {034006} (\bibinfo {year} {2014})},\ \Eprint {http://arxiv.org/abs/1401.0923} {arXiv:1401.0923 [hep-ph]} \BibitemShut {NoStop}%
\bibitem [{\citenamefont {Boer}\ \emph {et~al.}(2023)\citenamefont {Boer}, \citenamefont {Bor}, \citenamefont {Maxia}, \citenamefont {Pisano},\ and\ \citenamefont {Yuan}}]{Boer:2023zit}%
  \BibitemOpen
  \bibfield  {author} {\bibinfo {author} {\bibfnamefont {D.}~\bibnamefont {Boer}}, \bibinfo {author} {\bibfnamefont {J.}~\bibnamefont {Bor}}, \bibinfo {author} {\bibfnamefont {L.}~\bibnamefont {Maxia}}, \bibinfo {author} {\bibfnamefont {C.}~\bibnamefont {Pisano}}, \ and\ \bibinfo {author} {\bibfnamefont {F.}~\bibnamefont {Yuan}},\ }\href {\doibase 10.1007/JHEP08(2023)105} {\bibfield  {journal} {\bibinfo  {journal} {JHEP}\ }\textbf {\bibinfo {volume} {08}},\ \bibinfo {pages} {105} (\bibinfo {year} {2023})},\ \Eprint {http://arxiv.org/abs/2304.09473} {arXiv:2304.09473 [hep-ph]} \BibitemShut {NoStop}%
\bibitem [{\citenamefont {Celiberto}\ and\ \citenamefont {Fucilla}(2022)}]{Celiberto:2022dyf}%
  \BibitemOpen
  \bibfield  {author} {\bibinfo {author} {\bibfnamefont {F.~G.}\ \bibnamefont {Celiberto}}\ and\ \bibinfo {author} {\bibfnamefont {M.}~\bibnamefont {Fucilla}},\ }\href {\doibase 10.1140/epjc/s10052-022-10818-8} {\bibfield  {journal} {\bibinfo  {journal} {Eur. Phys. J. C}\ }\textbf {\bibinfo {volume} {82}},\ \bibinfo {pages} {929} (\bibinfo {year} {2022})},\ \Eprint {http://arxiv.org/abs/2202.12227} {arXiv:2202.12227 [hep-ph]} \BibitemShut {NoStop}%
\bibitem [{\citenamefont {Celiberto}(2022{\natexlab{a}})}]{Celiberto:2022keu}%
  \BibitemOpen
  \bibfield  {author} {\bibinfo {author} {\bibfnamefont {F.~G.}\ \bibnamefont {Celiberto}},\ }\href {\doibase 10.1016/j.physletb.2022.137554} {\bibfield  {journal} {\bibinfo  {journal} {Phys. Lett. B}\ }\textbf {\bibinfo {volume} {835}},\ \bibinfo {pages} {137554} (\bibinfo {year} {2022}{\natexlab{a}})},\ \Eprint {http://arxiv.org/abs/2206.09413} {arXiv:2206.09413 [hep-ph]} \BibitemShut {NoStop}%
\bibitem [{\citenamefont {Celiberto}(2024{\natexlab{b}})}]{Celiberto:2024omj}%
  \BibitemOpen
  \bibfield  {author} {\bibinfo {author} {\bibfnamefont {F.~G.}\ \bibnamefont {Celiberto}},\ }\href {\doibase 10.1140/epjc/s10052-024-12704-x} {\bibfield  {journal} {\bibinfo  {journal} {Eur. Phys. J. C}\ }\textbf {\bibinfo {volume} {84}},\ \bibinfo {pages} {384} (\bibinfo {year} {2024}{\natexlab{b}})},\ \Eprint {http://arxiv.org/abs/2401.01410} {arXiv:2401.01410 [hep-ph]} \BibitemShut {NoStop}%
\bibitem [{\citenamefont {Zhang}\ and\ \citenamefont {Ma}(2020)}]{Zhang:2020hoh}%
  \BibitemOpen
  \bibfield  {author} {\bibinfo {author} {\bibfnamefont {H.-F.}\ \bibnamefont {Zhang}}\ and\ \bibinfo {author} {\bibfnamefont {Y.-Q.}\ \bibnamefont {Ma}},\ }\href@noop {} {\  (\bibinfo {year} {2020})},\ \Eprint {http://arxiv.org/abs/2009.08376} {arXiv:2009.08376 [hep-ph]} \BibitemShut {NoStop}%
\bibitem [{\citenamefont {Bai}\ \emph {et~al.}(2025)\citenamefont {Bai}, \citenamefont {Huang},\ and\ \citenamefont {Sang}}]{Bai:2024flh}%
  \BibitemOpen
  \bibfield  {author} {\bibinfo {author} {\bibfnamefont {X.-W.}\ \bibnamefont {Bai}}, \bibinfo {author} {\bibfnamefont {Y.}~\bibnamefont {Huang}}, \ and\ \bibinfo {author} {\bibfnamefont {W.-L.}\ \bibnamefont {Sang}},\ }\href {\doibase 10.1103/PhysRevD.111.054006} {\bibfield  {journal} {\bibinfo  {journal} {Phys. Rev. D}\ }\textbf {\bibinfo {volume} {111}},\ \bibinfo {pages} {054006} (\bibinfo {year} {2025})},\ \Eprint {http://arxiv.org/abs/2411.19296} {arXiv:2411.19296 [hep-ph]} \BibitemShut {NoStop}%
\bibitem [{\citenamefont {Celiberto}\ \emph {et~al.}(2024)\citenamefont {Celiberto}, \citenamefont {Gatto},\ and\ \citenamefont {Papa}}]{Celiberto:2024mab}%
  \BibitemOpen
  \bibfield  {author} {\bibinfo {author} {\bibfnamefont {F.~G.}\ \bibnamefont {Celiberto}}, \bibinfo {author} {\bibfnamefont {G.}~\bibnamefont {Gatto}}, \ and\ \bibinfo {author} {\bibfnamefont {A.}~\bibnamefont {Papa}},\ }\href {\doibase 10.1140/epjc/s10052-024-13345-w} {\bibfield  {journal} {\bibinfo  {journal} {Eur. Phys. J. C}\ }\textbf {\bibinfo {volume} {84}},\ \bibinfo {pages} {1071} (\bibinfo {year} {2024})},\ \Eprint {http://arxiv.org/abs/2405.14773} {arXiv:2405.14773 [hep-ph]} \BibitemShut {NoStop}%
\bibitem [{\citenamefont {{\relax Moosavi}~Nejad}\ and\ \citenamefont {Amiri}(2022)}]{Nejad:2021mmp}%
  \BibitemOpen
  \bibfield  {author} {\bibinfo {author} {\bibfnamefont {S.~M.}\ \bibnamefont {{\relax Moosavi}~Nejad}}\ and\ \bibinfo {author} {\bibfnamefont {N.}~\bibnamefont {Amiri}},\ }\href {\doibase 10.1103/PhysRevD.105.034001} {\bibfield  {journal} {\bibinfo  {journal} {Phys. Rev. D}\ }\textbf {\bibinfo {volume} {105}},\ \bibinfo {pages} {034001} (\bibinfo {year} {2022})},\ \Eprint {http://arxiv.org/abs/2110.15251} {arXiv:2110.15251 [hep-ph]} \BibitemShut {NoStop}%
\bibitem [{\citenamefont {Celiberto}\ and\ \citenamefont {Papa}(2024)}]{Celiberto:2023rzw}%
  \BibitemOpen
  \bibfield  {author} {\bibinfo {author} {\bibfnamefont {F.~G.}\ \bibnamefont {Celiberto}}\ and\ \bibinfo {author} {\bibfnamefont {A.}~\bibnamefont {Papa}},\ }\href {\doibase 10.1016/j.physletb.2023.138406} {\bibfield  {journal} {\bibinfo  {journal} {Phys. Lett. B}\ }\textbf {\bibinfo {volume} {848}},\ \bibinfo {pages} {138406} (\bibinfo {year} {2024})},\ \Eprint {http://arxiv.org/abs/2308.00809} {arXiv:2308.00809 [hep-ph]} \BibitemShut {NoStop}%
\bibitem [{\citenamefont {Celiberto}(2021)}]{Celiberto:2020wpk}%
  \BibitemOpen
  \bibfield  {author} {\bibinfo {author} {\bibfnamefont {F.~G.}\ \bibnamefont {Celiberto}},\ }\href {\doibase 10.1140/epjc/s10052-021-09384-2} {\bibfield  {journal} {\bibinfo  {journal} {Eur. Phys. J. C}\ }\textbf {\bibinfo {volume} {81}},\ \bibinfo {pages} {691} (\bibinfo {year} {2021})},\ \Eprint {http://arxiv.org/abs/2008.07378} {arXiv:2008.07378 [hep-ph]} \BibitemShut {NoStop}%
\bibitem [{\citenamefont {Celiberto}(2022{\natexlab{b}})}]{Celiberto:2022rfj}%
  \BibitemOpen
  \bibfield  {author} {\bibinfo {author} {\bibfnamefont {F.~G.}\ \bibnamefont {Celiberto}},\ }\href {\doibase 10.1103/PhysRevD.105.114008} {\bibfield  {journal} {\bibinfo  {journal} {Phys. Rev. D}\ }\textbf {\bibinfo {volume} {105}},\ \bibinfo {pages} {114008} (\bibinfo {year} {2022}{\natexlab{b}})},\ \Eprint {http://arxiv.org/abs/2204.06497} {arXiv:2204.06497 [hep-ph]} \BibitemShut {NoStop}%
\bibitem [{\citenamefont {Celiberto}(2024{\natexlab{c}})}]{Celiberto:2024mrq}%
  \BibitemOpen
  \bibfield  {author} {\bibinfo {author} {\bibfnamefont {F.~G.}\ \bibnamefont {Celiberto}},\ }\href {\doibase 10.3390/sym16050550} {\bibfield  {journal} {\bibinfo  {journal} {Symmetry}\ }\textbf {\bibinfo {volume} {16}},\ \bibinfo {pages} {550} (\bibinfo {year} {2024}{\natexlab{c}})},\ \Eprint {http://arxiv.org/abs/2403.15639} {arXiv:2403.15639 [hep-ph]} \BibitemShut {NoStop}%
\bibitem [{\citenamefont {Celiberto}(2025{\natexlab{c}})}]{Celiberto:2025_TQ4Q11_AVT_suppl}%
  \BibitemOpen
  \bibfield  {author} {\bibinfo {author} {\bibfnamefont {F.~G.}\ \bibnamefont {Celiberto}},\ }\href@noop {} {\bibfield  {journal} {\bibinfo  {journal} {Supplemental material to this letter}\ } (\bibinfo {year} {2025}{\natexlab{c}})}\BibitemShut {NoStop}%
\bibitem [{\citenamefont {L\"u}\ \emph {et~al.}(2020)\citenamefont {L\"u}, \citenamefont {Chen},\ and\ \citenamefont {Dong}}]{Lu:2020cns}%
  \BibitemOpen
  \bibfield  {author} {\bibinfo {author} {\bibfnamefont {Q.-F.}\ \bibnamefont {L\"u}}, \bibinfo {author} {\bibfnamefont {D.-Y.}\ \bibnamefont {Chen}}, \ and\ \bibinfo {author} {\bibfnamefont {Y.-B.}\ \bibnamefont {Dong}},\ }\href {\doibase 10.1140/epjc/s10052-020-08454-1} {\bibfield  {journal} {\bibinfo  {journal} {Eur. Phys. J. C}\ }\textbf {\bibinfo {volume} {80}},\ \bibinfo {pages} {871} (\bibinfo {year} {2020})},\ \Eprint {http://arxiv.org/abs/2006.14445} {arXiv:2006.14445 [hep-ph]} \BibitemShut {NoStop}%
\bibitem [{\citenamefont {Yu}\ \emph {et~al.}(2023)\citenamefont {Yu}, \citenamefont {Li}, \citenamefont {Wang}, \citenamefont {Lu},\ and\ \citenamefont {Yan}}]{Yu:2022lak}%
  \BibitemOpen
  \bibfield  {author} {\bibinfo {author} {\bibfnamefont {G.-L.}\ \bibnamefont {Yu}}, \bibinfo {author} {\bibfnamefont {Z.-Y.}\ \bibnamefont {Li}}, \bibinfo {author} {\bibfnamefont {Z.-G.}\ \bibnamefont {Wang}}, \bibinfo {author} {\bibfnamefont {J.}~\bibnamefont {Lu}}, \ and\ \bibinfo {author} {\bibfnamefont {M.}~\bibnamefont {Yan}},\ }\href {\doibase 10.1140/epjc/s10052-023-11445-7} {\bibfield  {journal} {\bibinfo  {journal} {Eur. Phys. J. C}\ }\textbf {\bibinfo {volume} {83}},\ \bibinfo {pages} {416} (\bibinfo {year} {2023})},\ \Eprint {http://arxiv.org/abs/2212.14339} {arXiv:2212.14339 [hep-ph]} \BibitemShut {NoStop}%
\bibitem [{\citenamefont {Bertone}\ \emph {et~al.}(2014)\citenamefont {Bertone}, \citenamefont {Carrazza},\ and\ \citenamefont {Rojo}}]{Bertone:2013vaa}%
  \BibitemOpen
  \bibfield  {author} {\bibinfo {author} {\bibfnamefont {V.}~\bibnamefont {Bertone}}, \bibinfo {author} {\bibfnamefont {S.}~\bibnamefont {Carrazza}}, \ and\ \bibinfo {author} {\bibfnamefont {J.}~\bibnamefont {Rojo}},\ }\href {\doibase 10.1016/j.cpc.2014.03.007} {\bibfield  {journal} {\bibinfo  {journal} {Comput. Phys. Commun.}\ }\textbf {\bibinfo {volume} {185}},\ \bibinfo {pages} {1647} (\bibinfo {year} {2014})},\ \Eprint {http://arxiv.org/abs/1310.1394} {arXiv:1310.1394 [hep-ph]} \BibitemShut {NoStop}%
\bibitem [{\citenamefont {Buckley}\ \emph {et~al.}(2015)\citenamefont {Buckley}, \citenamefont {Ferrando}, \citenamefont {Lloyd}, \citenamefont {Nordstr\"om}, \citenamefont {Page}, \citenamefont {R\"ufenacht}, \citenamefont {Sch\"onherr},\ and\ \citenamefont {Watt}}]{Buckley:2014ana}%
  \BibitemOpen
  \bibfield  {author} {\bibinfo {author} {\bibfnamefont {A.}~\bibnamefont {Buckley}}, \bibinfo {author} {\bibfnamefont {J.}~\bibnamefont {Ferrando}}, \bibinfo {author} {\bibfnamefont {S.}~\bibnamefont {Lloyd}}, \bibinfo {author} {\bibfnamefont {K.}~\bibnamefont {Nordstr\"om}}, \bibinfo {author} {\bibfnamefont {B.}~\bibnamefont {Page}}, \bibinfo {author} {\bibfnamefont {M.}~\bibnamefont {R\"ufenacht}}, \bibinfo {author} {\bibfnamefont {M.}~\bibnamefont {Sch\"onherr}}, \ and\ \bibinfo {author} {\bibfnamefont {G.}~\bibnamefont {Watt}},\ }\href {\doibase 10.1140/epjc/s10052-015-3318-8} {\bibfield  {journal} {\bibinfo  {journal} {Eur. Phys. J. C}\ }\textbf {\bibinfo {volume} {75}},\ \bibinfo {pages} {132} (\bibinfo {year} {2015})},\ \Eprint {http://arxiv.org/abs/1412.7420} {arXiv:1412.7420 [hep-ph]} \BibitemShut {NoStop}%
\bibitem [{\citenamefont {Suzuki}(1977)}]{Suzuki:1977km}%
  \BibitemOpen
  \bibfield  {author} {\bibinfo {author} {\bibfnamefont {M.}~\bibnamefont {Suzuki}},\ }\href {\doibase 10.1016/0370-2693(77)90761-4} {\bibfield  {journal} {\bibinfo  {journal} {Phys. Lett. B}\ }\textbf {\bibinfo {volume} {71}},\ \bibinfo {pages} {139} (\bibinfo {year} {1977})}\BibitemShut {NoStop}%
\bibitem [{\citenamefont {Bjorken}(1978)}]{Bjorken:1977md}%
  \BibitemOpen
  \bibfield  {author} {\bibinfo {author} {\bibfnamefont {J.~D.}\ \bibnamefont {Bjorken}},\ }\href {\doibase 10.1103/PhysRevD.17.171} {\bibfield  {journal} {\bibinfo  {journal} {Phys. Rev. D}\ }\textbf {\bibinfo {volume} {17}},\ \bibinfo {pages} {171} (\bibinfo {year} {1978})}\BibitemShut {NoStop}%
\bibitem [{\citenamefont {Bodwin}\ and\ \citenamefont {Petrelli}(2002)}]{Bodwin:2002cfe}%
  \BibitemOpen
  \bibfield  {author} {\bibinfo {author} {\bibfnamefont {G.~T.}\ \bibnamefont {Bodwin}}\ and\ \bibinfo {author} {\bibfnamefont {A.}~\bibnamefont {Petrelli}},\ }\href {\doibase 10.1103/PhysRevD.66.094011} {\bibfield  {journal} {\bibinfo  {journal} {Phys. Rev. D}\ }\textbf {\bibinfo {volume} {66}},\ \bibinfo {pages} {094011} (\bibinfo {year} {2002})},\ \bibinfo {note} {[Erratum: Phys.Rev.D 87, 039902 (2013)]},\ \Eprint {http://arxiv.org/abs/hep-ph/0205210} {arXiv:hep-ph/0205210} \BibitemShut {NoStop}%
\bibitem [{\citenamefont {Ma}\ \emph {et~al.}(2015)\citenamefont {Ma}, \citenamefont {Qiu},\ and\ \citenamefont {Zhang}}]{Ma:2015yka}%
  \BibitemOpen
  \bibfield  {author} {\bibinfo {author} {\bibfnamefont {Y.-Q.}\ \bibnamefont {Ma}}, \bibinfo {author} {\bibfnamefont {J.-W.}\ \bibnamefont {Qiu}}, \ and\ \bibinfo {author} {\bibfnamefont {H.}~\bibnamefont {Zhang}},\ }\href {\doibase 10.1007/JHEP06(2015)021} {\bibfield  {journal} {\bibinfo  {journal} {JHEP}\ }\textbf {\bibinfo {volume} {06}},\ \bibinfo {pages} {021} (\bibinfo {year} {2015})},\ \Eprint {http://arxiv.org/abs/1501.04556} {arXiv:1501.04556 [hep-ph]} \BibitemShut {NoStop}%
\bibitem [{\citenamefont {Xu}\ \emph {et~al.}(2021)\citenamefont {Xu}, \citenamefont {Chen}, \citenamefont {Yao}, \citenamefont {Binosi}, \citenamefont {Cui},\ and\ \citenamefont {Roberts}}]{Xu:2021mju}%
  \BibitemOpen
  \bibfield  {author} {\bibinfo {author} {\bibfnamefont {Y.-Z.}\ \bibnamefont {Xu}}, \bibinfo {author} {\bibfnamefont {S.}~\bibnamefont {Chen}}, \bibinfo {author} {\bibfnamefont {Z.-Q.}\ \bibnamefont {Yao}}, \bibinfo {author} {\bibfnamefont {D.}~\bibnamefont {Binosi}}, \bibinfo {author} {\bibfnamefont {Z.-F.}\ \bibnamefont {Cui}}, \ and\ \bibinfo {author} {\bibfnamefont {C.~D.}\ \bibnamefont {Roberts}},\ }\href {\doibase 10.1140/epjc/s10052-021-09673-w} {\bibfield  {journal} {\bibinfo  {journal} {Eur. Phys. J. C}\ }\textbf {\bibinfo {volume} {81}},\ \bibinfo {pages} {895} (\bibinfo {year} {2021})},\ \Eprint {http://arxiv.org/abs/2107.03488} {arXiv:2107.03488 [hep-ph]} \BibitemShut {NoStop}%
\bibitem [{\citenamefont {Aaij}\ \emph {et~al.}(2024)\citenamefont {Aaij} \emph {et~al.}}]{LHCb:2024ybz}%
  \BibitemOpen
  \bibfield  {author} {\bibinfo {author} {\bibfnamefont {R.}~\bibnamefont {Aaij}} \emph {et~al.} (\bibinfo {collaboration} {LHCb}),\ }\href@noop {} {\  (\bibinfo {year} {2024})},\ \Eprint {http://arxiv.org/abs/2410.18018} {arXiv:2410.18018 [hep-ex]} \BibitemShut {NoStop}%
\bibitem [{\citenamefont {Faustov}\ \emph {et~al.}(2022)\citenamefont {Faustov}, \citenamefont {Galkin},\ and\ \citenamefont {Savchenko}}]{Faustov:2022mvs}%
  \BibitemOpen
  \bibfield  {author} {\bibinfo {author} {\bibfnamefont {R.~N.}\ \bibnamefont {Faustov}}, \bibinfo {author} {\bibfnamefont {V.~O.}\ \bibnamefont {Galkin}}, \ and\ \bibinfo {author} {\bibfnamefont {E.~M.}\ \bibnamefont {Savchenko}},\ }\href {\doibase 10.3390/sym14122504} {\bibfield  {journal} {\bibinfo  {journal} {Symmetry}\ }\textbf {\bibinfo {volume} {14}},\ \bibinfo {pages} {2504} (\bibinfo {year} {2022})},\ \Eprint {http://arxiv.org/abs/2210.16015} {arXiv:2210.16015 [hep-ph]} \BibitemShut {NoStop}%
\bibitem [{\citenamefont {Celiberto}\ \emph {et~al.}(2021{\natexlab{a}})\citenamefont {Celiberto}, \citenamefont {Fucilla}, \citenamefont {Ivanov},\ and\ \citenamefont {Papa}}]{Celiberto:2021dzy}%
  \BibitemOpen
  \bibfield  {author} {\bibinfo {author} {\bibfnamefont {F.~G.}\ \bibnamefont {Celiberto}}, \bibinfo {author} {\bibfnamefont {M.}~\bibnamefont {Fucilla}}, \bibinfo {author} {\bibfnamefont {D.~{\relax Yu}.}\ \bibnamefont {Ivanov}}, \ and\ \bibinfo {author} {\bibfnamefont {A.}~\bibnamefont {Papa}},\ }\href {\doibase 10.1140/epjc/s10052-021-09448-3} {\bibfield  {journal} {\bibinfo  {journal} {Eur. Phys. J. C}\ }\textbf {\bibinfo {volume} {81}},\ \bibinfo {pages} {780} (\bibinfo {year} {2021}{\natexlab{a}})},\ \Eprint {http://arxiv.org/abs/2105.06432} {arXiv:2105.06432 [hep-ph]} \BibitemShut {NoStop}%
\bibitem [{\citenamefont {Celiberto}\ \emph {et~al.}(2021{\natexlab{b}})\citenamefont {Celiberto}, \citenamefont {Fucilla}, \citenamefont {Ivanov}, \citenamefont {Mohammed},\ and\ \citenamefont {Papa}}]{Celiberto:2021fdp}%
  \BibitemOpen
  \bibfield  {author} {\bibinfo {author} {\bibfnamefont {F.~G.}\ \bibnamefont {Celiberto}}, \bibinfo {author} {\bibfnamefont {M.}~\bibnamefont {Fucilla}}, \bibinfo {author} {\bibfnamefont {D.~{\relax Yu}.}\ \bibnamefont {Ivanov}}, \bibinfo {author} {\bibfnamefont {M.~M.~A.}\ \bibnamefont {Mohammed}}, \ and\ \bibinfo {author} {\bibfnamefont {A.}~\bibnamefont {Papa}},\ }\href {\doibase 10.1103/PhysRevD.104.114007} {\bibfield  {journal} {\bibinfo  {journal} {Phys. Rev. D}\ }\textbf {\bibinfo {volume} {104}},\ \bibinfo {pages} {114007} (\bibinfo {year} {2021}{\natexlab{b}})},\ \Eprint {http://arxiv.org/abs/2109.11875} {arXiv:2109.11875 [hep-ph]} \BibitemShut {NoStop}%
\bibitem [{\citenamefont {Fadin}\ \emph {et~al.}(1975)\citenamefont {Fadin}, \citenamefont {Kuraev},\ and\ \citenamefont {Lipatov}}]{Fadin:1975cb}%
  \BibitemOpen
  \bibfield  {author} {\bibinfo {author} {\bibfnamefont {V.~S.}\ \bibnamefont {Fadin}}, \bibinfo {author} {\bibfnamefont {E.}~\bibnamefont {Kuraev}}, \ and\ \bibinfo {author} {\bibfnamefont {L.}~\bibnamefont {Lipatov}},\ }\href {\doibase 10.1016/0370-2693(75)90524-9} {\bibfield  {journal} {\bibinfo  {journal} {Phys. Lett. B}\ }\textbf {\bibinfo {volume} {60}},\ \bibinfo {pages} {50} (\bibinfo {year} {1975})}\BibitemShut {NoStop}%
\bibitem [{\citenamefont {Balitsky}\ and\ \citenamefont {Lipatov}(1978)}]{Balitsky:1978ic}%
  \BibitemOpen
  \bibfield  {author} {\bibinfo {author} {\bibfnamefont {I.}~\bibnamefont {Balitsky}}\ and\ \bibinfo {author} {\bibfnamefont {L.}~\bibnamefont {Lipatov}},\ }\href@noop {} {\bibfield  {journal} {\bibinfo  {journal} {Sov.\ J.\ Nucl.\ Phys.}\ }\textbf {\bibinfo {volume} {28}},\ \bibinfo {pages} {822} (\bibinfo {year} {1978})}\BibitemShut {NoStop}%
\bibitem [{\citenamefont {Colferai}\ and\ \citenamefont {Niccoli}(2015)}]{Colferai:2015zfa}%
  \BibitemOpen
  \bibfield  {author} {\bibinfo {author} {\bibfnamefont {D.}~\bibnamefont {Colferai}}\ and\ \bibinfo {author} {\bibfnamefont {A.}~\bibnamefont {Niccoli}},\ }\href {\doibase 10.1007/JHEP04(2015)071} {\bibfield  {journal} {\bibinfo  {journal} {JHEP}\ }\textbf {\bibinfo {volume} {04}},\ \bibinfo {pages} {071} (\bibinfo {year} {2015})},\ \Eprint {http://arxiv.org/abs/1501.07442} {arXiv:1501.07442 [hep-ph]} \BibitemShut {NoStop}%
\bibitem [{\citenamefont {Boussarie}\ \emph {et~al.}(2018)\citenamefont {Boussarie}, \citenamefont {Duclou\'e}, \citenamefont {Szymanowski},\ and\ \citenamefont {Wallon}}]{Boussarie:2017oae}%
  \BibitemOpen
  \bibfield  {author} {\bibinfo {author} {\bibfnamefont {R.}~\bibnamefont {Boussarie}}, \bibinfo {author} {\bibfnamefont {B.}~\bibnamefont {Duclou\'e}}, \bibinfo {author} {\bibfnamefont {L.}~\bibnamefont {Szymanowski}}, \ and\ \bibinfo {author} {\bibfnamefont {S.}~\bibnamefont {Wallon}},\ }\href {\doibase 10.1103/PhysRevD.97.014008} {\bibfield  {journal} {\bibinfo  {journal} {Phys. Rev. D}\ }\textbf {\bibinfo {volume} {97}},\ \bibinfo {pages} {014008} (\bibinfo {year} {2018})},\ \Eprint {http://arxiv.org/abs/1709.01380} {arXiv:1709.01380 [hep-ph]} \BibitemShut {NoStop}%
\bibitem [{\citenamefont {Celiberto}\ \emph {et~al.}(2015)\citenamefont {Celiberto}, \citenamefont {Ivanov}, \citenamefont {Murdaca},\ and\ \citenamefont {Papa}}]{Celiberto:2015yba}%
  \BibitemOpen
  \bibfield  {author} {\bibinfo {author} {\bibfnamefont {F.~G.}\ \bibnamefont {Celiberto}}, \bibinfo {author} {\bibfnamefont {D.~{\relax Yu}.}\ \bibnamefont {Ivanov}}, \bibinfo {author} {\bibfnamefont {B.}~\bibnamefont {Murdaca}}, \ and\ \bibinfo {author} {\bibfnamefont {A.}~\bibnamefont {Papa}},\ }\href {\doibase 10.1140/epjc/s10052-015-3522-6} {\bibfield  {journal} {\bibinfo  {journal} {Eur. Phys. J. C}\ }\textbf {\bibinfo {volume} {75}},\ \bibinfo {pages} {292} (\bibinfo {year} {2015})},\ \Eprint {http://arxiv.org/abs/1504.08233} {arXiv:1504.08233 [hep-ph]} \BibitemShut {NoStop}%
\bibitem [{\citenamefont {Ivanov}\ and\ \citenamefont {Papa}(2012)}]{Ivanov:2012iv}%
  \BibitemOpen
  \bibfield  {author} {\bibinfo {author} {\bibfnamefont {D.~{\relax Yu}.}\ \bibnamefont {Ivanov}}\ and\ \bibinfo {author} {\bibfnamefont {A.}~\bibnamefont {Papa}},\ }\href {\doibase 10.1007/JHEP07(2012)045} {\bibfield  {journal} {\bibinfo  {journal} {JHEP}\ }\textbf {\bibinfo {volume} {07}},\ \bibinfo {pages} {045} (\bibinfo {year} {2012})},\ \Eprint {http://arxiv.org/abs/1205.6068} {arXiv:1205.6068 [hep-ph]} \BibitemShut {NoStop}%
\bibitem [{\citenamefont {Bolognino}\ \emph {et~al.}(2021)\citenamefont {Bolognino}, \citenamefont {Celiberto}, \citenamefont {Fucilla}, \citenamefont {Ivanov},\ and\ \citenamefont {Papa}}]{Bolognino:2021mrc}%
  \BibitemOpen
  \bibfield  {author} {\bibinfo {author} {\bibfnamefont {A.~D.}\ \bibnamefont {Bolognino}}, \bibinfo {author} {\bibfnamefont {F.~G.}\ \bibnamefont {Celiberto}}, \bibinfo {author} {\bibfnamefont {M.}~\bibnamefont {Fucilla}}, \bibinfo {author} {\bibfnamefont {D.~{\relax Yu}.}\ \bibnamefont {Ivanov}}, \ and\ \bibinfo {author} {\bibfnamefont {A.}~\bibnamefont {Papa}},\ }\href {\doibase 10.1103/PhysRevD.103.094004} {\bibfield  {journal} {\bibinfo  {journal} {Phys. Rev. D}\ }\textbf {\bibinfo {volume} {103}},\ \bibinfo {pages} {094004} (\bibinfo {year} {2021})},\ \Eprint {http://arxiv.org/abs/2103.07396} {arXiv:2103.07396 [hep-ph]} \BibitemShut {NoStop}%
\bibitem [{\citenamefont {Celiberto}\ \emph {et~al.}(2018)\citenamefont {Celiberto}, \citenamefont {Gordo~G\'omez},\ and\ \citenamefont {Sabio~Vera}}]{Celiberto:2018muu}%
  \BibitemOpen
  \bibfield  {author} {\bibinfo {author} {\bibfnamefont {F.~G.}\ \bibnamefont {Celiberto}}, \bibinfo {author} {\bibfnamefont {D.}~\bibnamefont {Gordo~G\'omez}}, \ and\ \bibinfo {author} {\bibfnamefont {A.}~\bibnamefont {Sabio~Vera}},\ }\href {\doibase 10.1016/j.physletb.2018.09.045} {\bibfield  {journal} {\bibinfo  {journal} {Phys. Lett.}\ }\textbf {\bibinfo {volume} {B786}},\ \bibinfo {pages} {201} (\bibinfo {year} {2018})},\ \Eprint {http://arxiv.org/abs/1808.09511} {arXiv:1808.09511 [hep-ph]} \BibitemShut {NoStop}%
%%CITATION = ARXIV:1808.09511;%%
\bibitem [{\citenamefont {Bolognino}\ \emph {et~al.}(2018)\citenamefont {Bolognino}, \citenamefont {Celiberto}, \citenamefont {Ivanov},\ and\ \citenamefont {Papa}}]{Bolognino:2018rhb}%
  \BibitemOpen
  \bibfield  {author} {\bibinfo {author} {\bibfnamefont {A.~D.}\ \bibnamefont {Bolognino}}, \bibinfo {author} {\bibfnamefont {F.~G.}\ \bibnamefont {Celiberto}}, \bibinfo {author} {\bibfnamefont {D.~{\relax Yu}.}\ \bibnamefont {Ivanov}}, \ and\ \bibinfo {author} {\bibfnamefont {A.}~\bibnamefont {Papa}},\ }\href {\doibase 10.1140/epjc/s10052-018-6493-6} {\bibfield  {journal} {\bibinfo  {journal} {Eur. Phys. J.}\ }\textbf {\bibinfo {volume} {C78}},\ \bibinfo {pages} {1023} (\bibinfo {year} {2018})},\ \Eprint {http://arxiv.org/abs/1808.02395} {arXiv:1808.02395 [hep-ph]} \BibitemShut {NoStop}%
%%CITATION = ARXIV:1808.02395;%%
\bibitem [{\citenamefont {Ball}\ \emph {et~al.}(2018)\citenamefont {Ball}, \citenamefont {Bertone}, \citenamefont {Bonvini}, \citenamefont {Marzani}, \citenamefont {Rojo},\ and\ \citenamefont {Rottoli}}]{Ball:2017otu}%
  \BibitemOpen
  \bibfield  {author} {\bibinfo {author} {\bibfnamefont {R.~D.}\ \bibnamefont {Ball}}, \bibinfo {author} {\bibfnamefont {V.}~\bibnamefont {Bertone}}, \bibinfo {author} {\bibfnamefont {M.}~\bibnamefont {Bonvini}}, \bibinfo {author} {\bibfnamefont {S.}~\bibnamefont {Marzani}}, \bibinfo {author} {\bibfnamefont {J.}~\bibnamefont {Rojo}}, \ and\ \bibinfo {author} {\bibfnamefont {L.}~\bibnamefont {Rottoli}},\ }\href {\doibase 10.1140/epjc/s10052-018-5774-4} {\bibfield  {journal} {\bibinfo  {journal} {Eur. Phys. J.}\ }\textbf {\bibinfo {volume} {C78}},\ \bibinfo {pages} {321} (\bibinfo {year} {2018})},\ \Eprint {http://arxiv.org/abs/1710.05935} {arXiv:1710.05935 [hep-ph]} \BibitemShut {NoStop}%
%%CITATION = ARXIV:1710.05935;%%
\bibitem [{\citenamefont {Bacchetta}\ \emph {et~al.}(2020)\citenamefont {Bacchetta}, \citenamefont {Celiberto}, \citenamefont {Radici},\ and\ \citenamefont {Taels}}]{Bacchetta:2020vty}%
  \BibitemOpen
  \bibfield  {author} {\bibinfo {author} {\bibfnamefont {A.}~\bibnamefont {Bacchetta}}, \bibinfo {author} {\bibfnamefont {F.~G.}\ \bibnamefont {Celiberto}}, \bibinfo {author} {\bibfnamefont {M.}~\bibnamefont {Radici}}, \ and\ \bibinfo {author} {\bibfnamefont {P.}~\bibnamefont {Taels}},\ }\href {\doibase 10.1140/epjc/s10052-020-8327-6} {\bibfield  {journal} {\bibinfo  {journal} {Eur. Phys. J. C}\ }\textbf {\bibinfo {volume} {80}},\ \bibinfo {pages} {733} (\bibinfo {year} {2020})},\ \Eprint {http://arxiv.org/abs/2005.02288} {arXiv:2005.02288 [hep-ph]} \BibitemShut {NoStop}%
\bibitem [{\citenamefont {Bacchetta}\ \emph {et~al.}(2024)\citenamefont {Bacchetta}, \citenamefont {Celiberto},\ and\ \citenamefont {Radici}}]{Bacchetta:2024fci}%
  \BibitemOpen
  \bibfield  {author} {\bibinfo {author} {\bibfnamefont {A.}~\bibnamefont {Bacchetta}}, \bibinfo {author} {\bibfnamefont {F.~G.}\ \bibnamefont {Celiberto}}, \ and\ \bibinfo {author} {\bibfnamefont {M.}~\bibnamefont {Radici}},\ }\href {\doibase 10.1140/epjc/s10052-024-12927-y} {\bibfield  {journal} {\bibinfo  {journal} {Eur. Phys. J. C}\ }\textbf {\bibinfo {volume} {84}},\ \bibinfo {pages} {576} (\bibinfo {year} {2024})},\ \Eprint {http://arxiv.org/abs/2402.17556} {arXiv:2402.17556 [hep-ph]} \BibitemShut {NoStop}%
\bibitem [{\citenamefont {Celiberto}(2017)}]{Celiberto:2017ius}%
  \BibitemOpen
  \bibfield  {author} {\bibinfo {author} {\bibfnamefont {F.~G.}\ \bibnamefont {Celiberto}},\ }\emph {\bibinfo {title} {{High-energy resummation in semi-hard processes at the LHC}}},\ \href@noop {} {Ph.D. thesis},\ \bibinfo  {school} {Universit\`a della Calabria and INFN-Cosenza} (\bibinfo {year} {2017}),\ \Eprint {http://arxiv.org/abs/1707.04315} {arXiv:1707.04315 [hep-ph]} \BibitemShut {NoStop}%
\bibitem [{\citenamefont {Khachatryan}\ \emph {et~al.}(2016)\citenamefont {Khachatryan} \emph {et~al.}}]{Khachatryan:2016udy}%
  \BibitemOpen
  \bibfield  {author} {\bibinfo {author} {\bibfnamefont {V.}~\bibnamefont {Khachatryan}} \emph {et~al.} (\bibinfo {collaboration} {CMS}),\ }\href {\doibase 10.1007/JHEP08(2016)139} {\bibfield  {journal} {\bibinfo  {journal} {JHEP}\ }\textbf {\bibinfo {volume} {08}},\ \bibinfo {pages} {139} (\bibinfo {year} {2016})},\ \Eprint {http://arxiv.org/abs/1601.06713} {arXiv:1601.06713 [hep-ex]} \BibitemShut {NoStop}%
\bibitem [{\citenamefont {Kassabov}\ \emph {et~al.}(2023)\citenamefont {Kassabov}, \citenamefont {Ubiali},\ and\ \citenamefont {Voisey}}]{Kassabov:2022orn}%
  \BibitemOpen
  \bibfield  {author} {\bibinfo {author} {\bibfnamefont {Z.}~\bibnamefont {Kassabov}}, \bibinfo {author} {\bibfnamefont {M.}~\bibnamefont {Ubiali}}, \ and\ \bibinfo {author} {\bibfnamefont {C.}~\bibnamefont {Voisey}},\ }\href {\doibase 10.1007/JHEP03(2023)148} {\bibfield  {journal} {\bibinfo  {journal} {JHEP}\ }\textbf {\bibinfo {volume} {03}},\ \bibinfo {pages} {148} (\bibinfo {year} {2023})},\ \Eprint {http://arxiv.org/abs/2207.07616} {arXiv:2207.07616 [hep-ph]} \BibitemShut {NoStop}%
\bibitem [{\citenamefont {Harland-Lang}\ and\ \citenamefont {Thorne}(2019)}]{Harland-Lang:2018bxd}%
  \BibitemOpen
  \bibfield  {author} {\bibinfo {author} {\bibfnamefont {L.~A.}\ \bibnamefont {Harland-Lang}}\ and\ \bibinfo {author} {\bibfnamefont {R.~S.}\ \bibnamefont {Thorne}},\ }\href {\doibase 10.1140/epjc/s10052-019-6731-6} {\bibfield  {journal} {\bibinfo  {journal} {Eur. Phys. J. C}\ }\textbf {\bibinfo {volume} {79}},\ \bibinfo {pages} {225} (\bibinfo {year} {2019})},\ \Eprint {http://arxiv.org/abs/1811.08434} {arXiv:1811.08434 [hep-ph]} \BibitemShut {NoStop}%
\bibitem [{\citenamefont {Ball}\ \emph {et~al.}(2024{\natexlab{a}})\citenamefont {Ball} \emph {et~al.}}]{NNPDF:2024dpb}%
  \BibitemOpen
  \bibfield  {author} {\bibinfo {author} {\bibfnamefont {R.~D.}\ \bibnamefont {Ball}} \emph {et~al.} (\bibinfo {collaboration} {NNPDF}),\ }\href {\doibase 10.1140/epjc/s10052-024-12772-z} {\bibfield  {journal} {\bibinfo  {journal} {Eur. Phys. J. C}\ }\textbf {\bibinfo {volume} {84}},\ \bibinfo {pages} {517} (\bibinfo {year} {2024}{\natexlab{a}})},\ \Eprint {http://arxiv.org/abs/2401.10319} {arXiv:2401.10319 [hep-ph]} \BibitemShut {NoStop}%
\bibitem [{\citenamefont {Ball}\ \emph {et~al.}(2024{\natexlab{b}})\citenamefont {Ball}, \citenamefont {Candido}, \citenamefont {Cruz-Martinez}, \citenamefont {Forte}, \citenamefont {Giani}, \citenamefont {Hekhorn}, \citenamefont {Magni}, \citenamefont {Nocera}, \citenamefont {Rojo},\ and\ \citenamefont {Stegeman}}]{NNPDF:2023tyk}%
  \BibitemOpen
  \bibfield  {author} {\bibinfo {author} {\bibfnamefont {R.~D.}\ \bibnamefont {Ball}}, \bibinfo {author} {\bibfnamefont {A.}~\bibnamefont {Candido}}, \bibinfo {author} {\bibfnamefont {J.}~\bibnamefont {Cruz-Martinez}}, \bibinfo {author} {\bibfnamefont {S.}~\bibnamefont {Forte}}, \bibinfo {author} {\bibfnamefont {T.}~\bibnamefont {Giani}}, \bibinfo {author} {\bibfnamefont {F.}~\bibnamefont {Hekhorn}}, \bibinfo {author} {\bibfnamefont {G.}~\bibnamefont {Magni}}, \bibinfo {author} {\bibfnamefont {E.~R.}\ \bibnamefont {Nocera}}, \bibinfo {author} {\bibfnamefont {J.}~\bibnamefont {Rojo}}, \ and\ \bibinfo {author} {\bibfnamefont {R.}~\bibnamefont {Stegeman}} (\bibinfo {collaboration} {NNPDF}),\ }\href {\doibase 10.1103/PhysRevD.109.L091501} {\bibfield  {journal} {\bibinfo  {journal} {Phys. Rev. D}\ }\textbf {\bibinfo {volume} {109}},\ \bibinfo {pages} {L091501} (\bibinfo {year} {2024}{\natexlab{b}})},\ \Eprint {http://arxiv.org/abs/2311.00743} {arXiv:2311.00743 [hep-ph]} \BibitemShut {NoStop}%
\bibitem [{\citenamefont {Ball}\ \emph {et~al.}(2022)\citenamefont {Ball}, \citenamefont {Candido}, \citenamefont {Cruz-Martinez}, \citenamefont {Forte}, \citenamefont {Giani}, \citenamefont {Hekhorn}, \citenamefont {Kudashkin}, \citenamefont {Magni},\ and\ \citenamefont {Rojo}}]{Ball:2022qks}%
  \BibitemOpen
  \bibfield  {author} {\bibinfo {author} {\bibfnamefont {R.~D.}\ \bibnamefont {Ball}}, \bibinfo {author} {\bibfnamefont {A.}~\bibnamefont {Candido}}, \bibinfo {author} {\bibfnamefont {J.}~\bibnamefont {Cruz-Martinez}}, \bibinfo {author} {\bibfnamefont {S.}~\bibnamefont {Forte}}, \bibinfo {author} {\bibfnamefont {T.}~\bibnamefont {Giani}}, \bibinfo {author} {\bibfnamefont {F.}~\bibnamefont {Hekhorn}}, \bibinfo {author} {\bibfnamefont {K.}~\bibnamefont {Kudashkin}}, \bibinfo {author} {\bibfnamefont {G.}~\bibnamefont {Magni}}, \ and\ \bibinfo {author} {\bibfnamefont {J.}~\bibnamefont {Rojo}} (\bibinfo {collaboration} {NNPDF}),\ }\href {\doibase 10.1038/s41586-022-04998-2} {\bibfield  {journal} {\bibinfo  {journal} {Nature}\ }\textbf {\bibinfo {volume} {608}},\ \bibinfo {pages} {483} (\bibinfo {year} {2022})},\ \Eprint {http://arxiv.org/abs/2208.08372} {arXiv:2208.08372 [hep-ph]} \BibitemShut {NoStop}%
\bibitem [{\citenamefont {Guzzi}\ \emph {et~al.}(2023)\citenamefont {Guzzi}, \citenamefont {Hobbs}, \citenamefont {Xie}, \citenamefont {Huston}, \citenamefont {Nadolsky},\ and\ \citenamefont {Yuan}}]{Guzzi:2022rca}%
  \BibitemOpen
  \bibfield  {author} {\bibinfo {author} {\bibfnamefont {M.}~\bibnamefont {Guzzi}}, \bibinfo {author} {\bibfnamefont {T.~J.}\ \bibnamefont {Hobbs}}, \bibinfo {author} {\bibfnamefont {K.}~\bibnamefont {Xie}}, \bibinfo {author} {\bibfnamefont {J.}~\bibnamefont {Huston}}, \bibinfo {author} {\bibfnamefont {P.}~\bibnamefont {Nadolsky}}, \ and\ \bibinfo {author} {\bibfnamefont {C.~P.}\ \bibnamefont {Yuan}},\ }\href {\doibase 10.1016/j.physletb.2023.137975} {\bibfield  {journal} {\bibinfo  {journal} {Phys. Lett. B}\ }\textbf {\bibinfo {volume} {843}},\ \bibinfo {pages} {137975} (\bibinfo {year} {2023})},\ \Eprint {http://arxiv.org/abs/2211.01387} {arXiv:2211.01387 [hep-ph]} \BibitemShut {NoStop}%
\bibitem [{\citenamefont {Vogt}(2024)}]{Vogt:2024fky}%
  \BibitemOpen
  \bibfield  {author} {\bibinfo {author} {\bibfnamefont {R.}~\bibnamefont {Vogt}},\ }\href {\doibase 10.1103/PhysRevD.110.074036} {\bibfield  {journal} {\bibinfo  {journal} {Phys. Rev. D}\ }\textbf {\bibinfo {volume} {110}},\ \bibinfo {pages} {074036} (\bibinfo {year} {2024})},\ \Eprint {http://arxiv.org/abs/2405.09018} {arXiv:2405.09018 [hep-ph]} \BibitemShut {NoStop}%
\end{thebibliography}%
